\documentclass[%
 aip,
 amsmath,amssymb,
 reprint,%
]{revtex4-1}

\usepackage{graphicx}
\usepackage{dcolumn}
\usepackage{bm}

\usepackage{mathptmx}
\usepackage{etoolbox}

\usepackage{dsfont}
\usepackage[mathscr]{euscript}
\usepackage{amssymb,amsmath}
\usepackage{amsthm}
\usepackage{epstopdf}
\usepackage{color,xcolor}
\usepackage[english]{babel}
\usepackage{enumerate}
\usepackage{bbold}
\usepackage{leftidx}
\usepackage{mathrsfs}
\usepackage{mathtools}
\usepackage{array}
\usepackage{hyperref}

\definecolor{Blue}{rgb}{0.4765625, 0.4765625, 0.99609375}
\definecolor{Red}{rgb}{1, 0.25, 0.25}
\usepackage{tikz}
\usepackage{tikz-cd}
			
\usetikzlibrary{shapes.geometric,arrows.meta,positioning} 

\usetikzlibrary{fit,matrix,calc,backgrounds}
\usetikzlibrary{shapes.misc,decorations.pathreplacing,topaths}
\usetikzlibrary{graphs,scopes,chains}

\newtheorem{theorem}{Theorem}[section]
\newtheorem{lemma}{Lemma}[section]

\newcommand{\p}[1]{\left( #1 \right)}

\newcommand{\cb}[1]{\left\{ #1 \right\}}

\makeatletter
\def\@email#1#2{%
 \endgroup
 \patchcmd{\titleblock@produce}
  {\frontmatter@RRAPformat}
  {\frontmatter@RRAPformat{\produce@RRAP{*#1\href{mailto:#2}{#2}}}\frontmatter@RRAPformat}
  {}{}
}%
\makeatother

\begin{document}

\tikzset{   
            every picture/.style={thick},
            roundnode/.style={circle, draw=blue!100, fill=blue!5, thick, minimum size=10mm},
            specialnode/.style={circle, draw=red!100, fill=red!5, thick, minimum size=3mm},
            sigmanode/.style={circle, draw=orange!100, fill=orange!5,  minimum size=3mm, thick},
            spin/.style={circle, draw=black, inner sep=0mm, fill=black, ultra thick, minimum size=3mm},
            squarenode/.style={rectangle,minimum size=3mm,rounded corners=2mm,thick,draw=black!50,top color=white,bottom color=black!20},
            unitednode/.style={rectangle,minimum width=40mm, rounded corners=5mm,draw=red!100, fill=red!5, thick},
            ampersand replacement =\&}

\newcommand{\mrabc}{
\begin{figure*}
\centering
\begin{tikzpicture}
\matrix (rabc) [matrix of nodes, row sep=5mm,column sep=5mm, nodes in empty cells] {
$\dots$ \& |[spin]|\& |[spin]|\& |[spin]|\&|[spin]|\&|[spin]|\&|[spin]|\&|[spin]|\& |[spin]|\&|[spin]|\& $\dots$\\
$\dots$ \& |[roundnode]| $M$\&|[roundnode]| $M$\&|[roundnode]| $M$\&|[roundnode]| $M$\&|[roundnode]| $M$\&|[roundnode]| $M$\&|[roundnode]| $M$\&|[roundnode]| $M$\&|[roundnode]| $M$\& $\dots$ \\
\& \&|[roundnode]| $M$ \&|[roundnode]| $M$ \&|[roundnode]| $M$ \&|[roundnode]| $M$ \&|[roundnode]| $M$ \&|[roundnode]| $M$ \&|[roundnode]| $M$ \& \& \\
\& \&|[roundnode]| $M^\dagger$\&|[roundnode]| $M^\dagger$ \&|[roundnode]| $M^\dagger$ \&|[roundnode]| $M^\dagger$ \&|[roundnode]| $M^\dagger$ \&|[roundnode]| $M^\dagger$ \&|[roundnode]| $M^\dagger$ \& \& \\[3mm,between borders]

\& \&|[roundnode]| $M$\&|[roundnode]| $M$ \& \& $\tilde{M}_B$ \& \&|[roundnode]| $M$\&|[roundnode]| $M$\& \&\\
\& \&|[roundnode]| $M^\dagger$\&|[roundnode]| $M^\dagger$ \& \& $\tilde{M}_B^\dagger$ \& \&|[roundnode]| $M^\dagger$ \& |[roundnode]| $M^\dagger$\& \&\\
};

\begin{scope}[on background layer]
\node (bup) [unitednode,fit=(rabc-5-5)(rabc-5-6)(rabc-5-7),minimum height=10mm, behind path]{};
\node (bdn) [unitednode,fit=(rabc-6-5)(rabc-6-6)(rabc-6-7),minimum height=10mm, behind path]{};
\end{scope}

\path ($0.5*(rabc-3-3.west)+0.5*(rabc-4-3.west)+(-6mm,0)$) node (sigma)  [sigmanode] {$\sigma$};

\path ($0.5*(rabc-5-3.west)+0.5*(rabc-6-3.west)+(-6mm,0)$) node (sigma2)  [sigmanode] {$\sigma$};

\foreach \x [evaluate=\x as \xprev using \x-1] in {4,5,...,9} {
                            \draw   (rabc-3-\x.north) -- ++(0,3mm)
                                    (rabc-4-\x.south) -- ++(0,-3mm)
                                    (rabc-3-\x) -- (rabc-3-\xprev)
                                    (rabc-4-\x) -- (rabc-4-\xprev);}
                                    
\foreach \x [evaluate=\x as \xprev using \x-1] in {3,4,8,9} {
                            \draw   (rabc-5-\x.north) -- ++(0,3mm)
                                    (rabc-6-\x.south) -- ++(0,-3mm);}

\foreach \x in {5,6} {
                        \draw   (rabc-\x-3) -- (rabc-\x-4)
                                (rabc-\x-8) -- (rabc-\x-9)
                                (rabc-\x-3) -| (sigma2);}

\draw[-] (rabc-3-3.north) -- ++(0,3mm)
         (rabc-4-3.south) -- ++(0,-3mm)
         (rabc-3-9.east) -- ++ (4mm,0) |- (rabc-4-9.east);
         
\draw   (rabc-5-9.east) -- ++ (4mm,0) |- (rabc-6-9.east)
        (rabc-5-4) -- (bup) -- (rabc-5-8)
        (rabc-6-4) -- (bdn) -- (rabc-6-8);
        
\foreach \x in {5,6,7} {
                        \draw (rabc-5-\x |- bup.north) -- ++(0,3mm)
                        (rabc-6-\x |- bdn.south) -- ++(0,-3mm);}
         
\foreach \x in {3,4} {\draw (sigma)|-(rabc-\x-3.west);}

\foreach \x [evaluate=\x as \xprev using \x-1] in {3,4,...,10} {
    \draw   (rabc-2-\x) -- (rabc-2-\xprev)
            (rabc-2-\x.north) -- ++(0,3mm);
    }
\draw   (rabc-2-2.north)-- ++(0,3mm)
        (rabc-2-2.west)-- ++(-3mm,0)
        (rabc-2-10.east)-- ++(3mm,0);

\path ($(rabc-1-3)+(-6.5mm,4mm)$) coordinate (c1)
      ($(rabc-1-4)+(6.5mm,4mm)$) coordinate (c2)
      ($(rabc-1-5)+(-6.5mm,4mm)$) coordinate (c3)
      ($(rabc-1-7)+(6.5mm,4mm)$) coordinate (c4)
      ($(rabc-1-8)+(-6.5mm,4mm)$) coordinate (c5)
      ($(rabc-1-9)+(6.5mm,4mm)$) coordinate (c6);

\foreach \x in {1,2,...,6} {\draw[-,dashed] (c\x) -- ++(0,-90mm);}

\path ($(rabc-1-1)+(-7mm,7mm)$) node {$\mathrm{(a)}$}
      ($(rabc-2-1)+(-7mm,7mm)$) node (b) {$\mathrm{(b)}$}
      ($(b)+(0,-20mm)$) node (c) {$\mathrm{(c)}$}
      ($(c)+(0,-30mm)$) node {$\mathrm{(d)}$};

\draw [decorate,decoration={brace,amplitude=3mm},thick] ($(c1)+(0,2mm)$) -- ($(c2)+(0,2mm)$) node [black,midway,above,yshift=3mm] {$A$};
\draw [decorate,decoration={brace,amplitude=3mm},thick] ($(c3)+(0,2mm)$) -- ($(c4)+(0,2mm)$) node [black,midway,above,yshift=3mm] {$B$};
\draw [decorate,decoration={brace,amplitude=3mm},thick] ($(c5)+(0,2mm)$) -- ($(c6)+(0,2mm)$) node [black,midway,above,yshift=3mm] {$C$};

\end{tikzpicture}
\caption{Diagrammatic depictions of: (a) an infinite spin chain with an example of selection of regions $A$, $B$, and $C$, considered in this paper; (b) an iuMPS of the spin chain, generated by a tensor $M$ with $i,j=1,2,...,d_M$ and $s=1,2,...,d_s$; (c) the reduced density operator $\rho_{ABC}$ of the iuMPS in (a); the QMC $\tilde{\rho}_{ABC}$ approximating $\rho_{ABC}$, defined via the tensor $\tilde{M}_{B}$.}\label{dg:red_dens_op}
\end{figure*}
}

\newcommand{\mbtensor}{
\begin{figure}[t]
\begin{tikzpicture}
\node (mb) [unitednode,minimum height=10mm, minimum width=36mm]{$\tilde{M}_B$};

\node at ($(mb.north east)+(-17mm,2mm)$) {$\dots$};
\draw   (mb.west) to [edge label'=$i$,near end] ++ (-3mm,0)
        (mb.east) to [edge label=$j$,near end] ++ (3mm,0)
        ($(mb.north west)+(6mm,0)$) to [edge label=$s_1$,at end] ++ (0,3mm)
        ($(mb.north west)+(13mm,0)$) to [edge label=$s_2$,at end] ++ (0,3mm)
        ($(mb.north east)+(-6mm,0)$) to [edge label=$s_{|B|}$,at end] ++ (0,3mm);
        
\node at ($(mb.east)+(4mm,0)$) [anchor=west] {$=:\langle j|\tilde{M}^{s_1s_2\dots s_{|B|}}_B |i\rangle$};
\end{tikzpicture}
\caption{The tensor $\tilde{M}_B$ of the dimension 
$d_M\times d_s^{|B|}\times d_M$, and its representation as the matrix elements of the collection of $d_M\times d_M$ matrices $\tilde{M}^{s_1s_2...s_{|B|}}_B$, where $s_1,s_2,...,s_{|B|}=1,2,...,d_s$. 
}\label{dg:mb_tensor}
\end{figure}
}

\newcommand{\normalize}{
\begin{figure*}[t]
\centering
\begin{tikzpicture}

\node (up1) [roundnode] {$M$};
\path ($(up1)+(0,-20mm)$) node (dn1) [roundnode] {$M^\dagger$}
      ($0.5*(up1.west)+0.5*(dn1.west)+(-3mm,0)$) node (sigma)  [sigmanode] {$\sigma$}
      ($0.5*(up1.east)+0.5*(dn1.east)+(6mm,0)$) node (eq1) {{\LARGE $=$}}
      ($(eq1.east)+(1mm,0)$) node (sigma2) [anchor=west,sigmanode] {$\sigma$};

\node (lbla) at ($(up1.north)+(-12mm,0)$) {(b)};

\path ($(up1)-(38mm,0)$) node (up) [roundnode] {$M$}
      ($(up)+(0,-20mm)$) node (dn) [roundnode] {$M^\dagger$}
      ($0.5*(up.east)+0.5*(dn.east)+(6mm,0)$) node (eq) {{\LARGE $=$}}
      ($(eq.east)+(3mm,0)$) coordinate (id);

\path let \p1=($(up)-(up1)$) in node (lblb) at ($(lbla)+(\x1,0)+(0mm,0)$) {(a)};

\draw[-] (sigma)|-(up1)
         (sigma)|-(dn1)
         (up1.east)--++(3mm,0)
         (dn1.east)--++(3mm,0)
         (up1) -- (dn1)
         let \p1 = ($(dn1)-(sigma.south)$) in (sigma2.south) -- ++(0,\y1) -- ++(3mm,0)
         let \p1 = ($(up1)-(sigma.north)$) in (sigma2.north) -- ++(0,\y1) -- ++(3mm,0);

\foreach \x in {up,dn} {\draw[-] (\x.west)--++(-3mm,0);}
\draw[-] (up) -- (dn)
         (up.east) -- ++(3mm,0) |- (dn.east)
         let \p1 = ($(dn)-(id)$) in (id) -- ++(0,\y1) -- ++(-3mm,0)
         let \p1 = ($(up)-(id)$) in (id) -- ++(0,\y1) -- ++(-3mm,0);

\node (trmup) at ($(up1)+(55mm,0)$) [roundnode] {$M$};
\node (trmdn) at ($(trmup)+(0,-20mm)$) [roundnode] {$M^\dagger$};
\node [anchor=east] at ($0.5*(trmup)+0.5*(trmdn)+(-7mm,0)$) {$E_{jj';ii'}$ {\LARGE $=$}};
\draw (trmup)--(trmdn)
      (trmup.east) to [edge label=$j$] ++(3mm,0)                                 
      (trmup.west) to [edge label'=$i$] ++ (-3mm,0)
      (trmdn.east) to [edge label=$j'$] ++(3mm,0)                                 
      (trmdn.west) to [edge label'=$i'$] ++ (-3mm,0);
      
\path let \p1=($(trmup)-(up1)$) in node (lblc) at ($(lbla)+(\x1,0)+(-12mm,0)$) {(c)};      
      
\node (up2) [roundnode] at ($(trmup)+(35mm,0)$) {$M$};
\path ($(up2)+(0,-20mm)$) node (dn2) [roundnode] {$M^\dagger$};
\path ($0.5*(up2)+0.5*(dn2)+(-11mm,0)$) node (xin)  [squarenode,minimum height=30mm] {$X_k$};
\path ($0.5*(up2)+0.5*(dn2)+(10mm,0)$) node (eq) {{\LARGE $=$} $\nu_k$};
\path ($(eq.east)+(3mm,0)$) node (xout)  [squarenode,minimum height=30mm] {$X_k$};

\draw[-] (up2)--(dn2);
\foreach \x in {up2,dn2} {
                        \draw[-] (\x.east)--++(2mm,0);
                        \draw[-] (\x.west) -| (\x -| xin.east);
                        \draw[-] (\x -| xout.east)-- ++(3mm,0);
                        }

\node (lbld) at ($(lblc)+(40mm,0)$) {(d)};
\end{tikzpicture}
\caption{Diagrammatic depictions of: (a) the iuMPS in canonical form, as expressed in (\ref{eqn:condition_on_M}); (b) the density operator $\sigma$, the fixed point of the quantum channel $\mathcal{E}$, defined in (\ref{eqn:e_q_channel}); (c) the transfer matrix $E$ defined in (\ref{eqn:tr_matr}); (d) the (right) eigenvalue equation (\ref{eqn:tr_matr_eigpr}) for $E$.}\label{dg:norm_eigen}
\end{figure*}
}

\newcommand{\gramm}{
\begin{figure*}
\centering
\begin{tikzpicture}
    \matrix (contract) [matrix of nodes, row sep=5mm, column sep=3mm]
    {
    |[roundnode]| $M$ \& |[roundnode]| $M$ \& $\dots$ \& |[roundnode]| $M$ \& |[roundnode]| $M$ \\
    |[roundnode]| $M^\dagger$ \& |[roundnode]| $M^\dagger$ \& $\dots$ \& |[roundnode]| $M^\dagger$ \& |[roundnode]| $M^\dagger$ \\
    };
    
    \foreach \x [evaluate=\x as \xprev using \x-1] in {2,5} {
            \draw   (contract-1-\x) -- (contract-1-\xprev)
                    (contract-2-\x) -- (contract-2-\xprev);
    }
    \foreach \x in {1,2,4,5} {
            \draw (contract-1-\x) -- (contract-2-\x);
    }
    \draw   (contract-1-2.east) -- ++ (3mm,0)
            (contract-2-2.east) -- ++ (3mm,0)
            (contract-1-4.west) -- ++ (-3mm,0)
            (contract-2-4.west) -- ++ (-3mm,0)
            (contract-1-1.west) to [edge label'=$i$,at end] ++(-3mm,0)
            (contract-2-1.west) to [edge label'=$i'$,at end] ++(-3mm,0)
            (contract-1-5.east) to [edge label=$j$,at end] ++(3mm,0)
            (contract-2-5.east) to [edge label=$j'$,at end] ++(3mm,0);
            
    \path ($0.5*(contract-1-5)+0.5*(contract-2-5)+(12mm,0)$) node (eq1) {{\LARGE $=$}}
          ($(eq1)+(10mm,0)$) node (gram) [specialnode] {$G$}
          ($(gram)+(10mm,0)$) node (eq2) {{\LARGE $=$}}
          ($(eq2)+(10mm,0)$) node (diag) [diamond,draw=black!100,inner sep=1mm] {$\Sigma$}
          ($(diag.north)+(0,7mm)$) node (upis) [semicircle,draw=black!100,shape border rotate=180,inner sep=0.7mm,minimum size=5mm] {$W$}
          ($(diag.south)+(0,-7mm)$) node (dnis) [semicircle,draw=black!100,shape border rotate=0,inner sep=0.7mm,minimum size=5mm] {$W^\dagger$};
          
    \draw   (gram.65) to [edge label'=$j$,at end] ++(0,3mm)
            (gram.115) to [edge label=$i$,at end] ++(0,3mm)
            (gram.245) to [edge label'=$i'$,at end] ++(0,-3mm)
            (gram.295) to [edge label=$j'$,at end] ++(0,-3mm);
            
    \draw   (diag) to [edge label'=$l$] (upis)
            ($(upis.chord center)+(-3mm,0)$) to [edge label=$i$,at end] ++ (0,3mm)
            ($(upis.chord center)+(3mm,0)$) to [edge label'=$j$,at end] ++ (0,3mm)
            (diag) to [edge label=$l'$] (dnis)
            ($(dnis.chord center)+(-3mm,0)$) to [edge label'=$i'$,at end] ++ (0,-3mm)
            ($(dnis.chord center)+(3mm,0)$) to [edge label=$j'$,at end] ++ (0,-3mm);
\end{tikzpicture}
\caption{The first equality in the diagram depicts the definition of $G_{ii';jj'}$; the second equality depicts its eigendecomposition.}
\label{dg:numerical_methods_diagram_1}
\end{figure*}
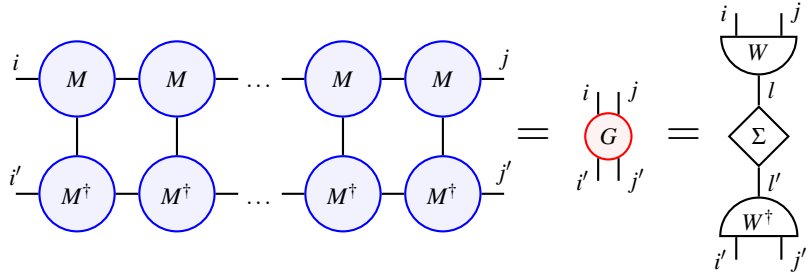
}

\newcommand{\figorthog}{
\begin{figure*}[t]
    \centering
    \begin{tikzpicture}
    \path   node (upinv) [diamond,draw=black!100,inner sep=1mm] {$\Sigma^{-\frac{1}{2}}$}
            ($(upinv.south)+(0,-2mm)$) node (isup) [semicircle,anchor=north,draw=black!100,shape border rotate=0,inner sep=0.7mm,minimum size=5mm] {$W^\dagger$};
    \matrix (contr) at ($(isup.south)+(0,-5mm)$) [matrix of nodes, row sep=5mm, column sep=3mm,anchor=north]
    {
    |[roundnode]| $M$ \& |[roundnode]| $M$ \& $\dots$ \& |[roundnode]| $M$ \& |[roundnode]| $M$ \\
    |[roundnode]| $M^\dagger$ \& |[roundnode]| $M^\dagger$ \& $\dots$ \& |[roundnode]| $M^\dagger$ \& |[roundnode]| $M^\dagger$ \\
    };
    
    \path   ($(contr.south)+(0,-5mm)$) node (isdn) [semicircle,anchor=north,draw=black!100,shape border rotate=180,inner sep=0.7mm,minimum size=5mm] {$W$}
            ($(isdn.south)+(0,-2mm)$) node (dninv) [anchor=north,diamond,draw=black!100,inner sep=1mm] {$\Sigma^{-\frac{1}{2}}$};
            
    \path ($(contr.east)+(7mm,0)$) node (eq2) {{\LARGE $=$}} ($(eq2)+(10mm,0)$) node (diag) [diamond,draw=black!100,inner sep=1mm] {$\Sigma$}
          ($(diag.north)+(0,7mm)$) node (upis) [semicircle,draw=black!100,shape border rotate=180,inner sep=0.7mm,minimum size=5mm] {$W$}
          ($(diag.south)+(0,-7mm)$) node (dnis) [semicircle,draw=black!100,shape border rotate=0,inner sep=0.7mm,minimum size=5mm] {$W^\dagger$}
          ($(upis.north)+(0,21mm)$) node (upinv1) [diamond,draw=black!100,inner sep=1mm] {$\Sigma^{-\frac{1}{2}}$}
          ($(upinv1.south)+(0,-2mm)$) node (isup1) [semicircle,anchor=north,draw=black!100,shape border rotate=0,inner sep=0.7mm,minimum size=5mm] {$W^\dagger$}
          ($(dnis.south)+(0,-5mm)$) node (isdn1) [semicircle,anchor=north,draw=black!100,shape border rotate=180,inner sep=0.7mm,minimum size=5mm] {$W$}
            ($(isdn1.south)+(0,-2mm)$) node (dninv1) [anchor=north,diamond,draw=black!100,inner sep=1mm] {$\Sigma^{-\frac{1}{2}}$};
            
    \foreach \x [evaluate=\x as \xprev using \x-1] in {2,5} {
            \draw   (contr-1-\x) -- (contr-1-\xprev)
                    (contr-2-\x) -- (contr-2-\xprev);
    }
    \foreach \x in {1,2,4,5} {
            \draw (contr-1-\x) -- (contr-2-\x);
    }
    \draw   (contr-1-2.east) -- ++ (3mm,0)
            (contr-2-2.east) -- ++ (3mm,0)
            (contr-1-4.west) -- ++ (-3mm,0)
            (contr-2-4.west) -- ++ (-3mm,0);
            
    \draw   (upinv) -- (isup)
            (dninv) -- (isdn)
            (upinv.north) to [edge label'=$l$] ++ (0,3mm)
            (dninv.south) to [edge label=$l'$] ++ (0,-3mm);
            
    \draw   (contr-1-1.west) -- ++ (-3mm,0)|-($0.5*(contr-1-1.north)+0.5*(isup.south)$)-|($(isup.chord center)+(-3mm,0)$)
            (contr-1-5.east) -- ++ (3mm,0)|-($0.5*(contr-1-5.north)+0.5*(isup.south)$)-|($(isup.chord center)+(3mm,0)$)
            (contr-2-1.west) -- ++ (-3mm,0)|-($0.5*(contr-2-1.south)+0.5*(isdn.north)$)-|($(isdn.chord center)+(-3mm,0)$)
            (contr-2-5.east) -- ++ (3mm,0)|-($0.5*(contr-2-5.south)+0.5*(isdn.north)$)-|($(isdn.chord center)+(3mm,0)$); 
            
    \draw   (upinv1) -- (isup1)
            (dninv1) -- (isdn1)
            (upis) -- (diag) -- (dnis)
            (upinv1.north) to [edge label'=$l$] ++ (0,3mm)
            (dninv1.south) to [edge label=$l'$] ++ (0,-3mm);
            
    \draw   ($(upis.chord center)+(-3mm,0)$) -- ($(isup1.chord center)+(-3mm,0)$)
            ($(upis.chord center)+(3mm,0)$) -- ($(isup1.chord center)+(3mm,0)$)
            ($(dnis.chord center)+(-3mm,0)$) -- ($(isdn1.chord center)+(-3mm,0)$)
            ($(dnis.chord center)+(3mm,0)$) -- ($(isdn1.chord center)+(3mm,0)$);
            
    \path   ($(diag.east)+(5mm,0)$) node (equal) {{\LARGE $=$}}
            ($(equal)+(5mm,0)$) coordinate (mid);

    \draw (mid) to [edge label'=$l$,very near end] ($(mid)+(0,15mm)$) 
          (mid) to [edge label=$l'$,very near end] ($(mid)+(0,-15mm)$);
\end{tikzpicture}
\caption{Verification of the property $P^\dagger P =\mathds{1}_{d_M ^2}$.}
\label{dg:numerical_methods_diagram_2}
\end{figure*}
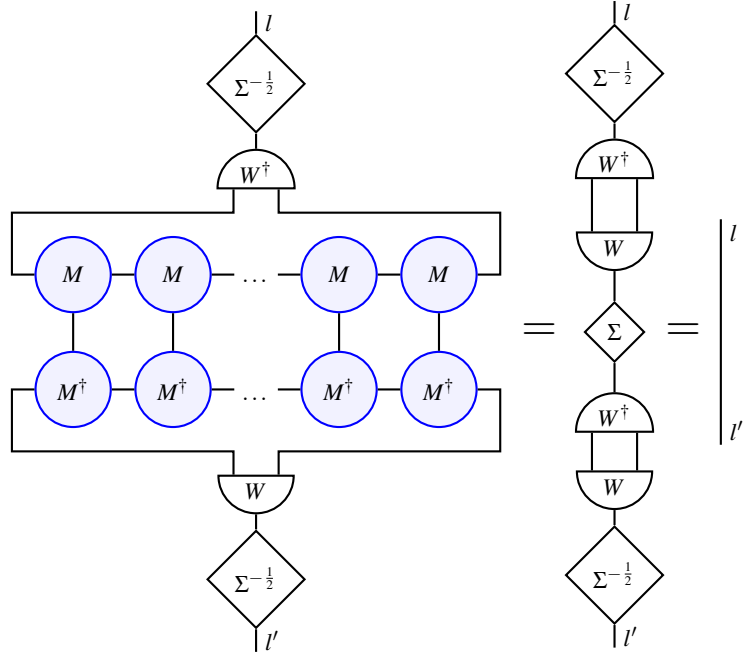
}

\newcommand{\projected}{
\begin{figure}
\centering
\begin{tikzpicture}
    \path   node (sigma) [sigmanode] {$\sigma$}
            ($(sigma.north east)+(10mm,8mm)$) node (upism) [semicircle,draw=black!100,shape border rotate=0,inner sep=0.7mm,minimum size=5mm] {$W^\dagger$}
            node (upsqrt) at ($(upism.north)+(0,2mm)$) [diamond,anchor=south,draw=black!100,inner sep=1mm] {$\Sigma^{\frac{1}{2}}$}
            ($(sigma.south east)+(10mm,-8mm)$) node (dnism) [semicircle,draw=black!100,shape border rotate=180,inner sep=0.7mm,minimum size=5mm] {$W$}
            node (dnsqrt) at ($(dnism.south)+(0,-2mm)$) [diamond,anchor=north,draw=black!100,inner sep=1mm] {$\Sigma^{\frac{1}{2}}$};
            
    \draw   (upism) -- (upsqrt)
            (dnism) -- (dnsqrt)
            (upsqrt.north) to [edge label'=$l$] ++ (0,3mm)
            (dnsqrt.south) to [edge label=$l'$] ++ (0,-3mm);
    
    \draw   ($(upism.chord center)+(-3mm,0)$) |- ($0.5*(sigma.north)+0.5*(upism.south)$) -|                    (sigma.north)
            ($(dnism.chord center)+(-3mm,0)$) |- ($0.5*(sigma.south)+0.5*(dnism.north)$) -| (sigma.south)
            ($(upism.chord center)+(3mm,0)$) |- ($0.5*(sigma.north)+0.5*(upism.south)+(15mm,0)$) |- ($0.5*(sigma.south)+0.5*(dnism.north)+(15mm,0)$)-|($(dnism.chord center)+(3mm,0)$);
\end{tikzpicture}
\caption{The diagram depicts $P^{\dagger} \rho_n P$, the density operator $\rho_n$ projected onto its support, where $P$ is defined by (\ref{eqn:numerical_methods_5}) and satisfies the property $P^\dagger \mathds{1}_{d_s ^n}P =\mathds{1}_{d_M ^2}$.}
\label{dg:numerical_methods_diagram_3}
\end{figure}
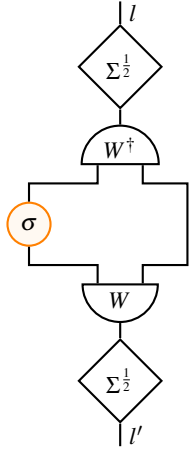
}

\title{Matrix product states and the decay of quantum conditional mutual information }
\author{Pavel Svetlichnyy}
\affiliation{School of Physics, Georgia Institute of Technology, Atlanta, GA, USA.}
\author{Shivan Mittal}
\affiliation{Department of Physics, The University of Texas at Austin, Austin, TX, USA.}
\author{T.A.B. Kennedy}
\affiliation{School of Physics, Georgia Institute of Technology, Atlanta, GA, USA.}

\begin{abstract}
A uniform matrix product state defined on a tripartite system of spins, denoted by $ABC,$ is shown to be an approximate quantum Markov chain when the size of subsystem $B,$ denoted $|B|,$ is large enough. The quantum conditional mutual information (QCMI) is investigated and proved to be bounded by a function proportional to $\exp(-q(|B|-K)+2K\ln|B|)$, with $q$ and $K$ computable constants. The properties of the bounding function are derived by a new approach, with a corresponding improved value given for its asymptotic decay rate $q$. We show the improved value of $q$ to be optimal. Numerical investigations of the decay of QCMI are reported for a collection of matrix product states generated by selecting the defining isometry with respect to Haar measure.
\end{abstract}

\maketitle

\section{Introduction}


In this article, we consider a one dimensional, homogeneous spin chain that is subdivided into a tripartite system with regions $A$ and $C$ separated by ``buffer region'' $B$.
The strong subadditivity of quantum entropy theorem states that for any density operator $\rho_{ABC}$ on Hilbert space $\mathcal H_A \otimes \mathcal H_B \otimes \mathcal H_C$, the quantum conditional mutual information (QCMI) of subsystems $A$ and $C$ given $B$, denoted $I(A:C|B)$. \footnote{The QCMI may be contrasted with the information mutually shared by regions $A$ and $C$ and described by the quantum mutual information (QMI) $I(A:C) := S(\rho_A) +S(\rho_C) - S(\rho_{AC}).$} is non-negative\cite{lieb-ruskai-short,lieb-ruskai-proof}
\begin{align*}
I(A:C | B) 
:= S(\rho_{AB}) + S(\rho_{CB}) - S(\rho_{ABC}) - S(\rho_{B}) \geq 0,
\end{align*}
where $S(\rho_R) = -\mbox{tr} \left( \rho_R \ln \rho_R \right)$ denotes the quantum entropy of the region $R$.
The states $\rho_{ABC}$ for which equality holds are the quantum Markov chains,\cite{petz-recovery} and these have the property that they may be recovered exactly from a reduced density operator; that is, there exists a completely positive map which induces the transformation $\rho_{AB} \mapsto \rho_{ABC}$,\cite{petz-1,petz-2,petz-recovery} by acting on the spins in the separating region $B$ only. While exact recovery is associated with the quantum states for which $I(A:C | B) = 0$,\cite{petz-recovery} approximate recovery requires investigation of states for which QCMI may be small $I(A:C|B) < g(\varepsilon),$ where $g$ is a known positive function and $\varepsilon := \| \rho_{ABC} - \mathcal R_{B \rightarrow BC} (\rho_{AB} )\|_1$ determines the recovery error for the recovery map $\mathcal R_{B \rightarrow BC}$ with domain the support of the reduced density operator $\rho_B$.\cite{universal-rec-1,universal-rec-2,universal-rec-3,book-sutter} Conversely, there exists a so-called universal recovery map $ \mathcal R^u_{B \rightarrow BC} $ which guarantees that when $I(A:C | B) $ is small, the recovery error is too \cite{universal-rec-1,universal-rec-2,universal-rec-3,book-sutter}
 $$
 \| \rho_{ABC} - \mathcal R^u_{B \rightarrow BC} (\rho_{AB} ) \|_1 \le \sqrt{(4 \ln 2) I(A:C | B)}.
 $$
The recovery of quantum states from a quantum memory that has been locally corrupted in region $C$, say, is thus bounded by the QCMI $I(A:C | B)$ when the recovery process implements the universal recovery map $\mathcal R^u_{B \rightarrow BC}(\rho_{AB})$. It follows, in the case that $I(A:C | B)$ is a rapidly decaying function of the size of the buffer region, the number of spins $|B|,$ that the quantum state may be globally recovered by experiments on the well-localized interface region $B$ and $C$.\cite{petz-1,petz-2,petz-recovery,universal-rec-1,universal-rec-2,universal-rec-3,storage,kato,eff-prep,swingle,approxerrcorrev}

The many-body ground states of spin lattices described by local gapped Hamiltonians are well approximated by matrix product states (MPS) in one dimension,
\cite{klumper,vidal-mps,mps-repr} and their generalization, the projected entangled pair states (PEPS) in higher dimensions.\cite{mpspepsvarrenorm} The MPS first effectively appeared as finitely correlated states (FCS) \cite{fcs,pgfcs} and were defined on infinite one-dimensional spin chains. (More precisely, MPS correspond to the purely generated FCS (pgFCS), which comprise a subclass of FCS.) The MPS provide an alternative description of the previously developed density matrix renormalization group algorithm,\cite{white-dmrg} and are also useful in time dependent studies of weakly entangled systems. In one dimension, MPS may be generalized to matrix product density operators (MPDO),\cite{mpdo-introd} which, in the translationally invariant infinite case correspond to the FCS. The MPDO may be used to approximate thermal states of local gapped Hamiltonians.\cite{mpdo-introd,kato,kuwahara,improved-thermal} Controlled experimental preparation techniques for MPS have been proposed in the literature.\cite{sequent-gen,lukin}

Given the significance of ground and thermal states of local gapped Hamiltonians, their quantum-information theoretic properties are of great interest.\cite{swingle,area-laws,exp-decay-area-law,efficient-mpdo} As are the properties of their approximants -- the MPS and MPDO, which are often easier to study, given their more explicit structure.\cite{kim-mpdo,fcs-mpdo,qcmi-decay}

In a recent paper Ref. \onlinecite{qcmi-decay} we proved that the pgFCS (infinite uniform MPS (iuMPS)), with reduced density operators $ \rho_{ABC}$, are approximate quantum Markov chains for the one-dimensional configuration space $ABC$ discussed above. Related to this result, the QCMI $I(A:C|B)$ decays and is bounded by a (sub)exponentially decaying function in the size of the separating region $B$.\cite{qcmi-decay} The goal of the current paper is to reexamine this problem and report an improved prediction for the exponential decay rate, differing by a factor of four from our earlier prediction. We also present numerical experiments in support of this result.

Our method of estimating the QCMI builds on earlier works Refs. \onlinecite{fcs,pgfcs,mps-repr,exp-decay-area-law} and our work Ref. \onlinecite{qcmi-decay}, where we construct a quantum Markov chain $\tilde{\rho}_{ABC}$ approximating $\rho_{ABC}$. Specifically, the method is based on (i) bounding the QCMI by the trace distance between the exact density operator $\rho_{ABC}$ and its approximant $\tilde \rho_{ABC}$, and (ii) bounding this trace distance by the distance in diamond norm between the induced quantum channels $\mathcal E^{|B|}$ and $\tilde{\mathcal E}^{|B|}$ (our notation is explained below). In our earlier work Ref. \onlinecite{qcmi-decay} the combination of Alicki-Fannes inequality \cite{alicki-fannes,tight-bounds} and Strinespring Continuity theorem \cite{stinespring,infdisturb} provided (i) a QCMI bound linear in trace distance, and (ii) the trace distance bounded by the square root of the channel distance. In the present work the QCMI bound is improved to a quadratic dependence on trace distance, while the latter is bounded by a linear function of the channel distance. To achieve these improvements, use of the Alicki-Fannes Inequality has been replaced by a bound based on the Taylor expansion of a logarithmic function representation of the density operator, with estimate of the remainder. The continuity of Stinespring dilation \cite{infdisturb} is no longer employed and instead a construction used in the proof of Choi's theorem \cite{choi} takes advantage of the nearness of the quantum channels, and provides a sharper tool in this context. Overall we go from a bound on QCMI which had square root dependence in the quantum channel distance, to one with quadratic dependence. This leads to a factor of four change in our predicted decay rate, a result which appears to be corroborated by our numerical experiments using tensor network methods.

The rest of the paper is organized as follows. In Section \ref{sec:theor_form} we introduce the framework for MPS and iuMPS. In Section \ref{sec:proof} we present and prove our main result, Theorem \ref{thrm:theorem_1}. In Section \ref{sec:stepwise_procedure_numerics} we describe and discuss the methods used in our numerical experiments, and present numerical results for comparison with Theorem \ref{thrm:theorem_1} in Section \ref{sec:numerical_verification}. Section \ref{sec:conclusion} contains concluding remarks. A series of appendices contains supporting technical information.

\normalize

\section{Theoretical formulation}\label{sec:theor_form}

A pure state vector for a system of $n$ spins, each with associated Hilbert space $\mathcal{H}_s$ and orthonormal basis $\{ |s\rangle : s=1,2,..,d_s \}$, may be expressed in the form
\begin{equation*}
    |\Psi\rangle=\sum_{s_1,\dots,s_n=1}^{d_s} a_{s_1s_2\dots s_n}|s_{n}\rangle\otimes\dots|s_2\rangle\otimes|s_1\rangle,
\end{equation*}
where $d_s$ is the dimension of $\mathcal{H}_s$.
For the class of states called uniform matrix product states (uMPS) each coefficient $a_{s_1s_2\dots s_n}$ has the form \cite{vidal-mps,mps-repr}
\begin{equation*}
    a_{s_1s_2\dots s_n}=\langle L|M^{s_n}\dots M^{s_2}M^{s_1}|R\rangle,
\end{equation*}
where each $M^{s_k}$ belongs to the set $\mathcal{S}:=\{M^s \in \mathbb{C}^{d_M \times d_M} \;|\; s=1,2\dots, d_s\}$ of complex matrices, and $|L\rangle,|R\rangle\in\mathbb{C}^{d_M}$. We will refer to the space $\mathcal{H}_M:=\mathbb{C}^{d_M}$ as virtual space, and to the space $\mathcal{H}_s$ as physical space. Hence, an uMPS $|\Psi\rangle$ may be expressed as 
\begin{widetext}
\begin{equation}\label{eqn:matr_prod_st}
    |\Psi\rangle=\sum_{s_1,s_2,\dots s_n=1}^{d_s} \langle L| M^{s_{n}}\dots M^{s_{2}}M^{s_{1}}|R\rangle\; |s_{n}\rangle\otimes\dots\otimes|s_{2}\rangle\otimes|s_1\rangle.
\end{equation}
\end{widetext}
Without loss of generality, we may assume that $|\Psi\rangle$ is in canonical form, so that $M^s$ satisfies the condition  diagrammatically depicted in Fig~\ref{dg:norm_eigen}(a), and by means of the equation
\begin{equation}\label{eqn:condition_on_M}
    \sum_{s=1}^{d_s}  M^{s\dagger}M^{s}=\mathbb{1}_M,
\end{equation}
which may be achieved by a suitable definition of $|L\rangle$ and $|R\rangle$.\cite{mps-repr}

Another, more compact, way to represent a MPS (\ref{eqn:matr_prod_st}) is by introducing the isometry $V:\mathcal{H}_M\rightarrow\mathcal{H}_s\otimes\mathcal{H}_M$,
\begin{equation*}
    V=\sum_{s=1}^{d_s} |s\rangle\otimes M^s,
\end{equation*}
satisfying $V^\dagger V=\mathbb{1}_{d_M}$. With the help of $V$, we may express (\ref{eqn:matr_prod_st}) as
\begin{equation*}
    |\Psi\rangle=\langle L|\underbrace{ V\cdots VV}_{n}|R\rangle,
\end{equation*}
where we assume the notational convention that each isometry $V$ acts non-trivially only on the Hilbert space factor $\mathcal{H}_M$, i.e., $VV:=(\mathbb{1}_{s} \otimes V)V$. We will henceforth denote the $n$-fold product $V_n:=\underbrace{ V\cdots VV}_{n}$. When $|R|$ is used to denote the number of spins in region $R$, we will write simply $V_R$ instead of $V_{|R|}$.

\mrabc

An infinite uniform MPS (iuMPS) is a suitable limit of the uniform MPS when the chain length tends to infinity.\cite{fcs,pgfcs}

The limiting state is fully characterized by the collection of reduced density operators $\rho_n$ for a continuous region of $n \in \mathbb N$ spins, and $\rho_n$ takes the form
\begin{widetext}
\begin{align}\label{eqn:eqn_5}
    \rho_n &=\qquad\sum_{\mathclap{\substack{s_1,s_2,\dots,s_n=1\\ s'_1,s'_2,\dots,s'_n=1}}}^{d_s}\; \mbox{tr}_M(M^{s_{n}}\dots M^{s_{2}}M^{s_{1}}\sigma M^{s'_1\dagger}M^{s'_2\dagger}\dots M^{s'_n\dagger})\; |s_{n}\rangle\langle s'_{n}|\otimes\dots\otimes|s_{2}\rangle\langle s'_2|\otimes|s_{1}\rangle\langle s'_1|, \\ \nonumber
    &= \mathrm{tr}_M(V\cdots VV\sigma V^\dagger\cdots V^\dagger V^\dagger)= \mathrm{tr}_M(V_n\sigma V^\dagger_n),
\end{align}
\end{widetext}
where $\mbox{tr} \rho_n = 1,$ and $\sigma$ is a fixed point of the quantum channel (completely positive trace-preserving (CPTP) map) $\mathcal{E}:\mathcal{B}(\mathcal{H}_M)\rightarrow \mathcal{B}(\mathcal{H}_M)$, with \cite{fcs,mps-repr}
\begin{equation}\label{eqn:e_q_channel}
    \mathcal{E}(X)=\sum_{s=1}^{d_s}M^s XM^{s\dagger}.
\end{equation}
The latter is conveniently represented as a linear operator and defines the transfer matrix $E\in\mathbb{C}^{d_M^2\times d_M^2}$,
\begin{equation}\label{eqn:tr_matr}
    E=\sum_{s=1}^{d_s} M^{s}\otimes \bar{M}^{s},
\end{equation}
where $\bar{M}^s$ is the complex conjugate of $M^s$. Without loss of generality, the density matrix $\sigma$ may be considered to be full-rank,\cite{fcs} which we will use in our derivations. The transfer matrix is depicted in Fig.\ref{dg:norm_eigen}(c), with the matrix elements $E_{jj';ii'}:=\langle j|\otimes\langle j'|E|i\rangle \otimes|i'\rangle$.
To uniquely define a iuMPS we must specify a family of $d_M\times d_M$ matrices $\{M^s\}_{s=1}^{d_s}$ (or, alternatively, the isometry $V$), and the fixed point density operator $\sigma$ that satisfies $\mathcal{E}(\sigma)=\sigma$ (refer to Fig.~\ref{dg:norm_eigen}(b)). 

\mbtensor

The eigenvalue problem, depicted in Fig.\ref{dg:norm_eigen}(d), and given by 
\begin{equation}\label{eqn:tr_matr_eigpr}
    E|X_k\rangle=\nu_k |X_k\rangle,\quad k=1,\dots,k_{\max},
\end{equation}
where $|X_k\rangle\in\mathcal{H}_M\otimes\mathcal{H}_M=\mathbb{C}^{d_M}\otimes\mathbb{C}^{d_M}$, determines the spectrum of the transfer matrix. Note that since $E$ is generally non-Hermitian, then $k_{\max}\leq d_M^2$. Nonetheless, the equation (\ref{eqn:tr_matr_eigpr}) provides all eigenvalues of $E$, which is sufficient at this step. Alternatively, we may formulate the eigenvalue problem for the quantum channel $\mathcal{E}$,
\begin{equation*}
    \mathcal{E}(X_k)=\nu_k X_k, \quad k=1,\dots,k_{\max},
\end{equation*}
where $X_k\in \mathcal{B}(\mathcal{H}_M)$.
For our purposes the most significant spectral parameter characterizing the transfer matrix is the largest absolute value of its non-peripheral eigenvalues,
\begin{equation*}
    \nu_{\mathrm{gap}}=\max_{|\nu_k|<1}|\nu_k|.
\end{equation*}
Note that $\nu_{\mathrm{gap}}<1$ for any $E$, even if several eigenvalues of $E$ have absolute value one. For an injective iuMPS, the quantity $1-\nu_{\mathrm{gap}}$ is the usually defined spectral gap, which motivates our choice of the subscript in $\nu_{\mathrm{gap}}$.

Using the notation introduced in (\ref{eqn:tr_matr}) we may express $\rho_n$ as
\begin{widetext}
\begin{equation}\label{eqn:general_rho_n}
    \rho_n=\sum_{\mathclap{s_1,s_2,\dots,s_n=1}}^{d_s}\;  \langle +|(M^{s_{n}}\otimes \bar{M}^{s'_{n}})\cdot\dots\cdot (M^{s_{2}}\otimes \bar{M}^{s'_{2}})(M^{s_{1}}\otimes \bar{M}^{s'_{1}})|\sigma\rangle\; |s_{n}\rangle\langle s'_{n}|\otimes\dots\otimes|s_{2}\rangle\langle s'_2|\otimes|s_{1}\rangle\langle s'_1|,
\end{equation}
\end{widetext}
where $|+\rangle=\sum_{i=1}^{d_M}|i\rangle\otimes|i\rangle$ and $|\sigma\rangle=\sum_{i=1}^{d_M}\sigma_i|i\rangle\otimes|i\rangle$.

In this paper we divide $n$ spins into three continuous subregions $A$, $B$, and $C$, and denote the reduced density operator $\rho_n$ as $\rho_{ABC}$. A diagrammatic representation of $\rho_{ABC}$ is given in Fig.\ref{dg:red_dens_op}(c). For $\rho_{ABC}$ we investigate the behavior of $I(A:C|B)$ as a function of the size of the buffer region $B$, denoted by $|B|$.

\section{Main Theorem and its proof} \label{sec:proof}

Consider the configuration $ABC$ shown in Fig.\ref{dg:red_dens_op}(a) and (b), and prepared in an iuMPS $\rho_{ABC}$, as illustrated in Fig.\ref{dg:red_dens_op}(c). For this system the following theorem on the asymptotic behavior of QCMI holds; it is a tighter version of the result given in Ref. \onlinecite{qcmi-decay}.
\begin{theorem}\label{thrm:theorem_1}
    For the described setup of system $ABC$, prepared in an infinite uniform matrix product state, the quantum conditional mutual information (QCMI) obeys the bound
    $$I(A:C|B)\leq Qe^{-q(|B|-K)+2K\ln|B|},$$
    for $|B|$ large enough.
    Here $q=-2\ln\nu_{\mathrm{gap}}$, $K$ is a non-negative integer, and $Q$ does not depend on $|B|$. The rate $q=-2\ln\nu_{\mathrm{gap}}$ cannot be further improved.
\end{theorem}
Compared to the similar bound in Ref. \onlinecite{qcmi-decay} with $q=-\frac{1}{2}\ln\nu_{\mathrm{gap}}$, the value predicted here is $q=-2\ln\nu_{\mathrm{gap}}.$

\noindent\textbf{Remarks:}\
\begin{enumerate}
    \item 
    For a given iuMPS $\rho_{ABC}$, the non-negative integer $K$ may be extracted from the Jordan decomposition of the corresponding transfer matrix (\ref{eqn:tr_matr}) : $K+1$ is the largest dimension of the Jordan blocks corresponding to eigenvalues of magnitude $\nu_{\mathrm{gap}}$. We provide the detailed computation of $K$ in Appendix \ref{app:constants}.
    \item 
    We prove that the decay rate $q=-2\ln\nu_{\mathrm{gap}}$ cannot be further improved in Appendix \ref{appendix:tight-bound}. There we provide an example of an injective iuMPS (with $K=0$) for which QCMI may be explicitly calculated to be $I(A:C|B)=O(e^{-q|B|})$. 
\end{enumerate}
We prove Theorem \ref{thrm:theorem_1}, building on the derivation of Theorem II.1 in Ref. \onlinecite{qcmi-decay}. For simplicity, in subsections \ref{subsec:qcmi_estmate} and \ref{subsec:tr_dist_estimate} we prove Theorem \ref{thrm:theorem_1} for the case of injective iuMPS \cite{mps-repr}, i.e., the case for which the transfer matrix (\ref{eqn:tr_matr}) has a unique eigenvector $|\sigma\rangle$  with eigenvalue of magnitude 1. The generalization to the case of an arbitrary iuMPS is straightforward, but more technical, and will be discussed in subsection \ref{subsec:generalization}. 

The key idea of the proof is identical to the one used in Theorem II.1 of Ref. \onlinecite{qcmi-decay} and involves the construction of a density operator $\tilde{\rho}_{ABC}$, that is simultaneously a quantum Markov chain and a good approximation to $\rho_{ABC}$ in trace norm. By comparison with Ref. \onlinecite{qcmi-decay}, however, we use different methods to estimate $I(A:C|B)$ and $\|\rho_{ABC}-\tilde{\rho}_{ABC}\|_1$, which allow for a tighter bound on $I(A:C|B)$.
The density operator $\tilde{\rho}_{ABC}$ is illustrated diagrammatically in Fig.\ref{dg:red_dens_op}(d)
with the related complex tensor $\tilde{M}_B$ of dimensions $d_M\times d_s^{|B|}\times d_M$, depicted in Fig.\ref{dg:mb_tensor}.

Throughout the Subsections \ref{subsec:qcmi_estmate}, \ref{subsec:tr_dist_estimate}, and \ref{subsec:generalization} we exploit the properties of the quantum channel $\tilde{\mathcal{E}}:\mathcal{B}(\mathcal{H}_M)\rightarrow\mathcal{B}(\mathcal{H}_M)$ (and its matrix representation $\tilde{E}:\mathcal{H}_M\otimes\mathcal{H}_M\rightarrow\mathcal{H}_M\otimes\mathcal{H}_M$), which corresponds to the peripheral spectrum of $\mathcal{E}$. It is straightforward to define $\tilde{\mathcal{E}}$, by first defining $\tilde{E}$ through
\begin{equation}\label{eqn:tr_matr_periph}
    \tilde{E}=PEP,
\end{equation}
where $P$ is the projector onto the subspace $\mathrm{span}\{|u_i\rangle\;|\; E|u_i\rangle=\nu_i|u_i\rangle,\, |\nu_i|=1\}$. The map $\tilde{\mathcal{E}}$ is then defined by 
\begin{equation*}
    \langle j|\mathcal{\tilde{E}}(|i\rangle\langle i'|)|j'\rangle :=(\langle j|\otimes\langle j'|)\tilde{E}(|i\rangle\otimes|i'\rangle).
\end{equation*}
It may be shown that the map $\tilde{\mathcal{E}}$ is indeed a quantum channel (see Lemma III.2 in Ref. \onlinecite{qcmi-decay}).

The key properties that make $\tilde{E}$ (and $\tilde{\mathcal{E}}$) useful, are the bounds on $\|E^n-\tilde{E}^n\|$ derived in Theorem III.2 of Ref. \onlinecite{wolf-szehr} and satisfied for large enough $n$ (specified in Appendix \ref{app:constants}), 
\begin{align}\label{eqn:22norm_leq_exp}
    c_1 n^{K}\nu_{\mathrm{gap}}^{n-K}&\leq\|E^n-\tilde{E}^n\|\\ \nonumber
    &=\|\mathcal{E}^n-\tilde{\mathcal{E}}^n\|_{2-2}\\ \nonumber
    &\leq c_2 n^{K}\nu_{\mathrm{gap}}^{n-K},  
\end{align}
where $n\in\mathbb{N}$, $K$ is a non-negative integer, and $c_1$ and $c_2$ are some constants. In Appendix \ref{app:constants} we obtain the values of $c_1$, $c_2$, and $K$ from the Jordan decomposition of $E$.

\subsection{A trace distance bound on QCMI.}\label{subsec:qcmi_estmate}
In this subsection we bound $I(A:C|B)$ in terms of $\|\rho_{ABC}-\tilde{\rho}_{ABC}\|_1$, with the assumption that $\tilde{\rho}_{ABC}$ is a quantum Markov chain. We will present the construction of $\tilde{\rho}_{ABC}$ that is indeed a quantum Markov chain in Subsection \ref{subsec:tilde_construction}. Here we will derive only the leading contribution to the bound of QCMI. We delegate the careful treatment of subleading contributions that arise from the expansion (\ref{eqn:log_expansion}) below to Appendix \ref{appendix:expansion}.

In QCMI
\begin{align}\label{eqn:qcmi_log}
I(A:C|B)=\mathrm{tr}&\left[\rho_{ABC}\left(\ln\rho_{ABC}+\ln\rho_{B}\right.\right.\\ \nonumber
&\left.\left.-\ln\rho_{AB}-\ln\rho_{BC}\right)\right],
\end{align}
we express 
\begin{equation*}
    \rho_{ABC}=\tilde{\rho}_{ABC}+\Delta\rho_{ABC},
\end{equation*}
with similar expressions for the marginals of $\rho_{ABC}$ obtained by suitable partial traces. Then, using the integral representation for the logarithm of an operator
$$
\ln \rho = \int_0^{\infty} ds \left(\frac{1}{1+s} - \frac{1}{\rho +s} \right),
$$
we expand (\ref{eqn:qcmi_log}) to second order in $\Delta\rho_{ABC}$ and its marginals. In particular for a typical density operator, and suppressing indices, we have
\begin{align}\label{eqn:log_expansion}
    \ln(\tilde{\rho}+\Delta\rho)=&\ln\tilde{\rho}\\ \nonumber &+\int_{0}^{+\infty}ds\left(\frac{1}{\tilde{\rho}+s\mathbb{1}}\Delta\rho\frac{1}{\tilde{\rho}+s\mathbb{1}}\right.\\ \nonumber
    &\left.-\frac{1}{\tilde{\rho}+s\mathbb{1}}\Delta\rho\frac{1}{\tilde{\rho}+s\mathbb{1}}\Delta\rho\frac{1}{\tilde{\rho}+s\mathbb{1}}\right)\\ \nonumber
    &+O\left(\|\Delta\rho\|_1^3\right).
\end{align}
Technically, while dealing with logarithm and inverse functions of operators we should keep track of the subspaces of support. In Appendix \ref{appendix:supports} we prove that the support of $\Delta\rho$ is contained within the support of $\tilde{\rho}$, i.e., $\mathrm{supp}(\Delta\rho)\subseteq\mathrm{supp}(\tilde{\rho})$ (for any continuous region of spins on which $\tilde{\rho}$ and $\Delta\rho$ are defined). 

The bound on the norm of the remainder term, which is indicated by $O\left(\|\Delta\rho\|_1^3\right)$, also contains the dependence on the minimum non-zero eigenvalue of $\tilde{\rho}$. The details may be found in Appendix \ref{appendix:expansion}.

We now consider the zeroth, first and second order terms in $\Delta \rho.$ Since $\tilde{\rho}_{ABC}$ is a quantum Markov chain, \cite{lieb-ruskai-proof,ruskai-ineq}
\begin{equation*}
    \ln\tilde{\rho}_{ABC}+\ln\tilde{\rho}_{B}-\ln\tilde{\rho}_{AB}-\ln\tilde{\rho}_{BC}=0,
\end{equation*}
and so the zeroth order term in the expansion of (\ref{eqn:qcmi_log}) vanishes,
\begin{align*}
    \mathrm{tr}&\left[\tilde{\rho}_{ABC}\left(\ln\tilde{\rho}_{ABC}+\ln\tilde{\rho}_{B}\right.\right.\\ \nonumber
&\left.\left.-\ln\tilde{\rho}_{AB}-\ln\tilde{\rho}_{BC}\right)\right]=0.
\end{align*}
The first order contribution has a term that vanishes for the same reason,
\begin{align*}
    \mathrm{tr}&\left[\Delta\rho_{ABC}\left(\ln\tilde{\rho}_{ABC}+\ln\tilde{\rho}_{B}\right.\right.\\ \nonumber
&\left.\left.-\ln\tilde{\rho}_{AB}-\ln\tilde{\rho}_{BC}\right)\right]=0.
\end{align*}

The other terms of first order in $\Delta\rho$ are all proportional to an expression of the following kind,
\begin{align*}
    &\mathrm{tr}\left(\tilde{\rho}\int_{0}^{+\infty}ds\frac{1}{\tilde{\rho}+s\mathbb{1}}\Delta\rho\frac{1}{\tilde{\rho}+s\mathbb{1}}\right)\\ \nonumber
    &=\mathrm{tr}\left(\int_{0}^{+\infty}ds\frac{1}{\tilde{\rho}+s\mathbb{1}}\tilde{\rho}\frac{1}{\tilde{\rho}+s\mathbb{1}}\Delta\rho\right)\\ \nonumber
    &=\mathrm{tr}\left(\left(\int_{0}^{+\infty}ds\frac{1}{(\tilde{\rho}+s\mathbb{1})^2}\right)\tilde{\rho}\Delta\rho\right)\\ \nonumber
    &=\mathrm{tr}\left(\tilde{\rho}^{-1}\tilde{\rho}\Delta\rho\right)\\ \nonumber
    &=\mathrm{tr}(\Delta\rho)\\ \nonumber
    &=0,
\end{align*}
since $\mathrm{tr}\Delta\rho_{ABC}=\mathrm{tr}\rho_{ABC}-\mathrm{tr}\tilde{\rho}_{ABC}=0$ (hence, $\mathrm{tr}\Delta{\rho}_{AB}=0$, etc.). 

Thus, the leading non-vanishing terms in the expansion of $I(A:C|B)$ are of the second order in $\Delta\rho$, coming from the terms of the type
\begin{align}\label{eqn:2nd_order}
    &\mathrm{tr}\left(\Delta\rho\int_{0}^{+\infty}ds\frac{1}{\tilde{\rho}+s\mathbb{1}}\Delta\rho\frac{1}{\tilde{\rho}+s\mathbb{1}}\right)\\ \nonumber
    &-\mathrm{tr}\left(\tilde{\rho}\int_{0}^{+\infty}ds\frac{1}{\tilde{\rho}+s\mathbb{1}}\Delta\rho\frac{1}{\tilde{\rho}+s\mathbb{1}}\Delta\rho\frac{1}{\tilde{\rho}+s\mathbb{1}}\right)\\ \nonumber
    &=\mathrm{tr}\left(\int_{0}^{+\infty}ds\frac{s}{\tilde{\rho}+s\mathbb{1}}\Delta\rho\frac{1}{\tilde{\rho}+s\mathbb{1}}\Delta\rho\frac{1}{\tilde{\rho}+s\mathbb{1}}\right),
\end{align}
which can be estimated as 
\begin{align}\label{eqn:2nd_order_bound}
    &\left|\mathrm{tr}\left(\int_{0}^{+\infty}ds\frac{s}{\tilde{\rho}+s\mathbb{1}}\Delta\rho\frac{1}{\tilde{\rho}+s\mathbb{1}}\Delta\rho\frac{1}{\tilde{\rho}+s\mathbb{1}}\right)\right|\\ \nonumber
    &\leq\int_{0}^{+\infty}ds s\left\|\frac{1}{\tilde{\rho}+s\mathbb{1}}\right\|^3\|\Delta\rho\|^2_1\\ \nonumber
    &\leq \int_{0}^{+\infty}ds \frac{s}{(\lambda_{\mathrm{min}}+s)^3}\|\Delta\rho\|^2_1\\ \nonumber
    &= \frac{\|\Delta\rho\|^2_1}{2\lambda_{\mathrm{min}}},
\end{align}
where $\lambda_{\mathrm{min}}$ is the smallest non-zero eigenvalue of $\tilde{\rho}$.

Expansion of each of the four terms in (\ref{eqn:qcmi_log}) contributes a term of the type (\ref{eqn:2nd_order_bound}) to the bound of $I(A:C|B)$. Moreover, for each of the four terms in (\ref{eqn:qcmi_log}), we need to account for the contribution emerging from the remainder term in the expansion (\ref{eqn:log_expansion}). This contribution, for large enough $|B|$, may also be bounded by the contribution resulting from terms of the second order (\ref{eqn:2nd_order_bound}), as shown on Appendix \ref{app:constants}. Thus, combining all contributions together, for $|B|$ large enough, we can guarantee the asymptotic bound  
\begin{equation}\label{eqn:qcmi_leq_11norm_lambda}
    I(A:C|B)\leq\frac{4}{\lambda_{\mathrm{min}}}\|\Delta\rho_{ABC}\|_1^2.
\end{equation}
We provide the detailed derivation of the bound (\ref{eqn:qcmi_leq_11norm_lambda}) in Appendix \ref{appendix:expansion}.

\noindent\textbf{Remark}:\ The inequality (\ref{eqn:qcmi_leq_11norm}) is an upper bound on QCMI quadratic in $\|\Delta\rho_{ABC}\|_1$, in contrast to the upper bound linear in $\|\Delta\rho_{ABC}\|_1$ that we used in Ref. \onlinecite{qcmi-decay}. In Subsection \ref{subsec:tr_dist_estimate} we will see that this improved bound is responsible for changing the rate of exponential decay of the QCMI bounding function by a factor of $2$, compared to that in Theorem II.1 in Ref. \onlinecite{qcmi-decay}.

\subsection{Construction of $\tilde{\rho}_{ABC}$ and evaluation of $\lambda_{\mathrm{min}}$.}\label{subsec:tilde_construction}

In this subsection we implicitly construct the tensor $\tilde{M}_B$, such that $\tilde{\rho}_{ABC}$ defined in Fig.\ref{dg:red_dens_op} becomes a quantum Markov chain, and we show that its smallest non-zero eigenvalue satisfies (for injective iuMPS)
\begin{align}
    \lambda_{\mathrm{min}}=\sigma^2_{\mathrm{min}}.
\end{align}
We represent the tensor $\tilde{M}_B$ as a collection of $d_s^{|B|}$ complex $d_M\times d_M$ matrices $\tilde{M}^{s_1s_2\dots s_{|B|}}_B$, where $s_k=1,2,\dots,d_s$ for $k=1,2,\dots,|B|$. The matrices $\tilde{M}^{s_1s_2\dots s_{|B|}}_B$ are defined by their matrix elements in Fig. \ref{dg:mb_tensor}.

For an injective iuMPS, the quantum channel $\tilde{\mathcal{E}}$ has the form \cite{fcs}
\begin{equation*}
\tilde{\mathcal{E}}(X)=\mathrm{tr}(X)\sigma.
\end{equation*}
Its matrix representation $\tilde{E}$ is then obtained with the use of (\ref{eqn:tr_matr}), 
\begin{equation*}
    \tilde{E}=\tilde{E}^n=|\sigma\rangle\langle +|.
\end{equation*}
In Section IV.A of Ref. \onlinecite{qcmi-decay} it was proved, that if we impose on $\tilde{M}_B$ the condition
\begin{align}
    \label{eqn:tilde_tr_matr}
\quad\;\mathclap{\sum_{s_1,\dots,s_{|B|}=1}^{d_s}}\quad \tilde{M}_B^{s_1\dots s_{|B|}}\otimes \overline{\tilde{M}}_B^{s_1\dots s_{|B|}}=\tilde{E}^{|B|}=|\sigma\rangle\langle +|,
\end{align}
or, equivalently, the condition
\begin{align}\label{eqn:tilde_channel}
    \tilde{\mathcal{E}}^{|B|}(X)&=\sum_{s_1,\dots,s_{|B|}=1}^{d_s} \tilde{M}_B^{s_1\dots s_{|B|}}X\tilde{M}_B^{s_1\dots s_{|B|}\dagger} \\ \nonumber
    &=\mathrm{tr}(X)\sigma,
\end{align}
then the induced $\tilde{\rho}_{ABC}$ is a quantum Markov chain. The $d_M\times d_M$ complex matrices $\overline{\tilde{M}}_B^{s_1\dots s_{|B|}}$ and $\tilde{M}_B^{s_1\dots s_{|B|}\dagger}$ are the element-wise complex conjugate and the complex conjugate transpose of $\tilde{M}_B^{s_1\dots s_{|B|}}$, respectively. We will now show that the conditions (\ref{eqn:tilde_tr_matr}) and (\ref{eqn:tilde_channel}) result in the desired bound (\ref{eqn:desired_bound}).

Notice that the isometries
\begin{widetext}
\begin{align*}
    V_B&=\sum_{s_1,\dots,s_{|B|}=1}^{d_s} |s_1\rangle\otimes|s_2\rangle\otimes\cdots\otimes|s_{|B|}\rangle\otimes M^{s_1}M^{s_2}\dots M^{s_{|B|}}, \\ \nonumber 
    \tilde{V}_B &=\sum_{s_1,\dots,s_{|B|}=1}^{d_s} |s_1\rangle\otimes|s_2\rangle\otimes\cdots\otimes|s_{|B|}\rangle\otimes\tilde{M}_B^{s_1\dots s_{|B|}},
\end{align*}
\end{widetext}
represent Stinespring dilations \cite{stinespring} for the quantum channels $\mathcal{E}^{|B|}$ and $\tilde{\mathcal{E}}^{|B|}$, respectively, with $\mathcal{H}_B$ serving as the dilating Hilbert space, i.e.,
\begin{align}\label{eqn:tilde_ch_fold_B}
    \mathcal{E}^{|B|}(X)&=\mathrm{tr}_B(V_BXV_B^\dagger), \\ 
    \tilde{\mathcal{E}}^{|B|}(X)&=\mathrm{tr}_B(\tilde{V}_BX\tilde{V}_B^\dagger)=\mathrm{tr}(X)\sigma. 
\end{align}
Moreover, we can conveniently reexpress $\tilde{M}_B$ in terms of the isometry $\tilde{V}_B$ as
\begin{equation}\label{eqn:tilde_m_b_via_isom}
    \tilde{M}_B^{s_1\dots s_{|B|}}=(\langle s_1|\otimes\langle s_2|\otimes\cdots\otimes\langle s_{|B|}|)\tilde{V}_B,
\end{equation}
which allows us to work with $\tilde{V}_B$ satisfying (\ref{eqn:tilde_ch_fold_B}) instead of $\tilde{M}_B$ satisfying (\ref{eqn:tilde_channel}).

We notice that we can express $\rho_{ABC}$ and $\tilde{\rho}_{ABC}$ in terms of $V_B$ and $\tilde{V}_B$ as 
\begin{align*}
    \rho_{ABC}&=\mathrm{tr}_M(V_CV_BV_A\sigma V^\dagger_AV_B^\dagger V^\dagger_C),\\ \nonumber
    \tilde{\rho}_{ABC}&=\mathrm{tr}_M(V_C\tilde{V}_BV_A\sigma V^\dagger_A\tilde{V}_B^\dagger V^\dagger_C).
\end{align*}

Now we estimate $\lambda_{\mathrm{min}}$ with the help of a purification argument. The density operator $\tilde{\rho}_{ABC}$ can be purified with the use of the auxiliary space $\mathcal{H}_{M'}$ (note  $\mathcal{H}_{M'}\cong\mathcal{H}_M$),
\begin{align*}
\tilde{\rho}_{ABC}=\mathrm{tr}_{MM'}\left((V_C\tilde{V}_BV_A\otimes\mathbb{1}_{M'})|\sqrt{\sigma}\rangle\langle \sqrt{\sigma}| \right. \\ \nonumber \times\left.(V_A^\dagger\tilde{V}_B^\dagger V_C^\dagger\otimes\mathbb{1}_{M'})\right),
\end{align*}
where $|\sqrt{\sigma}\rangle=\sum_{i=1}^{d_M}\sqrt{\sigma_i}|i\rangle\otimes|i\rangle\in\mathcal{H}_M\otimes\mathcal{H}_{M'}$, and $V_A$, $\tilde{V}_B$, and $V_C$ act on the space $\mathcal{H}_M$. This implies that $\tilde{\rho}_{ABC}$ has the same spectrum as \begin{align*}
&\mathrm{tr}_{ABC}\left((V_C\tilde{V}_BV_A\otimes\mathbb{1}_{M'})|\sqrt{\sigma}\rangle\langle \sqrt{\sigma}| \right.\\ \nonumber
&\qquad\qquad\qquad\qquad\times\left.(V_A^\dagger\tilde{V}_B^\dagger V_C^\dagger\otimes\mathbb{1}_{M'})\right)\\ \nonumber
&=(\mathcal{E}^{|C|}\circ\tilde{\mathcal{E}}^{|B|}\circ\mathcal{E}^{|A|}\otimes\mathrm{id}_{M'})\left(|\sqrt{\sigma}\rangle\langle\sqrt{\sigma}|\right),
\end{align*} 
and hence, assuming $\dim\mathcal{H}_{ABC}\geq\dim(\mathcal{H}_M\otimes\mathcal{H}_{M'})=d_M^2$, that
\begin{align}
    \tilde{\rho}_{ABC}=J(\mathcal{E}^{|C|}\circ\tilde{\mathcal{E}}^{|B|}\circ\mathcal{E}^{|A|}\otimes\mathrm{id}_{M'})\left(|\sqrt{\sigma}\rangle\langle\sqrt{\sigma}|\right)J^\dagger,
\end{align}
where $J:\mathcal{H}_M\otimes\mathcal{H}_{M'}\rightarrow\mathcal{H}_{ABC}$ is an isometry.

Now notice that $\tilde{\mathcal{E}}\circ\mathcal{E}=\tilde{\mathcal{E}}\circ\mathcal{E}=\tilde{\mathcal{E}}^2$, hence
\begin{equation}
    \tilde{\rho}_{ABC}=J(\tilde{\mathcal{E}}^{|ABC|}\otimes\mathrm{id}_{M'})\left(|\sqrt{\sigma}\rangle\langle\sqrt{\sigma}|\right)J^\dagger.
\end{equation}
Moreover, since any of the regions $ABC$, $AB$, $BC$, or $B$ contains the subregion $B$, then any $\tilde{\rho}_R$ may be expressed as 
\begin{equation}\label{eqn:rho_r_prime}
    \tilde{\rho}_{R}=J(\tilde{\mathcal{E}}^{|R|}\otimes\mathrm{id}_{M'})\left(|\sqrt{\sigma}\rangle\langle\sqrt{\sigma}|\right)J^\dagger,
\end{equation}
where $R$ may be any of $ABC$, $AB$, $BC$, or $B$, and $J:\mathcal{H}_M\otimes\mathcal{H}_{M'}\rightarrow\mathcal{H}_{R}$ is an isometry appropriate to region $R$.

For an injective iuMPS, $\tilde{\mathcal{E}}:\mathcal{B}(\mathcal{H}_M)\rightarrow\mathcal{B}(\mathcal{H}_M)$ acts as $\tilde{\mathcal{E}}(X)=\mathrm{tr}(X)\sigma$, and hence
\begin{equation}\label{eqn:rho_r_prime_inj}
    \tilde{\rho}_R=J\sigma\otimes\sigma J^\dagger,
\end{equation}
which implies that the smallest eigenvalue of $\tilde{\rho}_{R}$ is $\lambda_{\mathrm{min}}=\sigma^2_{\mathrm{min}},$ where
$
\sigma_{\mathrm{min}} := \mathrm{min}( \sigma_1, \cdots, \sigma_{d_M} ).
$
Recall that $\sigma$ is full-rank, hence $\sigma_{\mathrm{min}}>0$. With this information, we can update the bound (\ref{eqn:qcmi_leq_11norm_lambda}) to
\begin{equation}\label{eqn:qcmi_leq_11norm}
    I(A:C|B)\leq\frac{4}{\sigma^2_{\mathrm{min}}}\|\Delta\rho_{ABC}\|_1^2.
\end{equation}

\subsection{A (sub)exponential bound on trace distance.}\label{subsec:tr_dist_estimate}

We now show that $\|\Delta\rho_{ABC}\|_1$ in (\ref{eqn:qcmi_leq_11norm}) may be bounded as
\begin{align}\label{eqn:desired_bound}
    \|\Delta\rho_{ABC}\|_1 &\leq O(\|E^{|B|}-\tilde{E}^{|B|}\|) \\ \nonumber
    &\leq O(|B|^{K}\nu_{\mathrm{gap}}^{|B|-K}).
\end{align}

We first recall from Ref. \onlinecite{qcmi-decay} that
\begin{align}\label{eqn:11norm_leq_opnorm}
    \|\Delta\rho_{ABC}\|_1 \leq 2\|V_B-\tilde{V}_B\|.
\end{align}
Since the dimension of the dilating Hilbert space $\dim\mathcal{H}_B=d_s^{|B|}$ is exponentially large, an explicit construction of $\tilde{V}_B$ is not useful. Instead we notice that the quantum channels $\mathcal{E}^{|B|}$ and $\tilde{\mathcal{E}}^{|B|}$ may be dilated with a Hilbert space of dimension $d_M^2$, i.e. a space isomorphic to $\mathcal{H}_M\otimes\mathcal{H}_M$. We will thus explicitly construct the isometries $W,\tilde{W}:\mathcal{H}_M\rightarrow\mathcal{H}_M\otimes\mathcal{H}_M\otimes\mathcal{H}_M$, representing Stinespring dilations of $\mathcal{E}^{|B|}$ and $\tilde{\mathcal{E}}^{|B|}$, respectively. We may then express $V_B$ and $\tilde{V}_B$ in terms of these isometries, as we now show. Since $W$ and $V_B$ represent dilations of the same quantum channel $\mathcal{E}^{|B|}$, then, by Stinespring's theorem \cite{stinespring} and assuming $\dim\mathcal{H}_B\geq d_M^2$, \emph{there exists} an isometry $U:\mathcal{H}_M\otimes\mathcal{H}_M\rightarrow\mathcal{H}_B$, such that 
\begin{equation}\label{eqn:v_b_via_w}
    V_B=(U\otimes\mathbb{1}_M)W.
\end{equation} 
Using this isometry $U$, we may re-express $\tilde{V}_B$ as
\begin{equation}\label{eqn:tv_b_via_tw}
    \tilde{V}_B:=(U\otimes\mathbb{1}_M)\tilde{W}.
\end{equation} 
We will be able to construct $\tilde{W}$ explicitly, and thus $\tilde{V}_B$ follows from (\ref{eqn:tv_b_via_tw}) and $\tilde{M}_B$ follows from (\ref{eqn:tilde_m_b_via_isom}).

We may now express the bound (\ref{eqn:11norm_leq_opnorm}) in terms of the norm distance between $W$ and $\tilde{W}$,
\begin{align}\label{eqn:11norm_leq_opnorm_sm}
    \|\Delta\rho_{ABC}\|_1 &\leq 2\|V_B-\tilde{V}_B\|\\ \nonumber
    & =2\|(U\otimes\mathbb{1}_M)(W-\tilde{W})\| \\ \nonumber
    & =2\|W-\tilde{W}\|,
\end{align}
eliminating the need to explicitly specify $U$, which we will not do. 

We now construct $W$ and $\tilde{W}$ using Choi's methods,\cite{choi} which will enable us to show that $\|\Delta\rho_{ABC}\|_1\leq O(\|E^{|B|}-\tilde{E}^{|B|}\|)$. 

Let $C$ and $\tilde{C}$ be the Choi matrices of $\mathcal{E}^{|B|}$ and $\tilde{\mathcal{E}}^{|B|}$, respectively,
\begin{align}\label{eqn:c_tilde}
    C&=\left(\mathrm{id}\otimes\mathcal{E}^{|B|}\right)\left(|+\rangle\langle +|\right),\\ \nonumber
    \tilde{C}&=\left(\mathrm{id}\otimes\tilde{\mathcal{E}}^{|B|}\right)\left(|+\rangle\langle +|\right)=\sigma\otimes\mathbb{1}_M, 
\end{align}
where we have used (\ref{eqn:tilde_ch_fold_B}).
We further define 
\begin{equation}\label{eqn:delta_c}
    \Delta C:=C-\tilde{C}=\left(\mathrm{id}\otimes\left(\mathcal{E}^{|B|}-\tilde{\mathcal{E}}^{|B|}\right)\right)\left(|+\rangle\langle +|\right),
\end{equation}
from which it follows
\begin{align}\label{eqn:dc_norm_leq_exp}
    \|\Delta C\|_1 &\leq d_M^2\|\mathcal{E}^{|B|}-\tilde{\mathcal{E}}^{|B|}\|_{2\rightarrow 2} \\ \nonumber
    &\leq c_2d_M^2 |B|^{K}\nu_{\mathrm{gap}}^{|B|-K},\\ \nonumber
    \|\Delta C\| &\leq d_M \|\mathcal{E}^{|B|}-\tilde{\mathcal{E}}^{|B|}\|_{2\rightarrow 2}& \\ \nonumber
    &\leq c_2d_M |B|^{K}\nu_{\mathrm{gap}}^{|B|-K},
\end{align}
where we have used (\ref{eqn:22norm_leq_exp}). We aim to express $\|W-\tilde{W}\|$ in terms of $\|\Delta C\|_1$ and $\|\Delta C\|$, obtaining the desired bound (\ref{eqn:desired_bound}). 

We now employ Choi's construction.\cite{choi} Recall that $|+\rangle:=\sum_{i=1}^{d_M}|i\rangle\otimes|i\rangle$ in (\ref{eqn:c_tilde}) is defined in our chosen orthonormal product basis, which satisfies $(\mathbb{1}\otimes\sigma)|i\rangle\otimes|j\rangle=\sigma_j|i\rangle\otimes|j\rangle$ and $(\sigma\otimes\mathbb{1})|i\rangle\otimes|j\rangle=\sigma_i|i\rangle\otimes|j\rangle$ for all $i,j=1,2,\dots,d_M$. With respect to these basis, we decompose
\begin{align}\label{eqn:choi-tilde}
    \tilde{C}=\sum_{i,j=1}^{d_M} |\tilde{\psi}_{ij}\rangle\langle\tilde{\psi}_{ij}|,\\ \nonumber
    |\tilde{\psi}_{ij}\rangle:=\sqrt{\sigma_{i}}|i\rangle\otimes|j\rangle.
\end{align}
Notice that the vectors $|\tilde{\psi}_{ij}\rangle$ are pairwise orthogonal, $\langle\tilde{\psi}_{i'j'}|\tilde{\psi}_{ij}\rangle=\sqrt{\sigma_i\sigma_{i'}}\delta_{ii'}\delta_{jj'}$, and form a basis in $\mathcal{H}_M\otimes\mathcal{H}_M$. We define a set of projections $P_k:\mathcal{H}_M\otimes\mathcal{H}_M\rightarrow\mathcal{H}_M$ with $k=1,2,\dots,d_M$ by 
\begin{equation}\label{eqn:proj}
    P_k:=\mathbb{1}_M\otimes\langle k|.
\end{equation}
Now Choi's construction \cite{choi} provides the isometry $\tilde{W}$ dilating $\tilde{\mathcal{E}}^{|B|}$ explicitly in the form
\begin{align}\label{eqn:w_tilde_choi}
    \tilde{W}:=\sum_{k,i,j=1}^{d_M}\left(P_k|\tilde{\psi}_{ij}\rangle\otimes|i\rangle\otimes|j\rangle\right)\langle k|.
\end{align}

Now we assume a decomposition of $C$ of the form,
\begin{equation}\label{eqn:c_decomp}
    C=\sum_{i,j=1}^{d_M}|\psi_{ij}\rangle\langle\psi_{ij}|,
\end{equation}
where the vectors $|\psi_{ij}\rangle$ are not necessarily orthogonal. Recall that Choi's construction does not require the vectors $|\psi_{ij}\rangle$ in the decomposition (\ref{eqn:c_decomp}) to be orthogonal. Using the projectors $P_k$ defined in (\ref{eqn:proj}), we employ Choi's construction again, obtaining the isometry
\begin{align}\label{eqn:w_choi}
    W:=\sum_{k,i,j=1}^{d_M}\left(P_k|\psi_{ij}\rangle\otimes|i\rangle\otimes|j\rangle\right)\langle k|,
\end{align}
which dilates $\mathcal{E}^{|B|}$.

Since $|\tilde{\psi}_{ij}\rangle$ form a basis, we can express,
\begin{align}\label{eqn:psi_decomp}
    |\psi_{ij}\rangle&=|\tilde{\psi}_{ij}\rangle+\sum_{m,n=1}^{d_M}\gamma_{mn;ij}|\tilde{\psi}_{mn}\rangle\\ \nonumber
    &:=\left(\mathbb{1}_{M\otimes M}+\Gamma\right)|\tilde{\psi}_{ij}\rangle,
\end{align}
where $\gamma_{mn;ij}\in\mathbb{C}$ are some coefficients, and where we have defined
\begin{equation*}
    \Gamma:=\sum_{i,j,m,n=1}^{d_M}\gamma_{mn;ij}\sigma^{-1}_i|\tilde{\psi}_{mn}\rangle\langle\tilde{\psi}_{ij}|.
\end{equation*}
We express the distance in operator norm between the constructed isometries $W$ and $\tilde{W}$ in terms of $\Gamma$,
\begin{align*}
    \|W-\tilde{W}\|^2 &=\|(W-\tilde{W})^\dagger(W-\tilde{W})\| \\ \nonumber
    &\leq \|(W-\tilde{W})^\dagger(W-\tilde{W})\|_1 \\ \nonumber
    &= \mathrm{tr}\left((W-\tilde{W})^\dagger(W-\tilde{W})\right)\\ \nonumber
    &=\sum_{i,j,m,n=1}^{d_M} |\gamma_{mn;ij}|^2\sigma_m\\ \nonumber
    &=\mathrm{tr}\left(\Gamma (\sigma\otimes\mathbb{1}_M)\Gamma^\dagger\right)\\ \nonumber
    &=\mathrm{tr}\left(\Gamma \tilde{C}\Gamma^\dagger\right).
\end{align*}

We will construct such $\Gamma$, that (\ref{eqn:c_decomp}) is consistent with (\ref{eqn:psi_decomp}), and so that $\|W-\tilde{W}\|=O(\|\Gamma\|_1)=O(\|\Delta C\|_1)$. To find $\Gamma$, we proceed as follows, and note that from (\ref{eqn:choi-tilde}), (\ref{eqn:c_decomp}), and (\ref{eqn:psi_decomp}),
\begin{align}\label{eqn:gamma_eqn}
    \Delta C &= C-\tilde{C} \\ \nonumber
    &= \left(\mathbb{1}_{M\otimes M}+\Gamma\right) \tilde{C} (\mathbb{1}_{M\otimes M}+\Gamma^\dagger) -\tilde{C}\\ \nonumber
    &= \Gamma\tilde{C} + \tilde{C}\Gamma^\dagger + \Gamma\tilde{C}\Gamma^\dagger.
\end{align}
We make the definitions,
\begin{align}\label{eqn:substitutions}
    \Delta C'&:=\tilde{C}^{-\frac{1}{2}}\Delta C\tilde{C}^{-\frac{1}{2}} \\ \nonumber
    &= (\sigma^{-\frac{1}{2}}\otimes \mathbb{1}_M)\Delta C(\sigma^{-\frac{1}{2}}\otimes \mathbb{1}_M), \\ \nonumber
    \Gamma' &:=\tilde{C}^{-\frac{1}{2}}\Gamma\tilde{C}^{\frac{1}{2}}\\ \nonumber
    &= (\sigma^{-\frac{1}{2}}\otimes\mathbb{1}_M)\Gamma(\sigma^{\frac{1}{2}}\otimes\mathbb{1}_M),
\end{align}
transforming the equation (\ref{eqn:gamma_eqn}) into
\begin{equation}\label{eqn:gamma_prime_eqn}
    \Gamma' + \Gamma'^\dagger +\Gamma'\Gamma'^\dagger = \Delta C'.
\end{equation}
Notice that from (\ref{eqn:substitutions}) it follows that 
\begin{equation}\label{eqn:dcpr_leq_dc}
    \|\Delta C'\|_1\leq \frac{1}{\sigma_{\mathrm{min}}}\|\Delta C\|_1\leq \frac{c_2d^2_M}{\sigma_{\mathrm{min}}} |B|^{K}\nu_{\mathrm{gap}}^{|B|-K},
\end{equation}
where we have used (\ref{eqn:dc_norm_leq_exp}). This implies that $\Delta C'\rightarrow 0$ as $|B|\rightarrow\infty$. We wish to find $\Gamma'$ satisfying (\ref{eqn:gamma_prime_eqn}), such that $\|\Gamma'\|_1=O(\|\Delta C'\|_1)$, and via (\ref{eqn:dcpr_leq_dc}), $\|\Gamma\|_1= O(\|\Delta C\|_1)$. Since any solution of (\ref{eqn:gamma_prime_eqn}) satisfying this condition will suffice, and observing that $\Delta C'$ is Hermitian, we shall seek a Hermitian solution, $\Gamma'=\Gamma'^\dagger$. 

For $\Gamma'$ Hermitian, equation (\ref{eqn:gamma_prime_eqn}) reduces to a quadratic equation,
\begin{equation*}
    \Gamma'^2 + 2\Gamma' - \Delta C'=0,
\end{equation*}
which indeed has a Hermitian solution of small norm,
\begin{equation}\label{eqn:gamma_prime}
    \Gamma'=\sqrt{\mathbb{1}_{M\otimes M}+\Delta C'}-\mathbb{1}_{M\otimes M},
\end{equation}
provided that $\mathbb{1}_{M\otimes M}+\Delta C' \geq 0$, which is satisfied for large enough $|B|$.
Notice that through (\ref{eqn:substitutions}), (\ref{eqn:psi_decomp}), and (\ref{eqn:w_choi}), $\Gamma'$ explicitly defines $W$.

From (\ref{eqn:gamma_prime}) follows,
\begin{equation*}
    \Gamma'\Gamma'^\dagger=\Gamma'^2=2\mathbb{1}_{M\otimes M}+\Delta C'-2\sqrt{\mathbb{1}_{M\otimes M}+\Delta C'}.
\end{equation*}
Assuming $|B|$ large enough that $\|\Delta C'\|\leq 2^{-\frac{3}{2}}$, by Taylor expansion, we estimate
\begin{equation}\label{eqn:gamma_sq}
    0\leq \Gamma'^2 \leq \frac{\Delta C'^2}{2}. 
\end{equation}
Expressing $\|W-\tilde{W}\|$ in terms of $\Gamma'$, we find
\begin{equation*}
    \|W-\tilde{W}\|^2 \leq \mathrm{tr}\left(\Gamma \tilde{C}\Gamma^\dagger\right) =\mathrm{tr}\left(\tilde{C}\Gamma'\Gamma'^\dagger\right),
\end{equation*}
leading to
\begin{align}\label{eqn:opnorm_leq_22norm}
    \|W-\tilde{W}\|^2 & \leq \frac{1}{2} \mathrm{tr}\left(\tilde{C}\Delta C'^2\right) \\ \nonumber
    &=\frac{1}{2}\mathrm{tr}\left(\Delta C^2\tilde{C}^{-1}\right)\\ \nonumber
    &\leq \frac{1}{2}\|\tilde{C}^{-1}\|\|\Delta C\|\|\Delta C\|_1 \\ \nonumber
    &= \frac{1}{2\sigma_{\min}}\|\Delta C\|\|\Delta C\|_1 \\ \nonumber
    &\leq \frac{d_M^3}{2\sigma_{\min}}\|\mathcal{E}^{|B|}-\tilde{\mathcal{E}}^{|B|}\|_{2\rightarrow 2}^2,
\end{align}
where we have used (\ref{eqn:c_tilde}), (\ref{eqn:dc_norm_leq_exp}), (\ref{eqn:substitutions}), and $\|\tilde{C}^{-1}\|=\sigma^{-1}_{\mathrm{min}}$.

Combining (\ref{eqn:11norm_leq_opnorm}), (\ref{eqn:11norm_leq_opnorm_sm}), (\ref{eqn:opnorm_leq_22norm}), and (\ref{eqn:22norm_leq_exp}) together, we have
\begin{equation}\label{eqn:tr_dist_leq_exp}
    \|\Delta\rho_{ABC}\|_1\leq c_2d_M\sqrt{\frac{2d_M}{\sigma_{\mathrm{min}}}}|B|^{K}\nu_{\mathrm{gap}}^{|B|-K},
\end{equation}
proving that (an injective) iuMPS is an approximate quantum Markov chain.

Finally, using (\ref{eqn:qcmi_leq_11norm}), we obtain the desired bound on QCMI,
\begin{equation}\label{eqn:qcmi_triv_bound}
    I(A:C|B)\leq \frac{8d_M^3c_2^2}{\sigma_{\min}^3}e^{-2\ln\nu^{-1}_{\mathrm{gap}}\,(|B|-K)+2K\ln|B|},
\end{equation}
which corresponds to the statement of Theorem \ref{thrm:theorem_1} with $Q=\frac{8d_M^3c_2^2}{\sigma_{\min}^3}$, $q=2\ln\nu^{-1}_{\mathrm{gap}}$. This proves Theorem (\ref{thrm:theorem_1}) for the case of injective iuMPS. 

\noindent\textbf{Remarks}:\ 
\begin{enumerate}
    \item Notice that the inequality (\ref{eqn:tr_dist_leq_exp}) has the form $\|\Delta\rho_{ABC}\|_1\leq O(\|\mathcal{E}^{|B|}-\tilde{\mathcal{E}}^{|B|}\|_{2\rightarrow 2})$, in contrast to the bound $\|\Delta\rho_{ABC}\|_1\leq O(\sqrt{\|\mathcal{E}^{|B|}-\tilde{\mathcal{E}}^{|B|}\|_{2\rightarrow 2}})$ that we have obtained in Theorem II.1 of Ref. \onlinecite{qcmi-decay}. This improvement modifies the rate of exponential decay in (\ref{eqn:qcmi_triv_bound}) by a factor of $2$, in addition to the factor of $2$ discussed in Remark 1 in Subsection \ref{subsec:qcmi_estmate}. The rate of the exponential decay in the final bound (\ref{eqn:qcmi_triv_bound}) is thus four times larger than the one in Theorem II.1 of Ref. \onlinecite{qcmi-decay}. 
    \item The numerical results presented in Section \ref{sec:stepwise_procedure_numerics} are much closer to the new bound on the decay rate than the earlier value given in Ref. \onlinecite{qcmi-decay}. 
    \item Another improvement over Theorem II.1 of Ref. \onlinecite{qcmi-decay} is that the prefactor in (\ref{eqn:qcmi_triv_bound}) is independent of the size of any region of the spin chain. As a drawback, we gained the dependence on $\sigma^{-3}_{\mathrm{min}}$, which may be a large number. 
\end{enumerate}

\subsection{General case: infinite uniform matrix product states.}\label{subsec:generalization}
Now we outline the changes required to extend the proof to the general class of iuMPS. While the main steps of the proof remain the same, we have to account for the cases in which iuMPS is not injective. These entail changes in the peripheral part $\tilde{E}$ (\ref{eqn:tr_matr_periph}) of the transfer matrix $E$ (\ref{eqn:tr_matr}). Since the modifications we make involve the equations (\ref{eqn:rho_r_prime}) and (\ref{eqn:c_tilde}), formulated in terms of the quantum channel $\tilde{\mathcal{E}}$ (represented by $\tilde{E}$), we focus attention on $\tilde{\mathcal{E}}$, rather than $\tilde{E}$.

In the case of a general iuMPS, the quantum channel $\tilde{\mathcal{E}}:\mathcal{B}(\mathcal{H}_M)\rightarrow\mathcal{B}(\mathcal{H}_M)$ has the following properties \cite{fcs} (see also Propositions 1 and 2 in Ref. \onlinecite{qcmi-decay}):
\begin{itemize}
    \item The quantum channel $\tilde{\mathcal{E}}$ decomposes into a sum of components, $\tilde{\mathcal{E}}=\sum_{\alpha=1}^{N}\tilde{\mathcal{E}}_{\alpha}$, $\tilde{\mathcal{E}}_\alpha:\Pi^\alpha\mathcal{B}(\mathcal{H}_M)\Pi^\alpha\rightarrow \Pi^\alpha\mathcal{B}(\mathcal{H}_M)\Pi^\alpha$, defined by the set of projectors $\{\Pi^\alpha\}_{\alpha=1}^N$, $\sum_{\alpha=1}^N\tilde{\Pi}^{\alpha}=\mathbb{1}_M$. This implies that the fixed point $\sigma$ of $\mathcal{E}$ and $\tilde{\mathcal{E}}$ satisfies $\sigma=\sum_{\alpha=1}^N \Pi^\alpha \sigma \Pi^\alpha$.  
    \item For each $\alpha$, the component $\tilde{\mathcal{E}}_\alpha$ admits a further decomposition, $$\tilde{\mathcal{E}}_\alpha(X)=p_\alpha\sum_{k=0}^{p_\alpha-1}\mathrm{tr}\left(\Pi^\alpha_k X\Pi^\alpha_k\right)(\Pi^\alpha_{k+1}\sigma\Pi^\alpha_{k+1}),$$
    defined by the set of projectors $\Pi^\alpha_k$ with $k=0,\dots,p_\alpha-1$, satisfying $\sum_{k=0}^{p_\alpha-1}\Pi^\alpha_k=\Pi^\alpha$. The index $k$ in $\Pi^\alpha_k$ is taken modulo $p_\alpha$. This implies that $\sigma$ satisfies $\sigma=\sum_{\alpha=1}^N\sum_{k=0}^{p_\alpha-1} \Pi_k^\alpha \sigma \Pi_k^\alpha$.
\end{itemize}

We first consider the case when $N=1$, where there is a single component in the direct sum decomposition $\tilde{\mathcal{E}}=\bigoplus_{\alpha=1}^{N}\tilde{\mathcal{E}}_{\alpha}$. We allow, however, arbitrary period $p\geq 1$. Under these assumptions and using (\ref{eqn:c_tilde}),
\begin{equation*}
    \tilde{C}=p\sum_{k=0}^{p-1} \Pi_{k+|B|}\sigma \Pi_{k+|B|} \otimes \Pi_k.
\end{equation*}
While for $p>1$, the matrix $\tilde{C}$ is not full-rank, it still satisfies $\mathrm{supp}(\Delta C)\subseteq \mathrm{supp}(\tilde{C})$ and $\mathrm{supp}(C)\subseteq\mathrm{supp}(\tilde{C})$, which we prove in Appendix \ref{appendix:supports}. Hence, the derivations of Section \ref{subsec:tr_dist_estimate} may still be applied, if we consider expressions to be restricted to the support of $\tilde{C}$ where necessary. The inverse of $\tilde{C}$ should be considered to be defined on $\mathrm{supp}(\tilde{C})$. With this convention we may estimate,
\begin{equation*}
    \|\tilde{C}^{-1}\|=\frac{1}{p\sigma_{\min}}.
\end{equation*}
The bound (\ref{eqn:opnorm_leq_22norm}) is still valid if we update the value of $\|\tilde{C}^{-1}\|$,
\begin{equation*}
    \|W-\tilde{W}\|^2 \leq \frac{d_M^3}{2p\sigma_{\min}}\|\mathcal{E}^{|B|}-\tilde{\mathcal{E}}^{|B|}\|_{2\rightarrow 2}^2.
\end{equation*}

The density operator $\tilde{\rho}_{R}$, defined in (\ref{eqn:rho_r_prime}), changes from (\ref{eqn:rho_r_prime_inj}) to 
\begin{align*}
    &\tilde{\rho}_{R}=  p J\sum_{k=0}^{p-1}\Pi_k \sigma\Pi_k\otimes\sqrt{\sigma}\Pi_{k+|R|}\sqrt{\sigma}J^\dagger \\ \nonumber
    &= p J\sum_{k=0}^{p-1}\Pi_k \sigma\Pi_k\otimes\Pi_{k+|R|}\sigma\Pi_{k+|R|} J^\dagger\\ \nonumber
    &= p J\sum_{k=0}^{p-1}\left(\Pi_k \otimes\Pi_{k+|R|}\right)(\sigma\otimes\sigma)\\ \nonumber
    &\qquad\qquad\qquad\qquad \times\left(\Pi_k \otimes\Pi_{k+|R|}\right) J^\dagger,
\end{align*}
where $J:\mathcal{H}_M\otimes\mathcal{H}_M\rightarrow\mathcal{H}_R$ is an isometry, and where we have used $\sigma=\sum_{k=0}^{p-1}\Pi_k\sigma\Pi_k$. This implies that for any $\tilde{\rho}_R$
its smallest eigenvalue $\lambda_{\mathrm{min}}$ is bounded from below by 
\begin{equation*}
    \lambda_{\mathrm{min}}\geq p\sigma^2_{\mathrm{min}}.
\end{equation*}
This modifies the pre-factor in the bound (\ref{eqn:qcmi_leq_11norm}) from $\frac{4}{\sigma^2_{\mathrm{min}}}$ to $\frac{4}{p\sigma^2_{\mathrm{min}}}$. Thus, the overall bound (\ref{eqn:qcmi_triv_bound}) gains a factor $p^{-2}\leq 1$,
\begin{equation}\label{eqn:qcmi_per_bound}
    I(A:C|B)\leq \frac{8d_M^3c_2^2}{p^2\sigma_{\min}^3}e^{-2\ln\nu^{-1}_{\mathrm{gap}}\,(|B|-K)+2K\ln|B|},
\end{equation}
proving Theorem \ref{thrm:theorem_1} for the case $N=1$ and $p\geq 1$.
 
Generalizing to the case of $N\geq 1$ we notice that $\tilde{C}$ obtains a direct sum structure,
\begin{equation}\label{eqn:C_tilde_gen}
    \tilde{C}=\bigoplus_{\alpha=1}^N p_\alpha\sum_{k=0}^{p_\alpha-1} \Pi^\alpha_{k+|B|}\sigma \Pi^\alpha_{k+|B|} \otimes \Pi^\alpha_k.
\end{equation}  

The density operator $\tilde{\rho}_{R}$ also gets a direct sum structure,
\begin{multline}\label{eqn:rho_prime_gen}
        \tilde{\rho}_{R}=J\bigoplus_{\alpha=1}^N  p_\alpha\sum_{k=0}^{p_\alpha-1}\left(\Pi^\alpha_k \otimes\Pi^\alpha_{k+|R|}\right) \\ (\sigma\otimes\sigma)\left(\Pi^\alpha_k \otimes\Pi^\alpha_{k+|R|}\right) J^\dagger.
\end{multline}

Proceeding as in the case of $N=1$, the equations (\ref{eqn:C_tilde_gen}) and (\ref{eqn:rho_prime_gen}) imply that the bound (\ref{eqn:tr_dist_leq_exp}) is modified to
\begin{align}\label{eqn:gen_tr_dist_leq_exp}
    &\|\Delta\rho_{ABC}\|_1\leq \\ \nonumber
    &\qquad \max_\alpha\left( \sqrt{\frac{d^3_{M_\alpha}}{p_\alpha}}\right)c_2\sqrt{\frac{2}{\sigma_\mathrm{min}}}|B|^{K}\nu_{\mathrm{gap}}^{|B|-K},
\end{align}
and the bound (\ref{eqn:qcmi_per_bound}) to
\begin{align}\label{eqn:qcmi_gen_bound}
    &I(A:C|B)\leq \\ \nonumber
    &\quad \max_\alpha \left(\frac{d_{M_\alpha}^3}{p_\alpha^2}\right)\frac{8c_2^2}{\sigma_{\min}^3}e^{-2\ln\nu^{-1}_{\mathrm{gap}}\,(|B|-K)+2K\ln|B|},
\end{align}
where $d_{M_\alpha}=\dim (\Pi^\alpha\mathcal{H}_M)$. The bound (\ref{eqn:gen_tr_dist_leq_exp}) implies that any iuMPS is an approximate quantum Markov chain for sufficiently large $|B|$. We note that since $p_\alpha\geq 1$ and $d_{M_\alpha}\leq d_M$, the bounds (\ref{eqn:tr_dist_leq_exp}) and (\ref{eqn:qcmi_triv_bound}) hold for the general class of iuMPS.
\vfill
\section{Numerical calculation of QCMI \label{sec:stepwise_procedure_numerics}}

\subsection{Constructing iuMPS \label{subsec:constructing_iumps}}

\gramm

Recall that an iuMPS is completely characterized by reduced density operators $\rho_n$ on finite continuous regions of $n$ sites (refer to (\ref{eqn:general_rho_n})). In turn, $\rho_n$ is characterized by the fixed point of the quantum channel $\mathcal{E}$ and the choice of the set $\mathcal{S}$. The quantum channel $\mathcal{E}$ is represented by a transfer matrix $E$ that may have a single or multiple fixed point(s) (eigenvectors corresponding to eigenvalue 1 of $E$). There are three cases that account for all iuMPS:\par

\noindent \textit{Case~1.}\ When $E$ has a single fixed point, and,
\begin{align*}
	\mathcal{S} := \cb{ \left. M^s \right\rvert M^s \in \mathbb{C}^{d_M \times d_M}, \forall s \in \cb{1, 2, \dots, d_s}},
\end{align*} 

\noindent \textit{Case~2.}\ When $E$ has two fixed points and $\rho_n$ is of period 1, and,
\begin{multline*}
	\mathcal{S} :=  \\
    \left\{\left.  M^s := \begin{bmatrix}
	M_1 ^s & 0 \\
	0 & M_2 ^s
	\end{bmatrix} \right\rvert M_1 ^s , M_2 ^s \in \mathbb{C}^{\frac{d_M}{2} \times \frac{d_M}{2}}, \right. \\
	\left. M^s \in \mathbb{C}^{d_M \times d_M}, \forall s \in \cb{1, 2, \dots, d_s} \right.\Bigg\},
\end{multline*}

\noindent \textit{Case~3.}\ When $E$ has two fixed points and $\rho_n$ is of period 2,
\begin{multline*}
	\mathcal{S} := \\
    \left\{\left. M^s := \begin{bmatrix}
	0 & M_1 ^s \\
	M_2 ^s & 0
	\end{bmatrix} \right\rvert M_1 ^s , M_2 ^s \in \mathbb{C}^{\frac{d_M}{2} \times \frac{d_M}{2}}, \right. \\
	\left. M^s \in \mathbb{C}^{d_M \times d_M}, \forall s \in \cb{1, 2, \dots, d_s} \right.\Bigg\},
\end{multline*}

The matrices $M^s$ in the set $\mathcal{S}$ in \textit{Case~1} are constructed as follows,
\begin{align}\label{eqn:numerical_methods_1}
    \p{M^s}_{ij} = \sum_{s' = 1} ^{d_s} \sum_{j' = 1} ^{d_M} U_{ij'} ^{s s'} \Psi_{j'j} ^{s'},
    \end{align}
where the unitary $U \in \mathbb{C}^{d_s d_M\times d_s d_M}$ is selected randomly with respect to Haar measure and $\Psi \in \mathbb{C}^{d_s d_M \times d_M}$ is given by,
\begin{align}\label{eqn:numerical_methods_2}
	\Psi := \frac{1}{\sqrt{3}} \begin{bmatrix} 1 \\ 1 \\ 1 \end{bmatrix} \otimes \mathds{1}_{ d_M},
\end{align}
where we fix $d_s = 3$ and $d_M = 4$ in our numerical calculations. In \textit{Case~2} and \textit{Case~3}, each of the two matrices $M_1 ^s$ and $M_2 ^s$ is constructed using an equation of the form (\ref{eqn:numerical_methods_1}) and (\ref{eqn:numerical_methods_2}), but with $d_M$ replaced by $\frac{d_M}{2}$, and different and independent choices of the unitary $U$.
    
After constructing $\mathcal{S}$, the left and right fixed points of the transfer matrix $E$ denoted by $|+\rangle$ and $|\sigma \rangle$ in (\ref{eqn:general_rho_n}), respectively, remain to be found to completely specify the iuMPS. Recall, fixed points of $E$ are its left and right eigenvectors corresponding to a left and right eigenvalue of 1, respectively. If there are multiple fixed points of $E$, in other words, if (left and/or right) eigenvalue of 1 of $E$ is degenerate, then the fixed point used to define iuMPS is the uniform linear combination of the multiple fixed points. That is, the left and/or right eigenvectors are defined as the normalized sum of all left and/or right eigenvectors, respectively.

\subsection{Calculating entropies}\label{sec:numerical_methods_results}

\figorthog

\projected

QCMI is computed from von Neumann entropies of the states on the continuous subregions $AB$, $BC$, $B$ and $ABC$ that are in turn computed from the eigenvalues of the reduced density operators for the respective subregions. A numerical check of Theorem~\ref{thrm:theorem_1} requires QCMI for increasing size, $\left\lvert B \right\rvert$, of the separating region $B$. Direct diagonalization is computationally inefficient, because even linear growth of $\left\lvert B \right\rvert$ would imply exponential growth of the Hilbert space dimension of any subregion that includes $B$. To overcome this fundamental issue of dimensionality, we use tensor network contractions to project reduced density operators on to their respective supports.\footnote{The benefit of finding eigenvalues of $\rho_n$ by projecting $\rho_n$ onto its support versus direct diagonalization is revealed by comparing how the memory required to implement the two schemes scales with $n$. Direct diagonalization requires knowledge of all elements of $\rho_{n}$, that is $O\p{d_s ^n}$ real numbers. Projecting $\rho_n$ onto its support requires storing and multiplying $2n$ copies of $M$, $n$ copies each of $W$ and $W^\dagger$ and 2 copies of $\Sigma^{-\frac{1}{2}}$ (refer to (\ref{eqn:numerical_methods_4})), that is, $O\p{2n d_s d_M ^2 + 2d_M ^4} = O\p{n}$ real numbers. Therefore, the projection method is exponentially more efficient than direct diagonalization.}

Consider an iuMPS, $\rho_n$ given by (\ref{eqn:general_rho_n}). The support of $\rho_n$ is constructed as follows. Define $G \in \mathbb{C}^{d_M ^2 \times d_M ^2}$ by,
\begin{multline}\label{eqn:numerical_methods_3}
    G := E^n= \sum_{s_1, s_2, \dots, s_n = 1} ^{d_s} \p{M^{s_n} \otimes \bar{M}^{s_n}} \dots \\
    \dots \p{M^{s_2} \otimes \bar{M}^{s_2}} \p{M^{s_1} \otimes \bar{M}^{s_1}},
\end{multline}
which can be numerically computed efficiently. Then, express the matrix elements of the permutation of $G$, defined by $\tilde{G}_{ij;i'j'} := G_{jj';ii'}$, in terms of its spectral decomposition,
\begin{align}\label{eqn:numerical_methods_4}
    \tilde{G}_{ij;i'j'} := \sum_{l, l' = 1} ^{d_M ^2} W_{ij;l} \Sigma_{l;l'} \overline{W_{i'j';l'}},
\end{align}
which can also be computed efficiently, and where $i, j, i', j' \in \cb{1, 2, \dots d_M}$, unitary $W \in \mathbb{C}^{d_M ^2 \times d_M ^2}$, and $\Sigma$ is the diagonal matrix of eigenvalues of $\tilde{G}$. Equations (\ref{eqn:numerical_methods_3}) and (\ref{eqn:numerical_methods_4}) are diagrammatically depicted in Fig.~\ref{dg:numerical_methods_diagram_1}. The equation (\ref{eqn:numerical_methods_4}) implies that the support of $\rho_n$, denoted as $\mathrm{supp}(\rho_n)$, is isomorphic to $\mathcal{H}_M\otimes\mathcal{H}_M$. We define isometry $P:\mathcal{H}_M\otimes\mathcal{H}_M\rightarrow\mathcal{H}^{\otimes n}_s$, with $\mathrm{range}(P)=\mathrm{supp}(\rho_n)$, defined by its matrix elements, 
\begin{multline}\label{eqn:numerical_methods_5}
    P_{s_1, s_2, \dots s_n;l} := 
    \sum_{i, j = 1} ^{d_M} \sum_{k = 1} ^{d_M ^2} \overline{\p{M^{s_n} \dots M^{s_2} M^{s_1}}_{ij}} \times \\ 
    \times W_{ij; k} \Sigma^{-\frac{1}{2}}_{k;l}.
\end{multline}
The isometry $P$ satisfies
\begin{align}\label{eqn:numerical_methods_7}
    P^\dagger \mathds{1}_{d_s ^n}P =\mathds{1}_{d_M ^2},
\end{align}
as diagrammatically verified in Fig.~\ref{dg:numerical_methods_diagram_2}. The projection of $\rho_n$ onto its support is given by
\begin{equation}\label{eqn:rho_n_projected}
    P^\dagger\rho_n P=\Sigma^{\frac{1}{2}}W^\dagger(\sigma\otimes\mathbb{1}_{d_M})W\Sigma^{\frac{1}{2}},
\end{equation}
a $d_M^2\times d_M^2$ matrix, with non-zero eigenvalues identical to those of $\rho_n$. Figure~\ref{dg:numerical_methods_diagram_3} depicts $P^\dagger \rho_n P$. From (\ref{eqn:rho_n_projected}) we observe that the $d_M^2\times d_M^2$ matrix $P^\dagger \rho_n P$ can be efficiently computed from $W$ and $\Sigma$, without explicitly computing $P$, which would be prohibitively expensive. The von Neumann entropy, denoted by $S\p{\rho_n}$, is then found by computing the eigenvalues $\lambda_i$ of $P^\dagger\rho_n P$, and is given by
\begin{align}\label{eqn:numerical_methods_8}
    S\p{\rho_n} := -\sum_{i = 1} ^{d_M ^2} \lambda_i \ln \lambda_i.
\end{align}
Finally, we repeat the above projection procedure, finding the non-zero eigenvalues and entropies for each of the four reduced density operators appearing in the definition of QCMI (\ref{eqn:qcmi_log}).

\noindent\textbf{Remark:}\\
Alternatively, to calculate the entropy $S(\rho_R)$ for the reduced density operator on the continuous region $R$, we could have used the relation analogous to (\ref{eqn:rho_r_prime}), realising that $\rho_R$ and 
\begin{align}\label{eqn:rho_r_isom}
    J^\dagger \rho_R J &= (\mathcal{E}^{|R|}\otimes \mathrm{id}_{M'})(|\sqrt{\sigma}\rangle\langle\sqrt{\sigma}|))\\ \nonumber
    &\in \mathcal{B}(\mathcal{H}_M\otimes\mathcal{H}_M)
\end{align}
have the same eigenvalues and, thus, entropies. The numerical calculation would be of comparable complexity to the one we have chosen to employ. 

\section{Numerical verification of Theorem~\ref{thrm:theorem_1} \label{sec:numerical_verification}}

In this section, we present numerical calculations of QCMI for a randomly generated collection of iuMPS. The main result is given in Fig.~\ref{fig:numerical_methods_figure_2}, which (1) shows that the QCMI decays approximately exponentially as a function of $|B|$, and (2) compares its numerical decay rate with the decay rate $q = 2\ln\p{\nu_{\mathrm{gap}} ^{-1}}$ of the bounding function $$I(A:C|B)\leq Qe^{-q(|B|-K)+2K\ln|B|},$$
derived in Theorem~\ref{thrm:theorem_1}. Before presenting Fig.~\ref{fig:numerical_methods_figure_2}, we discuss the choice of parameters for the numerical calculations and the results of benchmarking our tensor network algorithm against the analytical example in Appendix~\ref{appendix:a}.

\subsection{Choice of parameters \label{subsec:choice_of_parameters}}
There are 4 parameters in the numerical calculations: sizes of subregions $A$ and $C$ and the dimensions of the physical and virtual spaces $d_s$ and $d_M$, respectively. We fix $d_s = 3$ and $d_M = 4$, and the sizes of subregions $A$ and $C$ are held fixed at 1 lattice site each. $\left\lvert B \right\rvert$ is varied from 2 to 40 in multiples of 2 (to decrease the complexity of the tensor network algorithm for projecting the reduced density operators onto their supports).

\subsection{Benchmarking the numerical calculations}

\begin{figure}
    \centering
    \begin{tikzpicture}
        \node (image) at (0,0)
        {\includegraphics[width=0.4\textwidth]{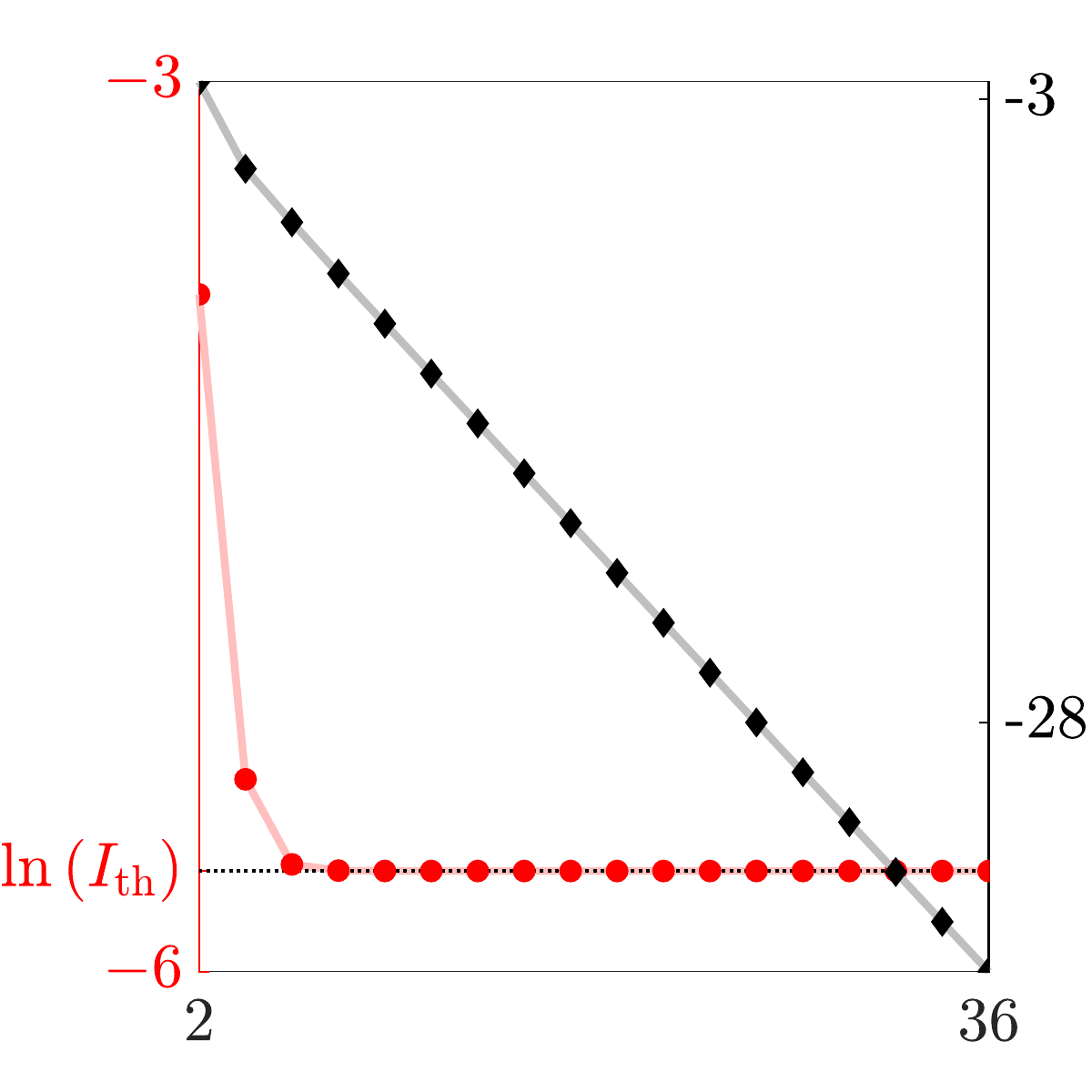}};
        \node[below=of image, node distance=0cm, yshift=0.4in] {$\lvert B \rvert$};
        \node[left=of image, node distance=0cm, rotate=90, anchor=center,yshift=-0.625in]{{\color{Red}$\ln\p{I\p{A : C}}$}};
        \node[right=of image, node distance=0cm, rotate=90, anchor=center,yshift=0.35in]{$\ln\p{I\p{A : C \left\lvert B\right.}}$};
    \end{tikzpicture}
    \caption{The numerically computed QMI (in red circles) and QCMI (in black diamonds) for the example in Appendix~\ref{appendix:a} that converge to a finite value and exponentially to 0, respectively. Left (red) and right (black) hand side vertical axes represent the natural logarithms of QMI and QCMI, respectively. The horizontal axes represents the size of the separating region, $\lvert B\rvert$. The solid light red and gray lines connecting the red circles and black diamonds guide the readers' eyes towards the trends in QMI and QCMI, respectively. The theoretically computed QMI value for $\left \lvert B \right \rvert \rightarrow \infty$ is denoted by $I_{\mathrm{th}}$ (dotted black line) and is given in the text. The choice of parameter values is given in Sec.~\ref{subsec:choice_of_parameters}.}
    \label{fig:numerical_methods_figure_1}
\end{figure}
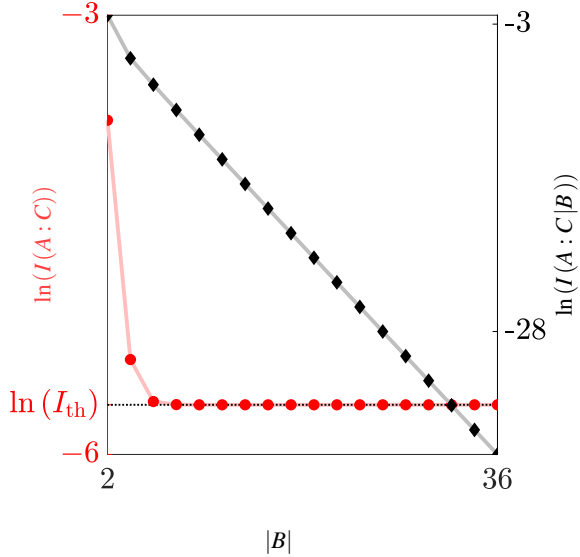

In the example discussed in Appendix~\ref{appendix:a}, quantum mutual information (QMI), defined by $$I(A:C)=S(\rho_A)+S(\rho_C)-S(\rho_{AC}),$$ is analytically shown to converge to the non-zero finite value, $I_{\mathrm{th}} := \lim_{\left \lvert B\right \rvert \rightarrow \infty} I\p{A:C}$, given by, 
\begin{align*}
    \nonumber
    I_{\mathrm{th}} &= 17 \left. \ln\p{2} \right / 16 - 9 \left. \ln \p{3} \right / 8 + 5 \left. \ln \p{5} \right / 16, \\ 
    \nonumber
    &\approx 3.5\times 10^{-3}.
\end{align*}
In Fig.~\ref{fig:numerical_methods_figure_1}, we show the corresponding numerically computed QMI as a function of $|B|$ in red. We observe that already for $\left \lvert B \right \rvert = 26$ the QMI is identical to $I_{\mathrm{th}}$ down to machine precision ($O(10^{-15})$). This agreement between the analytical and numerical QMI values benchmarks our tensor network algorithm. With the same algorithm, we compute the QCMI for the considered example and show the result in black in Fig.~\ref{fig:numerical_methods_figure_1}. We observe that the QCMI decays exponentially, which is consistent with Theorem~\ref{thrm:theorem_1}.

\begin{figure}
    \centering
    \begin{tikzpicture}
        \node (image) at (0,0)
        {\includegraphics[width=0.40\textwidth]{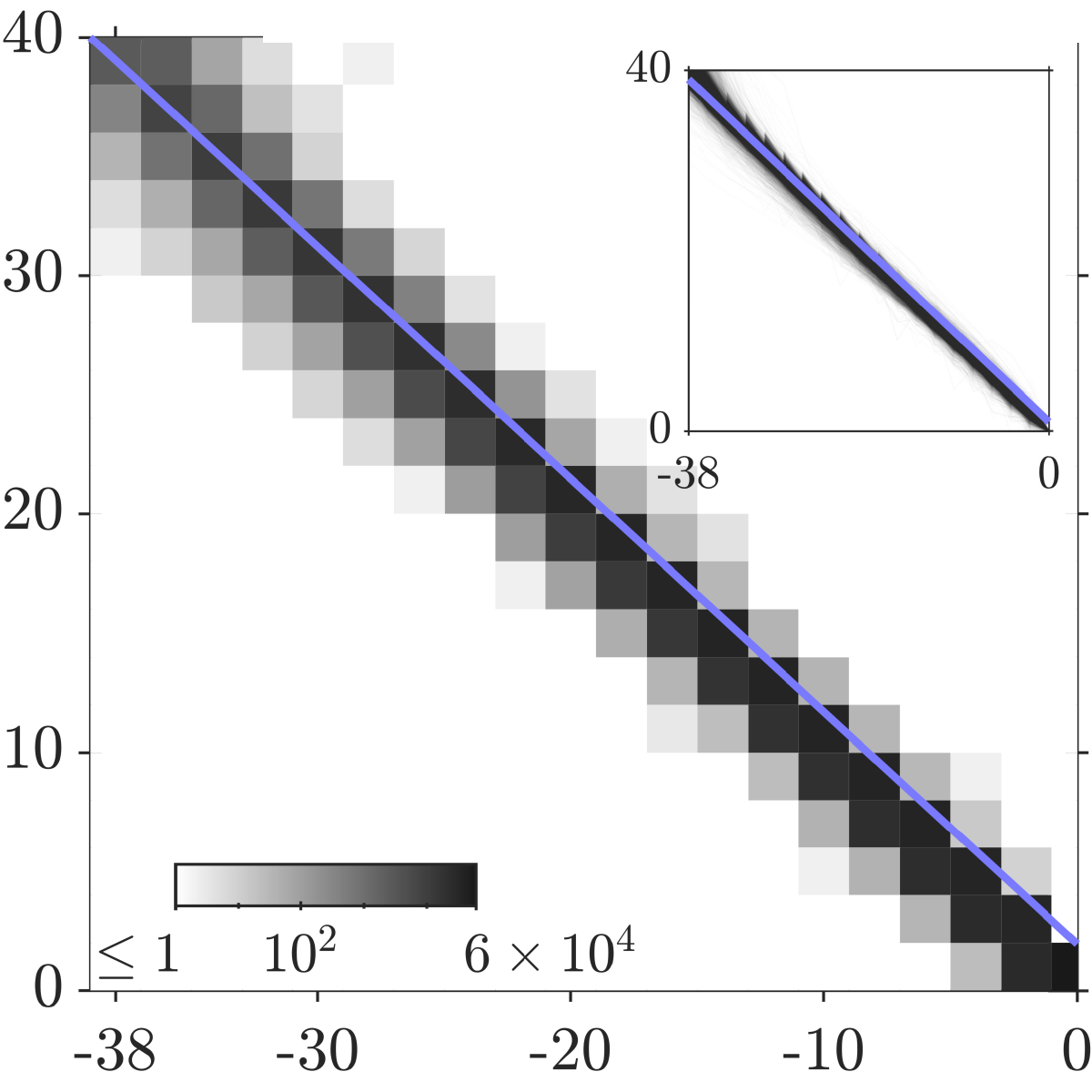}};
        \node[below=of image, node distance=0cm, yshift=0.40in, xshift=0.125in] {$|B| - |B_i|$};
        \node[left=of image, node distance=0cm, rotate=90, anchor=center,yshift=-0.25in]{$f\p{|B|} - f\p{|B_i|}$};
    \end{tikzpicture}
    \caption{This figure illustrates the graphs $G_i \in \Gamma$ given by (\ref{eqn:plot_graph}) as line plots (inset panel) and as a 2D histogram (main panel). Refer to text in Sec.~\ref{subsec:main_numerical_result} for the description of the figure.}
    \label{fig:numerical_methods_figure_2}
\end{figure}

\subsection{Main Numerical Result \label{subsec:main_numerical_result}}

Let $\mathcal F = \{1,2,...,N \}$ denote a subset of integers, and take $i \in \mathcal{F}$ to label each instance of $N$ randomly generated iuMPS constructed using the methods described in Sec.~\ref{sec:stepwise_procedure_numerics}; here $ N = 6 \times 10^4$. Define the function $f_i : \mathcal{B}_i \rightarrow \mathbb{R},$ where $i \in \mathcal{F}$ and $\mathcal{B}_i := \{0, 2, 4, \cdots, |B_i|\},$ such that,
\begin{align}\label{eqn:numerical_qcmi}
f_i(|B|) := \ln I_{i}(A:C | B) / 2 \ln( \nu_{\mathrm{gap}} ^{-1} ). 
\end{align} 
Then the inset panel in Fig.~\ref{fig:numerical_methods_figure_2} displays the collection of graphs $\Gamma = \{G_i : i \in \mathcal F \}$ where,
\begin{multline}\label{eqn:plot_graph}
    G_i := \{\left(x - x_i,f_i(x) - f_i(x_i)\right) : x = |B|, \\ x_i = |B_i| \}.
\end{multline}
Note that for graph $G_i$ the values of $|B| \leq |B_i|$, which is the maximum size of region $B$. Several graphs could have the same value of $|B_i|$, i.e. the function $i \mapsto |B_i|$ is many to one. In practice $|B_i|$ is found by the first instance that the condition $\exp\p{2\ln\p{\nu_{\mathrm{gap}} ^{-1}} f_i(|B_i|)} > 10^{-k}$ does not hold, where $k=12$ is determined independently by errors in numerical eigenvalue computations as discussed in Appendix~\ref{appendix:error_eigenvalue}. In Fig.~\ref{fig:numerical_methods_figure_2} (in both panels), we plot blue lines of slope $-1$ to help compare graph data with the theorem. The origin $\p{0, 0}$ of both the inset and the main panels in Fig.~\ref{fig:numerical_methods_figure_2} corresponds to the maximum dimension $|B| = |B_i|$. By comparing the graphs (black lines) against the blue line of slope $-1$ in the inset panel, we see that the slopes of most graphs in $\Gamma$ are eventually less than or equal to $-1$. This indicates, according to Theorem~\ref{thrm:theorem_1}, that for most graphs in $\Gamma$, $|B| = |B_i|$ explores the asymptotic regime in the numerical calculations.
In the main panel of Fig.~\ref{fig:numerical_methods_figure_2}, we display the same graph data as a 2D histogram, where the bins of the histogram $H_{i, j} \subset \mathbb R^2$ are given by, $$ H_{i, j} = \p{-\p{2i + 1}, -\p{2i - 1}} \times \p{2j, 2j + 2},$$ for $i, j \in \cb{0, 1, 2, \dots, 19}$. We observe that for a large fraction of graphs in $\Gamma$, the slopes are less than or equal to $-1$ and the occupancy of the bins just below the diagonal of slope $-1$ increases as $|B| \rightarrow |B_i|$.

\begin{figure}[t]
    \centering
    \begin{tikzpicture}
        \node (image) at (0,0)
        {\includegraphics[width=0.4\textwidth]{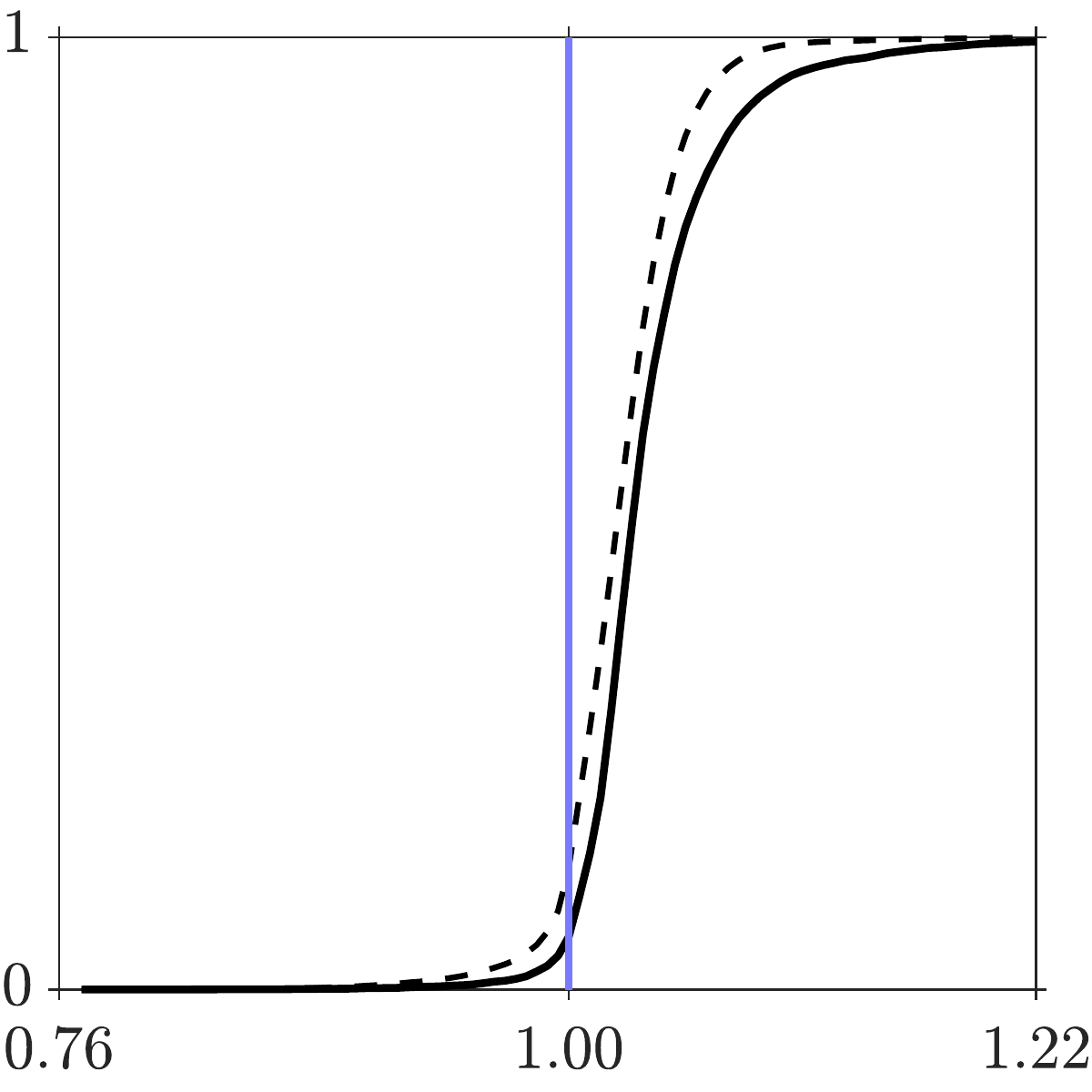}};
        \node[below=of image, node distance=0cm, yshift=0.40in, xshift=0.125in] {Normalized decay rate};
    \end{tikzpicture}
    \caption{
        This figure depicts the cumulative distribution of the QCMI decay rates normalized by $2 \ln{\nu_{\mathrm{gap}} ^{-1}}$ computed (i) for all $G_i \in \Gamma$, illustrated by the dashed black line, and (ii) for subset of $\Gamma$ corresponding to $|B_i| = 40$, illustrated by the solid black line. The vertical solid blue line corresponds to the slope of the solid blue line in Fig.~\ref{fig:numerical_methods_figure_2}. Note that the cumulative distribution extends further than 1.22 on the horizontal axis, but is not significant to our discussion and is omitted.
        }
    \label{fig:numerical_methods_figure_3}
\end{figure}

In Fig.~\ref{fig:numerical_methods_figure_3}, we further probe the decay of QCMI by plotting the cumulative distribution of the QCMI decay rates. The decay rates are extracted by linear regression from the graphs $G_i \in \Gamma$. Note that for all the iuMPS in our generated sample, the constant $K$ is zero, and the logarithmic correction is absent, which we verify during the computation. This may be observed from the fact, that in the bound given in Theorem \ref{thrm:theorem_1}, the constant $K$ is non-zero, only if the Jordan decomposition of the transfer matrix contains non-trivial Jordan blocks (i.e., contains nilpotent operators) for some eigenvalues with magnitude less than $1$. (For the eigenvalues of magnitude $1$, Jordan blocks are trivial, as shown, for example, in the proof of Proposition 3.1 of Ref. \onlinecite{fcs}.) For randomly generated states described in Cases 1-3 of Section \ref{subsec:constructing_iumps}, non-trivial Jordan blocks will typically be absent, since the eigenvalues (of magnitude less than 1 for Cases 2 and 3) are typically non-degenerate. Since we compute the eigenvalues of the transfer matrix, we can check that the eigenvalues are indeed non-degenerate for all the generated iuMPS. 

The dashed black line represents the distribution for all $G_i \in \Gamma,$ and the solid black line represents the distribution for graphs $G_i$ in the subset of $\Gamma$ corresponding to  $|B_i| = \mathrm{max}_j |B_j| = 40$, the predefined maximum explored in our numerical calculations. Recall that in other cases $|B_i|$ is defined as the value of $|B|$ two units prior to the value of $|B|$ at which the QCMI reaches the numerical QCMI error. The latter is determined primarily by errors accrued in calculating the eigenvalues of the transfer matrix (refer to Appendix~\ref{appendix:error_eigenvalue}). In other words, the dashed black line includes those graphs that reach QCMI error for $|B_i| < 40$, whereas, the solid black line corresponds to only those graphs that reach QCMI error for $|B_i| \geq 40$. As in Fig.~\ref{fig:numerical_methods_figure_2}, a large fraction of all graphs in $\Gamma$ are consistent with Theorem \ref{thrm:theorem_1} (dashed black line), because they have decay rates that obey the asymptotic lower bound. A still higher fraction of graphs with $|B_i| = 40$ (solid black line) is consistent with the theorem. Therefore, with the limited choice of numerical parameters: $\left \lvert B_i \right \rvert \le 40$ and $(d_s,d_M) = (3,4),$  Fig.~\ref{fig:numerical_methods_figure_2} and Fig.~\ref{fig:numerical_methods_figure_3} taken together provide supporting evidence for Theorem~\ref{thrm:theorem_1}. In particular Fig.~\ref{fig:numerical_methods_figure_3} suggests that those example which appear to contradict the theorem reach the QCMI error for $|B_i| < 40$, before entering their asymptotic regime, for which Theorem \ref{thrm:theorem_1} holds. We suggest that the examples with $|B_i|\geq 40$ that contradict the theorem have also not yet reached their asymptotic regime. 

\section{Conclusion}\label{sec:conclusion}

In this paper we considered a system $ABC$ as shown in Fig.\ref{dg:red_dens_op}(a) and (b), prepared in an infinite uniform matrix product state with density operator $\rho_{ABC}$, illustrated in Fig.\ref{dg:red_dens_op}(c). We proved that the QCMI, denoted $I(A:C|B)$, is bounded by an exponentially decaying function in the asymptotic limit of large size $|B|$ of the separating region. By taking advantage of the more direct methods of estimation than in Ref. \onlinecite{qcmi-decay}, we have improved the exponent of the asymptotic bounding function for QCMI by a factor of 4. We showed that the newly obtained exponent cannot be further improved. We have also shown that $\rho_{ABC}$ is close to an exact quantum Markov chain $\tilde{\rho}_{ABC}$ in the trace distance, which is also bounded by an exponentially decaying function. Compared to Ref. \onlinecite{qcmi-decay}, the exponent of the latter function is improved by a factor of 2. The theory was supported by numerical experiments in which the QCMI is computed for a collection of matrix product states generated by sampling their generating isometries with respect to Haar measure.

\section*{Acknowledgement}

We are grateful to Glen Evenbly for useful discussions on the numerical methods involving tensor networks. We are especially thankful to one of referees for pointing out a missing term in the expansion of QCMI in the earlier version of Appendix \ref{appendix:expansion} and for for comments that led to the inclusion of Appendix \ref{appendix:supports}. 

\section*{Data Availability Statement}
The data that support the findings of this study are available from the corresponding author upon reasonable request.

\newpage
\appendix

\section{An example: convergence of QMI to a non-zero value\label{appendix:a}}

In this appendix we specify the parameters of the iuMPS used in Fig.\ref{fig:numerical_methods_figure_1} and Fig.\ref{fig:numerical_methods_figure_2}, and the asymptotic limit of its QMI $$I(A:C)=S(\rho_A)+S(\rho_C)-S(\rho_{AC}),$$ which we use to benchmark our numerical computations. 

The iuMPS is generated by
\begin{equation*}
    M^s=\begin{bmatrix}
	M_1 ^s & 0 \\
	0 & M_2 ^s
	\end{bmatrix} ,
\end{equation*}
i.e., belongs to  \textit{Case 2} defined in Sec.~\ref{subsec:constructing_iumps}, with submatrices specified by,
\begin{align}
    \nonumber
    &M_1 ^1 = M_2 ^3 = 
    \begin{bmatrix}
    0 & -\frac{1}{\sqrt{2}} \\
    0 & 0
    \end{bmatrix}, \\
    \label{eqn:numerical_methods_9}
    &M_1 ^2  = M_2 ^2 = 
    \begin{bmatrix}
    -\frac{1}{\sqrt{2}} & 0 \\
    0 & \frac{1}{\sqrt{2}}
    \end{bmatrix},\\
    \nonumber
    &M_1 ^3 = M_2 ^1 = 
    \begin{bmatrix}
    0 & 0 \\
    \frac{1}{\sqrt{2}} & 0
    \end{bmatrix}, 
\end{align}
and by the fixed point $|\sigma\rangle=(1/2,1/2,1/2,1/2)^T$.

As was shown in Appendix D of Ref. \onlinecite{qcmi-decay}, if we set $|A|=|C|=1$, then, when $|B|\rightarrow\infty$, the reduced density operators $\rho_{AC}$, $\rho_{A}$, and $\rho_C$ converge to 
\begin{align}
    \label{eqn:numerical_methods_10}
    \rho_{AC}&=\frac{1}{32}\begin{pmatrix}
                            1 & 0 & 0\\
                            0 & 2 & 0\\
                            0 & 0 & 1
                          \end{pmatrix}
                          \otimes
                          \begin{pmatrix}
                            1 & 0 & 0\\
                            0 & 2 & 0\\
                            0 & 0 & 1
                          \end{pmatrix} \\ \nonumber
                          &+
                          \frac{1}{32}
                          \begin{pmatrix}
                            1 & 0 & 0\\
                            0 & 1 & 0\\
                            0 & 0 & 2
                          \end{pmatrix}
                          \otimes
                          \begin{pmatrix}
                            1 & 0 & 0\\
                            0 & 1 & 0\\
                            0 & 0 & 2
                          \end{pmatrix}, \\ \nonumber
    \rho_A=&\frac{1}{8}\begin{pmatrix}
                            2 & 0 & 0\\
                            0 & 3 & 0\\
                            0 & 0 & 3
                          \end{pmatrix}, \;
    \rho_C=\frac{1}{8}\begin{pmatrix}
                            2 & 0 & 0\\
                            0 & 3 & 0\\
                            0 & 0 & 3
                          \end{pmatrix}.
\end{align}
By direct calculation, the QMI  $I(A:C)$, converges to the value $I_{\mathrm{th}} = 17 \left. \ln\p{2} \right / 16 - 9 \left. \ln \p{3} \right / 8 + 5 \left. \ln \p{5} \right / 16 \approx 3.5\times 10^{-3}$. In contrast, as is proven by Theorem \ref{thrm:theorem_1} or by Theorem II.1 of Ref. \onlinecite{qcmi-decay}, the QCMI converges to zero.

\section{Taylor expansion with remainder}\label{appendix:expansion}
In this appendix we provide details of the derivation of the bound (\ref{eqn:qcmi_leq_11norm}), that we have omitted in Subsection \ref{subsec:qcmi_estmate}.

Recall that we are aiming to bound the QCMI
\begin{align}\label{eqn:qcmi_app}
    I(A:C|B)=\mathrm{tr}_M\left(\rho_{ABC}\left(\ln\rho_{ABC}+\ln\rho_{B}-\ln\rho_{AB}-\ln\rho_{BC}\right)\right)
\end{align}
in terms of $\|\Delta\rho_{ABC}\|_1$, where $\Delta\rho_{ABC}=\rho_{ABC}-\tilde{\rho}_{ABC}$. We achieve this by making
the Taylor expansion of all the logarithms, which, with an explicit remainder term, takes the form \cite{cartan},
\begin{widetext}
\begin{align}\label{eqn:log_expansion_full}
    \ln(\tilde{\rho}+\Delta\rho)=\ln\tilde{\rho} &+\int_{0}^{+\infty}ds\left(\frac{1}{\tilde{\rho}+s\mathbb{1}}\Delta\rho\frac{1}{\tilde{\rho}+s\mathbb{1}}-\frac{1}{\tilde{\rho}+s\mathbb{1}}\Delta\rho\frac{1}{\tilde{\rho}+s\mathbb{1}}\Delta\rho\frac{1}{\tilde{\rho}+s\mathbb{1}}\right.\\ \nonumber
    &\quad+\left.\int_0^1 dt (1-t)^3\frac{1}{\tilde{\rho}+t\Delta\rho+s\mathbb{1}}\Delta\rho\frac{1}{\tilde{\rho}+t\Delta\rho+s\mathbb{1}}\Delta\rho\frac{1}{\tilde{\rho}+t\Delta\rho+s\mathbb{1}}\Delta\rho\frac{1}{\tilde{\rho}+t\Delta\rho+s\mathbb{1}}\right),
\end{align}
\end{widetext}
and where we dropped index $R$ that can be any of $ABC$, $AB$, $BC$, or $B$.

We have already shown in Subsection \ref{subsec:qcmi_estmate}, that the contributions to $I(A:C|B)$ of the zeroth and the first order in $\|\Delta\rho_{ABC}\|_1$ vanish. We have also estimated the contribution of any of the four terms of the second order in (\ref{eqn:2nd_order_bound}).

Now we estimate the contributions of the two remaining terms of order greater than two, which are
\begin{widetext}
\begin{align*}
    &\mathrm{tr}\left(\Delta\rho\int^{+\infty}_0 ds\frac{1}{\tilde{\rho}+s\mathbb{1}}\Delta\rho\frac{1}{\tilde{\rho}+s\mathbb{1}}\Delta\rho\frac{1}{\tilde{\rho}+s\mathbb{1}}\right)\\ \nonumber &+\mathrm{tr}\left((\tilde{\rho}+\Delta\rho)\int_{0}^{+\infty}ds\int_0^1 dt(1-t)^3\frac{1}{\tilde{\rho}+t\Delta\rho+s\mathbb{1}}\Delta\rho\frac{1}{\tilde{\rho}+t\Delta\rho+s\mathbb{1}}\Delta\rho\frac{1}{\tilde{\rho}+t\Delta\rho+s\mathbb{1}}\Delta\rho\frac{1}{\tilde{\rho}+t\Delta\rho+s\mathbb{1}}\right).
\end{align*} 
\end{widetext}

The first of these may be estimated as
\begin{align}\label{eqn:reminder_estimate_1}
&\left|\mathrm{tr}\left(\Delta\rho\int^{+\infty}_0 ds\frac{1}{\tilde{\rho}+s\mathbb{1}}\Delta\rho\frac{1}{\tilde{\rho}+s\mathbb{1}}\Delta\rho\frac{1}{\tilde{\rho}+s\mathbb{1}}\right)\right| \\ \nonumber 
&\qquad\qquad\leq \int^{+\infty}_0 ds \left\|\frac{1}{\tilde{\rho}+s\mathbb{1}}\right\|^3\|\Delta\rho\|^3_1 \\ \nonumber
&\qquad\qquad\leq \int^{+\infty}_0 ds \frac{1}{(\lambda_{\mathrm{min}}+s)^3}\|\Delta\rho\|^3_1 \\ \nonumber
&\qquad\qquad\leq\frac{\|\Delta\rho\|^3_1}{2\lambda_{\mathrm{min}}^2}.
\end{align}

The second term may be estimated as
\begin{widetext}
\begin{align}\label{eqn:reminder_estimate_2}
&\left|\mathrm{tr}\left((\tilde{\rho}+\Delta\rho)\int_{0}^{+\infty}ds\int_0^1 dt(1-t)^3\frac{1}{\tilde{\rho}+t\Delta\rho+s\mathbb{1}}\Delta\rho\frac{1}{\tilde{\rho}+t\Delta\rho+s\mathbb{1}}\Delta\rho\frac{1}{\tilde{\rho}+t\Delta\rho+s\mathbb{1}}\Delta\rho\frac{1}{\tilde{\rho}+t\Delta\rho+s\mathbb{1}}\right)\right|\\ \nonumber
&\qquad\qquad\leq \|\tilde{\rho}+\Delta\rho\|_1\left\|\int_{0}^{+\infty}ds\int_0^1 dt(1-t)^3\frac{1}{\tilde{\rho}+t\Delta\rho+s\mathbb{1}}\Delta\rho\frac{1}{\tilde{\rho}+t\Delta\rho+s\mathbb{1}}\Delta\rho\frac{1}{\tilde{\rho}+t\Delta\rho+s\mathbb{1}}\Delta\rho\frac{1}{\tilde{\rho}+t\Delta\rho+s\mathbb{1}}\right\| \\ \nonumber
&\qquad\qquad\leq\int_{0}^{+\infty}ds\int^1_0 dt (1-t)^3\left\|\frac{1}{\tilde{\rho}+t\Delta\rho+s\mathbb{1}}\right\|^4\|\Delta\rho\|^3 \\ \nonumber    &\qquad\qquad\leq\int_{0}^{+\infty}ds\left\|\frac{1}{\tilde{\rho}+t'\Delta\rho+s\mathbb{1}}\right\|^4\|\Delta\rho\|^3 \\ \nonumber
    &\qquad\qquad\leq \int_{0}^{+\infty}\frac{ds}{(\lambda'_{\mathrm{min}}+s)^4}\|\Delta\rho\|^3 \\ \nonumber
    &\qquad\qquad= \frac{\|\Delta\rho\|^3}{3\lambda'^3_{\mathrm{min}}}\\ \nonumber
    &\qquad\qquad\leq \frac{\|\Delta\rho\|_1^3}{3\lambda'^3_{\mathrm{min}}},
\end{align}
\end{widetext}
where $0\leq t' \leq 1$ and $\lambda'_{\mathrm{min}}$ is the smallest eigenvalue of $\tilde{\rho}+t'\Delta\rho$.

We may estimate $\lambda'_{\mathrm{min}}$ using the inequality
\begin{align}\label{eqn:min_eig_estimate_t}
    |\lambda'_{\mathrm{min}} -\lambda_{\mathrm{min}}| &\leq  \|\tilde{\rho}_{ABC}+t'\Delta\rho_{ABC}-\tilde{\rho}_{ABC}\|_1 \\ \nonumber
    &\leq \|\Delta\rho_{ABC}\|_1 \\ \nonumber
    &\leq c_2d_M\sqrt{\frac{2d_M}{\sigma_{\mathrm{min}}}}|B|^{K}\nu_{\mathrm{gap}}^{|B|-K}.
\end{align}

Combining (\ref{eqn:2nd_order_bound}), (\ref{eqn:reminder_estimate_1}), and (\ref{eqn:reminder_estimate_2}), and recalling that there are four logarithms contributing to $I(A:C|B)$ (\ref{eqn:qcmi_app}), we estimate, 
\begin{widetext}
\begin{align*}
    I(A:C|B) \leq 4\left(\frac{\|\Delta\rho_{ABC}\|^2_1}{2\lambda_{\mathrm{min}}}+\frac{\|\Delta\rho_{ABC}\|^3_1}{2\lambda^2_{\mathrm{min}}}+\frac{\|\Delta\rho_{ABC}\|^3_1}{3\lambda'^3_{\mathrm{min}}}\right)  =\frac{2\|\Delta\rho_{ABC}\|^2_1}{\lambda_{\mathrm{min}}}\left(1+\left(\frac{1}{\lambda_{\mathrm{min}}}+\frac{2\lambda_{\mathrm{min}}}{3\lambda'^3_{\mathrm{min}}}\right)\|\Delta\rho_{ABC}\|_1\right),
\end{align*}
\end{widetext}
where we have redefined $\lambda_{\mathrm{min}}$ to be the smallest eigenvalue of all $\tilde{\rho}_R$, where $R$ can be any of $ABC$, $AB$, $BC$, or $B$. We have similarly redefined $\lambda'_{\mathrm{min}}$, with $\tilde{\rho}_R+t\Delta\rho_R$ replacing $\tilde{\rho}_R$. We have also used the inequality $\|\Delta\rho_{R}\|_1\leq\|\Delta\rho_{ABC}\|_1$.

From (\ref{eqn:min_eig_estimate_t}) we observe that $\lambda'_{\mathrm{min}}$ converges to $\lambda_{\mathrm{min}}$, while (from (\ref{eqn:tr_dist_leq_exp})) the trace distance $\|\Delta\rho_{ABC}\|_1$ vanishes as $|B|\rightarrow\infty$. Then, for large enough $|B|$,
\begin{equation}\label{eqn:size_req_2}
    \left(\frac{1}{\lambda_{\mathrm{min}}+\frac{2\lambda_{\mathrm{min}}}{3\lambda'^3_{\mathrm{min}}}}\right)\|\Delta\rho_{ABC}\|_1\leq 1.
\end{equation}
This leads to the bound
\begin{equation*}
    I(A:C|B) \leq \frac{4\|\Delta\rho_{ABC}\|^2_1}{\lambda_{\mathrm{min}}}.
\end{equation*}

Recalling that from (\ref{eqn:rho_r_prime_inj}) $\lambda_{\mathrm{min}}=\sigma^2_{\mathrm{min}}$,
\begin{equation*}
    I(A:C|B) \leq \frac{4}{\sigma^2_{\mathrm{min}}}\|\Delta\rho_{ABC}\|^2_1.
\end{equation*}

\section{Sufficient conditions on $|B|$}

In this appendix we combine together all the requirements imposed on the size $|B|$ of the separating region in order to get a unified sufficient condition.

First, we require $|B|$ to be large enough, that the bound (\ref{eqn:22norm_leq_exp}) is satisfied.
We derive the following conditions on $|B|$ in Appendix \ref{app:constants}, 
\begin{align*}
        |B|&>2K, \\ \nonumber
        \left(\frac{|B|}{K}\right)^K\nu_{\mathrm{gap}}^{|B|-K}&\geq (K_{\nu_i}+1)\left(\frac{|B|e}{K_{\nu_i}}\right)^{K_{\nu_i}}|\nu_{i}|^{|B|-K_{\nu_i}},\\ 
        &\qquad\qquad\; \mathrm{for \; all \;}  \nu_i\in \Upsilon\backslash\Upsilon_*.
\end{align*}
The constants $K$ and $K_{\nu_i}$ may be extracted from the Jordan decomposition of $E$, as described in Appendix \ref{app:constants}.

Second, we consider the requirement imposed on $|B|$ in Appendix \ref{appendix:expansion} in (\ref{eqn:size_req_2}). Using (\ref{eqn:min_eig_estimate_t}) we may estimate
\begin{equation*}
    \frac{\lambda'_{\mathrm{min}}}{\lambda_{\mathrm{min}}}  \geq 1-\frac{\|\Delta\rho_{ABC}\|_1}{\lambda_{\mathrm{min}}}.
\end{equation*}
Now we impose the condition
\begin{equation}\label{eqn:lambda_cond}
    \frac{\|\Delta\rho_{ABC}\|_1}{\lambda_{\mathrm{min}}} \leq \frac{1}{2},
\end{equation}
which leads to
\begin{equation*}
    \frac{\lambda_{\mathrm{min}}}{\lambda'_{\mathrm{min}}}  \leq 2.
\end{equation*}
This allows to estimate the left-hand side of (\ref{eqn:size_req_2}) as 
\begin{widetext}
\begin{equation*}
    \left(1+\frac{2\lambda^2_{\mathrm{min}}}{3\lambda'^3_{\mathrm{min}}}\right)\frac{\|\Delta\rho_{ABC}\|_1}{\lambda_{\mathrm{min}}} \leq \left(1+\frac{16}{3\lambda_{\mathrm{min}}}\right)\frac{\|\Delta\rho_{ABC}\|_1}{\lambda_{\mathrm{min}}} \leq \frac{19}{3}\frac{\|\Delta\rho_{ABC}\|_1}{\lambda^2_{\mathrm{min}}}.
\end{equation*}
\end{widetext}
To satisfy the requirement (\ref{eqn:size_req_2}) we impose 
\begin{equation*}
    \|\Delta\rho_{ABC}\|_1 \leq \frac{3}{19}\lambda^2_{\mathrm{min}},
\end{equation*}
which supersedes the condition (\ref{eqn:lambda_cond}). Using (\ref{eqn:11norm_leq_opnorm}) we arrive at the condition
\begin{equation*}
    c_2d_M\sqrt{\frac{2d_M}{\sigma_{\mathrm{min}}}}|B|^{K}\nu_{\mathrm{gap}}^{|B|-K} \leq \frac{3}{19}\lambda^2_{\mathrm{min}}
\end{equation*}
Recalling that $\lambda_{\mathrm{min}}=\sigma^2_{\mathrm{min}}$, we re-express the above inequality in the form
\begin{equation}\label{eqn:1st_cond}
    |B|^{K}\nu_{\mathrm{gap}}^{|B|-K} \leq \frac{3}{19\sqrt{2}}\frac{\sigma^{\frac{9}{2}}_{\mathrm{min}}}{c_2 d_M^{\frac{3}{2}}}
\end{equation}

We imposed a further condition on $|B|$, namely $\|\Delta C'\| \leq 1$, in order to derive (\ref{eqn:gamma_prime}), with $\Delta C'$ defined by (\ref{eqn:delta_c}) and (\ref{eqn:substitutions}). It was superseded by the stronger condition $\|\Delta C'\| \leq 2^{-3/2}$, which we used to guarantee the inequality (\ref{eqn:gamma_sq}). By using (\ref{eqn:dc_norm_leq_exp}), this condition leads to the inequality
\begin{equation*}
    \frac{1}{\sigma_{\mathrm{min}}}c_2d_M |B|^{K}\nu^{|B|}_{\mathrm{gap}} \leq \frac{1}{2\sqrt{2}}.
\end{equation*}
Expressing this inequality in the same form as (\ref{eqn:1st_cond}), we arrive at the condition
\begin{equation*}
    |B|^{K}\nu^{|B|}_{\mathrm{gap}} \leq \frac{1}{2\sqrt{2}} \frac{\sigma_{\mathrm{min}}}{c_2d_M}.
\end{equation*}
This condition is, however, redundant, since it is strictly weaker than (\ref{eqn:1st_cond}).


We have imposed one more condition on $|B|$, in Subsection \ref{subsec:qcmi_estmate}, by requiring $\dim\mathcal{H}_B\geq d^2_M$,
\begin{equation*}
    |B|\geq 2\frac{\ln d_M}{\ln d_s} .
\end{equation*}

Thus, the size $|B|$ should be large enough to satisfy the conditions
\begin{align}\label{eqn:final_conditions}
        |B|&>2K, \\ \nonumber 
        \left(\frac{|B|}{K}\right)^K\nu_{\mathrm{gap}}^{|B|-K}&\geq (K_{\nu_i}+1)\left(\frac{|B|e}{K_{\nu_i}}\right)^{K_{\nu_i}}|\nu_{i}|^{|B|-K_{\nu_i}}, \; \\ \nonumber
        &\qquad\qquad\mathrm{for \; all \;} \nu_i\in \Upsilon\backslash\Upsilon_*, \\ \nonumber
        |B|^{K}\nu_{\mathrm{gap}}^{|B|-K} & \leq \frac{3}{19\sqrt{2}}\frac{\sigma^{\frac{9}{2}}_{\mathrm{min}}}{c_2 d_M^{\frac{3}{2}}}, \\ \nonumber
        |B|&\geq 2\frac{\ln d_M}{\ln d_s}.
\end{align}
The constant $c_2$ is specified in Appendix \ref{app:constants}.

We remind the reader that the constraints (\ref{eqn:final_conditions}) should not be taken as necessary to prove Theorem \ref{thrm:theorem_1}, and are only sufficient (and convenient) to guarantee validity of the steps in the proof.

We performed the derivation above for the case of an injective iuMPS. If we repeat the same derivation for a general case of iuMPS, we will have to replace (with respect to the definitions made in Section \ref{subsec:generalization}) the factors $d^{-3/2}_M\mapsto \max_{\alpha}\{d^{-3/2}_{M_\alpha}p^{5/2}_\alpha\}$ and $\ln d_M \mapsto \max_{\alpha}\ln d_{M_\alpha}$ in the third and in the fourth lines of (\ref{eqn:final_conditions}), respectively. Alternatively, we can use (\ref{eqn:final_conditions}) without any changes, since the described alternations lead to weaker conditions.

\section{An example of errors in eigenvalue calculations} \label{appendix:error_eigenvalue}

Errors resulting from the numerical computation of eigenvalues of  reduced density operators dictate the smallest value of QCMI that we can reliably compute. In turn, this restricts the maximum size of the separating region $|B|$ we can explore. Since $I(A:C|B)=S(\rho_{AB})+S(\rho_{BC})-S(\rho_{ABC})-S(\rho_B)$, the error incurred in calculating QCMI is estimated as
\begin{equation*}
    \delta I(A:C|B)\sim \delta S,    
\end{equation*}
where,
\begin{equation*}
    \delta S := \mathrm{max}\left\{\delta S\p{\rho_{ABC}}, \delta S\p{\rho_{B}}, \delta S\p{\rho_{AB}}, \delta S\p{\rho_{BC}} \right\},
\end{equation*} 
is the maximum error in a calculation of the entropies. We cannot reliably compute values of QCMI smaller than $\delta I$, and thus are forced to stop calculations when the size of the separating region $|B|$ becomes large enough that $I(A:C|B)$ decays to values of the order $O(\delta I)$.

As discussed in Section \ref{sec:numerical_methods_results}, each of the density operators $\rho_{ABC}$, $\rho_{AB}$, $\rho_{BC}$, and $\rho_{B}$ is supported on a subspace of dimension $d_M^2$ so
\begin{align*}
    S = -\sum_{i=1}^{d_M^2} \lambda_i \ln \lambda_i.
\end{align*}
Hence, we may estimate the error in calculating entropies as 
\begin{align*}
    \delta S &\leq \sum_{i=1}^{d_M^2} \delta \lambda \ln \lambda_i ^{-1} - \delta \lambda \lesssim \sum_{i=1}^{d_M^2} \delta \lambda \ln \lambda_i ^{-1}\lesssim d_M ^2 \delta \lambda \ln \lambda_{\mathrm{min}} ^{-1},
\end{align*}
where $\lambda_{\mathrm{min}}:=\min_i \lambda_i$, and $\delta \lambda >0$ is the maximum error in computing the eigenvalues. In practice, we use the MATLAB's built in eigenvalue calculation function ``eig'' to compute eigenvalues, with an error of order $O\p{10^{-14}},$ as explained in the following paragraph. Hence we find
\begin{equation}\label{eqn:app-s-error}
   \delta I \sim \delta S \sim 4^2 \cdot 10^{-14}\cdot\ln 10^{14} \sim 10^{-12}.
\end{equation}

In our numerical calculations, refering to (\ref{eqn:numerical_qcmi}) and below (\ref{eqn:plot_graph}), for each $i \in \mathcal{F}$, we denote $\left \lvert B_i \right \rvert$ to be the maximum size of the separating region such that $\exp\p{2\ln\p{\nu_{\mathrm{gap}} ^{-1}} f_i(|B_i|)} > 10^{-k}.$ From the above discussion, we can reliably compute QCMI as small as $O\p{10^{-12}}$, and, thus, we take $k = 12$ for our numerical calculations.

To gauge the size of errors incurred in eigenvalue calculations we compare numerically computed values against an exact result. We consider the transfer matrix eigenvalue problem of \textit{Case~1} in Sec.~\ref{subsec:constructing_iumps} and recall that one of the eigenvalues $\nu_1 = 1.$ The numerically computed value incurs an error of $O(10^{-14})$. This estimate of the error is derived empirically from Fig.~\ref{fig:numerical_methods_figure_4}(a), which depicts the absolute difference between $\nu_1$, and its numerically computed value for $2\times 10^4$ different instances of transfer matrices.

\section{Spectrum of the transfer matrix}\label{appendix:spectrum}
In this appendix we discuss the spectrum for the transfer matrix of an iuMPS, which is sampled with respect to the Haar measure (as defined in Section \ref{sec:stepwise_procedure_numerics}). 

For the purpose of comparing theoretically expected eigenvalues of the transfer matrix to numerically computed eigenvalues, both denoted by $\nu_i$ and delineated by context in what follows, we assume the ordering $|\nu_1| \geq |\nu_2| \geq \dots \geq |\nu_{d_M ^2}|$. Recall that we define
\begin{equation*}
    \nu_{\mathrm{gap}}=\max_{|\nu_i|<1}|\nu_i|.
\end{equation*}

\begin{figure*}[!ht]
    \centering
    \begin{tikzpicture}
        \node (image) at (0,0)
        {\includegraphics[trim={0, 0, 12in, 0}, clip, height=0.35\columnwidth]{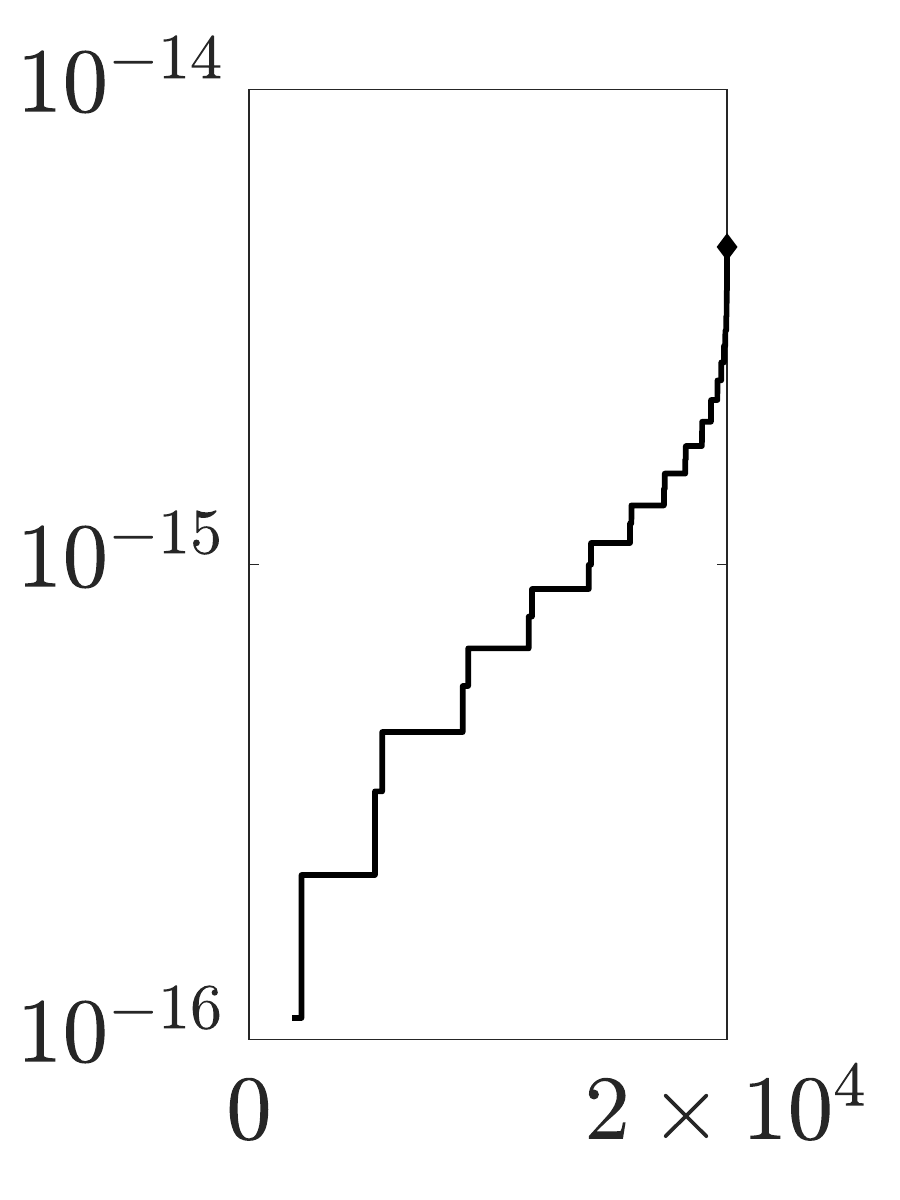}};
        \node[below=of image, node distance=0cm, yshift=0.50in, xshift=0in] {Instance};
        \node[left=of image, node distance=0cm, rotate=90, anchor=center,yshift=-0.375in]{$\left\lvert 1 - \left \lvert \nu_1 \right\rvert \right\rvert$};
        \node[above=of image, node distance=0cm, yshift=-0.5in, xshift=0in] {$\p{\mathrm{a}}$};
    \end{tikzpicture}
    \begin{tikzpicture}
        \node (image) at (0,0)
        {\includegraphics[trim={6in, 0, 6.25in, 0}, clip, height=0.35\columnwidth]{numerical_figures_4.pdf}};
        \node[below=of image, node distance=0cm, yshift=0.50in, xshift=0in] {Instance};
        \node[left=of image, node distance=0cm, rotate=90, anchor=center,yshift=-0.375in]{$\left\lvert \left\lvert\nu_1 \right\rvert - \left\lvert \nu_2 \right\rvert \right\rvert$};
        \node[above=of image, node distance=0cm, yshift=-0.5in, xshift=0in] {$\p{\mathrm{b}}$};
    \end{tikzpicture}
    \begin{tikzpicture}
        \node (image) at (0,0)
        {\includegraphics[trim={11.75in, 0, 0, 0}, clip, height=0.35\columnwidth]{numerical_figures_4.pdf}};
        \node[below=of image, node distance=0cm, yshift=0.50in, xshift=0in] {Instance};
        \node[left=of image, node distance=0cm, rotate=90, anchor=center,yshift=-0.375in]{$\left\lvert \left\lvert\nu_2 \right\rvert - \left\lvert \nu_3 \right\rvert \right\rvert$};
        \node[above=of image, node distance=0cm, yshift=-0.5in, xshift=0in] {$\p{\mathrm{c}}$};
    \end{tikzpicture}
    \caption{This figure depicts absolute differences between (a) 1 and $\left\lvert \nu_1 \right\rvert$, (b) $\left\lvert \nu_1 \right\rvert$ and $\left \lvert \nu_2 \right \rvert$, and (c) $\left\lvert \nu_2 \right\rvert$ and $\left \lvert \nu_3 \right \rvert$, where $\left \lvert \nu_1 \right \rvert$, $\left \lvert \nu_2 \right \rvert$ and $\left \lvert \nu_3 \right \rvert$ denote the numerically computed first, second and third largest absolute eigenvalues, respectively, of the transfer matrices of \textit{Case~1} in Sec.~\ref{subsec:constructing_iumps}. The $2\times 10^4$ points on each graph are plotted in ascending order of their y-coordinate values. The black diamonds denote the maxima of the graphs. The black circle in (b) represents the minimum of the graph. The black squares in (c) represent the bounds on the region around $10^{-9}$ where no numerically sampled differences are observed.} 
    \label{fig:numerical_methods_figure_4}
\end{figure*}

In Fig.~\ref{fig:numerical_methods_figure_4}(a) we depict the magnitude of difference between 1, the known largest eigenvalue, and $|\nu_1|$, the largest numerically computed value. As mentioned in Appendix \ref{appendix:error_eigenvalue}, we infer from the figure that eigenvalue computations incur a numerical error of the order $O(10^{-14})$.

In Fig.~\ref{fig:numerical_methods_figure_4}(b), the magnitude of the difference between $|\nu_1|$ and $|\nu_2|$ is shown. We observe a gap between the computed eigenvalue magnitudes that clearly exceeds the numerical error. The gap is, however, an artifact of the finite sampling of Haar random isometries used to generate the iuMPS's, as a simple example illustrates. Consider the family of quantum channels $\mathcal{E}:\mathbb{C}^{4\times 4}\rightarrow\mathbb{C}^{4\times 4}$, generated by the dilating isometry $V:\mathbb{C}^4\rightarrow\mathbb{C}^3\otimes\mathbb{C}^4$, defined by
\begin{align*}
    V=&\sqrt{\beta}|1\rangle\otimes|-\rangle\langle+|\otimes\mathbb{1}_2+\sqrt{1-\beta}|0\rangle\otimes\mathbb{1}_2\otimes\mathbb{1}_2 \\ \nonumber
    &-\sqrt{\beta}|-1\rangle\otimes|+\rangle\langle-|\otimes\mathbb{1}_2,
\end{align*}
where we have decomposed $\mathbb{C}^4\cong\mathbb{C}^2\otimes\mathbb{C}^2$, the bases of $\mathbb{C}^2$ and $\mathbb{C}^3$ are denoted by $\{|+\rangle,|-\rangle\}$ and $\{|1\rangle,|0\rangle,|-1\rangle\}$, respectively, and where $0\leq\beta\leq 1$. The isometry induces the quantum channel
\begin{equation*}
    \mathcal{E}=\mathrm{tr}_3(VXV^\dagger)=(1-\beta)\,\mathrm{id}_2\otimes\mathrm{id}_2+\beta\,\mathcal{E}_1\otimes\mathrm{id}_2,
\end{equation*}
where $\mathrm{id}_2:\mathbb{C}^{2\times 2}\rightarrow\mathbb{C}^{2\times 2}$ is the identity channel, and the quantum channel $\mathcal{E}_1:\mathbb{C}^{2\times 2}\rightarrow\mathbb{C}^{2\times 2}$ is defined by
\begin{align*}
    \mathcal{E}_1(|+\rangle\langle +|)&=|-\rangle\langle -|, \\ \nonumber
    \mathcal{E}_1(|-\rangle\langle -|)&=-|+\rangle\langle +|, \\ \nonumber
    \mathcal{E}_1(|+\rangle\langle-|)&=\mathcal{E}_1(|-\rangle\langle+|)=0.
\end{align*}
For this quantum channel, we may readily compute
\begin{equation*}
    |\nu_1|-|\nu_2|=2\beta,
\end{equation*}
which may be arbitrarily close to $0$, thus closing the gap in Fig.~\ref{fig:numerical_methods_figure_4}(b). These channels belong to the family represented by \emph{Case 1} in Section \ref{subsec:constructing_iumps}. Our numerical results, however, suggest that the probability of sampling channels of this kind is small. The spectrum of random quantum channels (Haar-distributed or otherwise) has been studied in previous works Ref. \onlinecite{haar-rand-gap,mps-typ-gap,gener-rand-q-chan,univer-spectr-int-q,expanders-hastings}, including the gap in the spectrum of quantum channels. The presence of a gap in a typical - with a particular sense of this word depending on the theoretical study - quantum channel was proven, for example, for Gaussian distributed quantum channels,\cite{mps-typ-gap} and Haar-distributed quantum channels,\cite{haar-rand-gap} in an asymptotic limit of large dimensions. These  settings are closely related to this paper, but our numerical computations involve iuMPS with the virtual dimension $d_M=4$ and physical dimension $d_s=3$, both of which are too small for the results of Ref. \onlinecite{haar-rand-gap} to be applicable. Nevertheless, our numerical experiments indicate that the gap in the spectrum of Haar-random quantum channels is present even for such small dimensions. 

In Fig.~\ref{fig:numerical_methods_figure_4}(c), we show the magnitude of the numerically computed difference between $|\nu_2|$ and $|\nu_3|$. We observe that, in contrast to the difference $||\nu_{1}|-|\nu_2||$ in Fig.~\ref{fig:numerical_methods_figure_4}(b), for a non-zero fraction of the sampled transfer matrices the second and the third largest eigenvalue magnitudes are equal (or differ by less than numerical error). We speculate that these occurrences might be explained as follows. Recall that eigenvalues of a transfer matrix come in complex conjugate pairs. Hence, if we sample a transfer matrix with complex eigenvalues, we will necessarily find complex conjugate pairs of equal modulus. Thus, together, Figs.~\ref{fig:numerical_methods_figure_4}(b) and (c) indicate that, using the Haar measure, it is possible to sample with non-vanishing probability transfer matrices that have complex eigenvalues of magnitude strictly less than 1 (Fig.~\ref{fig:numerical_methods_figure_4}(c)); the probability of sampling transfer matrices with more than one eigenvalue of magnitude one (including cases of non-real eigenvalues) is, however, vanishingly small (Fig.~\ref{fig:numerical_methods_figure_4}(b)). (See Ref. \onlinecite{univer-spectr-int-q} for a relevant study of distributions of eigenvalues of random quantum channels.)

The gap in the sampled values of $||\nu_2|-|\nu_3||$ is, as in the case of $||\nu_1|-|\nu_2||$, a numerical artifact. For the collection of transfer matrices generated by the one-parameter family of isometries
\begin{align*}
    V=&\left((1-\beta)\sqrt{\frac{2}{3}}+\beta\frac{\sqrt{3}}{2}\right)|1\rangle\otimes|-\rangle\langle +|\otimes\mathbb{1}_2 \\ \nonumber
    &-\left((1-\beta)\frac{1}{\sqrt{3}}+\beta\frac{1}{2}\right)|0\rangle\otimes|+\rangle\langle +|\otimes \mathbb{1}_2 \\ \nonumber
    &+\left((1-\beta)\frac{1}{\sqrt{3}}+\beta\frac{1}{2}\right)|0\rangle\otimes|-\rangle\langle -|\otimes\mathbb{1}_2 \\ \nonumber
    &-\left((1-\beta)\sqrt{\frac{2}{3}}+\beta\frac{\sqrt{3}}{2}\right)|-1\rangle\otimes|+\rangle\langle -|\otimes\mathbb{1}_2,
\end{align*}
with $0\leq\beta\leq 1$, the analytically calculated value,
\begin{equation*}
    |\nu_2|-|\nu_3|=\frac{\sqrt{6}-2}{\sqrt{3}}\beta+O(\beta^2),
\end{equation*} 
may be arbitrarily close to zero.

\section{Values of constants}\label{app:constants}
In this appendix we specify the constants $K$, $c_1$, and $c_2$ in the bound (\ref{eqn:22norm_leq_exp}), as well as the alternative value for $c_2$, denoted as $c_3$. 

First, we determine the values of $c_1$ and $K$, and the value of $c_2$, that follow from Theorem III.2 of Ref. \onlinecite{wolf-szehr}. We perform the Jordan decomposition of the transfer matrix $E$,   
\begin{equation}\label{eqn:app_jord_decomp_init}
    E = S\bigoplus_{\nu_i}(\nu_i\tilde{P}_{\nu_i}+\tilde{N}_{\nu_i})S^{-1},
\end{equation}
where $S$ is an invertible matrix, $\nu_i$ is an eigenvalue, $\tilde{P}_{\nu_i}$ is the projector onto the subspace corresponding to the eigenvalue $\nu_i$, and $\tilde{N}_{\nu_i}$ is a matrix with the elements immediately above the main diagonal equal to $1$, and all other elements equal to $0$, possessing the index of nilpotency equal to $(\tilde{K}_{\nu_i}+1)$ (i.e.,  $\tilde{N}_{\nu_i}^{\tilde K_{\nu_i}+1}=0$).  Let us set
\begin{align*}
    K := \max_{|\nu_i|=\nu_{\mathrm{gap}}} \tilde{K}_{\nu_i}.
\end{align*}
Let us introduce the set $\Upsilon$ containing all distinct values of $\nu_i$.

The Jordan decomposition (\ref{eqn:app_jord_decomp_init}) can contain terms corresponding to equal eigenvalues. For each distinct eigenvalue, let us combine such the terms together, arriving to the decomposition 
\begin{equation}\label{eqn:app_jord_decomp}
    E = S \bigoplus_{\nu_i\in \Upsilon}(\nu_iP_{\nu_i}+N_{\nu_i}) S^{-1},
\end{equation}
where $P_{\nu_i}$ is still a projector, and $N_{\nu_i}$ is still a nilpotent matrix with the index of nilpotency $K_{\nu_i}$ being the largest one among the nilpotency indices $\tilde{K}_{\nu_i}$ of the combined operators $\tilde{N}_{\nu_i}$. We define $\Upsilon_*:=\{\nu_i\in\Upsilon\;|\; K_{\nu_i}=K\}$, the subset of $\Upsilon$.

Then, by Theorem III.2 in Ref. \onlinecite{wolf-szehr}, the bound (\ref{eqn:22norm_leq_exp}) is satisfied for $n$
large enough that
\begin{equation*}
    \|\bigoplus_{\nu_i\in \Upsilon}(\nu_iP_{\nu_i}+N_{\nu_i})^n\|=\|\max_{\nu_*\in\Upsilon_*}(\nu_*P_{\nu_*}+N_{\nu_*})^n\|,
\end{equation*}
which may be guaranteed by satisfying the conditions,
\begin{align*}
        n &> 2K, \\
        \left(\frac{n}{K}\right)^{K}\nu_{\mathrm{gap}}^{n-K} &\geq (K_{\nu_i}+1)\left(\frac{ne}{K_{\nu_i}}\right)^{K_{\nu_i}}|\nu_{i}|^{n-K_{\nu_i}},\\ \nonumber
        &\qquad\qquad\mathrm{\; for \; all \;}  \nu_i\in \Upsilon\backslash\Upsilon_*,
\end{align*}
with the constants
\begin{align*}
    c_1 & := \|S\|^{-1}\|S^{-1}\|^{-1}, \\ \nonumber
    c_2 &:= (K+1)\left(\frac{e}{K}\right)^K\|S\|\|S^{-1}\|.
\end{align*}

The Theorem IV.3 of Ref. \onlinecite{wolf-szehr} provides an alternative bound for $\|E^n-\tilde{E}^n\|$,
\begin{align}\label{eqn:app-iv3-bound}
    \|E^n-\tilde{E}^n\|&\leq \nu_{\mathrm{gap}}^{n+1}\frac{4e^2\sqrt{D}(D+1)}{n\left(1-(1+\frac{1}{n})\nu_{\mathrm{gap}}\right)^{\frac{3}{2}}}\\ \nonumber
    &\qquad\qquad\times \sup_{|z|=(1+\frac{1}{n})\nu_{\mathrm{gap}}}\prod_{\nu_i\in \Upsilon} \left|\frac{1-\bar{\nu_i}z}{z-\nu_i}\right|^{K_{\nu_i}+1}, 
\end{align}
which is satisfied for $n>\nu_{\mathrm{gap}}(1-\nu_{\mathrm{gap}})^{-1}$. Here $D=\sum_{\nu_i\in\Upsilon}(K_{\nu_i}+1)$, and may be bounded as $D\leq d_M^2$. By defining the function
\begin{align*}
    g(n)&=\frac{4e^2\sqrt{D}(D+1)}{\left(1-(1+\frac{1}{n})\nu_{\mathrm{gap}}\right)^{\frac{3}{2}}} \left(\frac{\nu_{\mathrm{gap}}}{n}\right)^{K+1}\\ \nonumber
    &\qquad\qquad\times\sup_{|z|=(1+\frac{1}{n})\nu_{\mathrm{gap}}}\prod_{\nu_i\in \Upsilon} \left|\frac{1-\bar{\nu_i}z}{z-\nu_i}\right|^{K_{\nu_i}+1},
\end{align*}
we may express (\ref{eqn:app-iv3-bound}) as 
\begin{equation*}
    \|E^n-\tilde{E}^n\|\leq g(n) n^K\nu_{\mathrm{gap}}^{n-K}.
\end{equation*}
Notice that $g(n)$ is a bounded function, since
for large enough $n$ the supremum factor in (\ref{eqn:app-iv3-bound}) is (i) attained for $z$ in the vanishingly small neighborhood of some $\nu_*\in\Upsilon_*$, and (ii) grows asymptotically  in proportion to $(n/\nu_{\mathrm{gap}})^{K+1}$. Hence, for large enough $n$, we may estimate
\begin{widetext}
\begin{align}\label{eqn:alt-bound-aux}
    \sup_{|z|=(1+\frac{1}{n})\nu_{\mathrm{gap}}}\prod_{\nu_i\in \Upsilon} \left|\frac{1-\bar{\nu_i}z}{z-\nu_i}\right|^{K_{\nu_i}+1} & \leq 2 \left(\frac{n}{\nu_{\mathrm{gap}}}\right)^{K+1} \max_{\nu_*\in\Upsilon_*}\prod_{\nu_*\neq\nu_i\in \Upsilon} \left|\frac{1-\bar{\nu_i}\nu_*}{\nu_*-\nu_i}\right|^{K_{\nu_i}+1}, \\ \nonumber
    1-\left(1+\frac{1}{n}\right)\nu_{\mathrm{gap}}  &> \frac{1-\nu_{\mathrm{gap}}}{2}.
\end{align}
\end{widetext}
The precise condition on how large $n$ should be for the first inequality in (\ref{eqn:alt-bound-aux}) to be satisfied is not trivial to specify in a reasonable closed expression. We choose not to do this, since our primary goal is to simply illustrate that using the bound of Theorem IV.3 of Ref. \onlinecite{wolf-szehr} leads to the qualitatively same conclusion as using the bound of Theorem III.2 of Ref. \onlinecite{wolf-szehr}. With (\ref{eqn:alt-bound-aux}), we may bound the $g(n)$ from above, 
\begin{align*}
    g(n)\leq \frac{16e^2\sqrt{D}(D+1)}{\sqrt{2}\left(1-\nu_{\mathrm{gap}}\right)^{\frac{3}{2}}}  \left(1-\nu_{\mathrm{gap}}^2\right)^{K+1} \left(\frac{2}{\Delta}\right)^{D-K-1},
\end{align*}
where 
\begin{equation*}
    \Delta=\min_{\nu_*\in\Upsilon_*}\min_{\nu_*\neq \nu_i\in \Upsilon}|\nu_i-\nu_*|\leq 2.
\end{equation*}
Thus, for large enough $n$, we may bound $g(n)\leq c_3$, which leads to $\|E^n-\tilde{E}^n\|\leq c_3 n^{K}\nu_{\mathrm{gap}}^{n-K}$, with
\begin{align*}
     c_3 &:=\frac{16e^2\sqrt{D}(D+1)}{\sqrt{2}\left(1-\nu_{\mathrm{gap}}\right)^{\frac{3}{2}}}  \left(1-\nu_{\mathrm{gap}}^2\right)^{K+1} \left(\frac{2}{\Delta}\right)^{D-K-1} \\ \nonumber
    &\leq \frac{16e^2d_M(d_M^2+1)}{\sqrt{2}\left(1-\nu_{\mathrm{gap}}\right)^{\frac{3}{2}}}  \left(1-\nu_{\mathrm{gap}}^2\right)^{K+1} \left(\frac{2}{\Delta}\right)^{d_M^2-K-1}.
\end{align*}

\section{Containment of supports}\label{appendix:supports}
In this appendix we show that for $R$ any of the regions $ABC$, $AB$, $BC$, or $B$, 
\begin{align*}
\mathrm{supp}(\rho_R) &\subseteq \mathrm{supp}(\tilde{\rho}_R), \\ \nonumber
\mathrm{supp}(C) &\subseteq \mathrm{supp}(\tilde{C}), 
\end{align*}
with $C$ and $\tilde{C}$ defined in (\ref{eqn:c_tilde}), for a general iuMPS (both injective and non-injective cases). We separate the latter statement and its proof into Lemma \ref{lemma:supp_lemma}, which might be of independent interest for the reader, since the lemma is valid for any quantum channel with the same input and output spaces, and not only for the quantum channels to which it is applied in this paper.

In the case when an iuMPS has a component with period $p_\alpha>1$, the Choi matrix $\tilde{C}$ is not of full-rank. It follows, as a consequence of Lemma \ref{lemma:supp_lemma} proved below, that the Choi matrix $C$ is also not of full-rank. To accommodate for such a general case of iuMPS, we must modify the decomposition (\ref{eqn:choi-tilde}) in which the index pair $(i,j)$ of the vector $|\tilde{\psi}_{ij}\rangle$, lies in the Cartesian product $ \tilde{\mathcal{O}}:=\{1,2,\cdots,d_M^2 \} \times \{1,2,\cdots,d_M^2 \}$. We now need, however, to limit the choice of pairs $(i,j)$ to the set \cite{qcmi-decay}
\begin{align*}
\mathcal{O}:=\left\{(i,j)\in\tilde{\mathcal{O}}\;|\; \exists\alpha\in\{1,\dots,N\},\exists k_\alpha\in\{1,\dots,p_\alpha\}:\right.\\ \nonumber \left.|i\rangle\in\Pi^\alpha_{k_\alpha+|B|}\mathcal{H}_M, |j\rangle\in\Pi^\alpha_{k_\alpha}\mathcal{H}_M\right\}.
\end{align*}  
(The particular definition of the set $\mathcal{O}$ does not play a role in any further arguments and is included for the sake of completeness.) Thus, the decomposition (\ref{eqn:choi-tilde}) becomes 
\begin{align}\label{eqn:c_tilde_decomp_copy}
    \tilde{C}&=\sum_{(i,j)\in \mathcal{O}}|\tilde{\psi}_{ij}\rangle\langle\tilde{\psi}_{ij}|,
\end{align}
where $|\tilde{\psi}_{ij}\rangle$ are mutually orthogonal. Lemma \ref{lemma:supp_lemma}, permits us to decompose $C$ similarly to (\ref{eqn:c_decomp}) in the form, 
\begin{align}\label{eqn:c_decomp_copy}
    C & =\sum_{(i,j)\in \mathcal{O}}|\psi_{ij}\rangle\langle\psi_{ij}|.
\end{align}
Now we are ready to show that $\mathrm{supp}(\rho_R)\subseteq \mathrm{supp}(\tilde{\rho}_R)$ follows from $\mathrm{supp}(C)\subseteq \mathrm{supp}(\tilde{C})$. Consider, for example,
\begin{align*}
    \rho_{ABC}&=\mathrm{tr}_M(V_CV_BV_A\sigma V^\dagger_AV_B^\dagger V^\dagger_C),\\ \nonumber
    \tilde{\rho}_{ABC}&=\mathrm{tr}_M(V_C\tilde{V}_BV_A\sigma V^\dagger_A\tilde{V}_B^\dagger V^\dagger_C).
\end{align*}
Using (\ref{eqn:v_b_via_w}) and (\ref{eqn:tv_b_via_tw}), we may express $\rho_{ABC}$ and $\tilde{\rho}_{ABC}$ as
\begin{align}\label{eqn:rho_w_explicit}
    \rho_{ABC}&=\mathrm{tr}_M(V_C(U\otimes\mathbb{1}_M)WV_A\sigma V^\dagger_AW^\dagger(U^\dagger\otimes\mathbb{1}_M) V^\dagger_C),\\ \nonumber
    \tilde{\rho}_{ABC}&=\mathrm{tr}_M(V_C(U\otimes\mathbb{1}_M)\tilde{W}V_A\sigma V^\dagger_A\tilde{W}^\dagger(U^\dagger\otimes\mathbb{1}_M) V^\dagger_C),
\end{align}
where the isometries $W,\tilde{W}:\mathcal{H}_M\rightarrow\mathcal{H}_M\otimes\mathcal{H}_M\otimes\mathcal{H}_M$ are defined similarly to (\ref{eqn:w_tilde_choi}) and (\ref{eqn:w_choi}), but with a modification needed to account for Choi matrices $C$ and $\tilde{C}$ not being full-rank,
\begin{align}\label{eqn:w_copy}
W&=\sum_{k=1}^{d_M}\sum_{(i,j)\in \mathcal{O}}\left(P_k|\psi_{ij}\rangle\otimes|i\rangle\otimes|j\rangle\right)\langle k|,\\ \nonumber
\tilde{W}&=\sum_{k=1}^{d_M}\sum_{(i,j)\in \mathcal{O}}\left(P_k|\tilde{\psi}_{ij}\rangle\otimes|i\rangle\otimes|j\rangle\right)\langle k|. 
\end{align}
The projectors $P_k$ and the isometry $U$ were defined in Section \ref{subsec:tr_dist_estimate}. 

From (\ref{eqn:rho_w_explicit}) and (\ref{eqn:w_copy}) it is clear that if $\mathrm{span}\{|\psi_{ij}\rangle\}\subseteq\mathrm{span}\{|\tilde{\psi}_{ij}\rangle\}$, then $\mathrm{supp}(\rho_{ABC})\subseteq \mathrm{supp}(\tilde{\rho}_{ABC})$. Moreover, since $\mathrm{span}\{|\psi_{ij}\rangle\}\subseteq\mathrm{supp}(C)$ and $\mathrm{span}\{|\tilde{\psi}_{ij}\rangle\}=\mathrm{supp}(\tilde{C})$, it suffices to prove that $\mathrm{supp}(C)\subseteq\mathrm{supp}(\tilde{C})$, which we do in Lemma \ref{lemma:supp_lemma} below. Besides $R=
ABC$, we may also perform a similar argument for $R=AB,BC$, and $B.$  

We will be considering a quantum channel $\mathcal{T}$, which is an automorphism of the linear space $\mathcal{B}(\mathcal{K})$, i.e.,  $\mathcal{T}:\mathcal{B}(\mathcal{K})\rightarrow\mathcal{B}(\mathcal{K})$, where $\mathcal{K}$ is a finite-dimensional Hilbert space of dimension $d:=\dim\mathcal{K}$. The Lemma \ref{lemma:supp_lemma} is valid for any quantum channel of this type. For a quantum channel $\mathcal{T}$ one can define its peripheral part $\tilde{\mathcal{T}}$ as follows. The linear space $\mathcal{B}(\mathcal{K})$ may be considered a Hilbert space with respect to a Hilbert-Schmidt inner product. We may choose some basis $\{|i\rangle\}$ in $\mathcal{K}$ and vectorize $\mathcal{B}(\mathcal{K})$ through the usual process of identifying $|i\rangle\langle j| \leftrightarrow |i\rangle\otimes|j\rangle$.\cite{watrouse} In the vectorized space $\mathcal{K}\otimes \mathcal{K}$, the quantum channel $\mathcal{T}$ will be represented by a (a generally non-Hermitian) matrix $T:\mathcal{K}\otimes \mathcal{K}\rightarrow \mathcal{K}\otimes \mathcal{K}$. For the matrix $T$ we may perform the Jordan decomposition
\begin{equation*}
    T=\sum_{k=1}^{k_{\max}} (\nu_k D_k + N_k),
\end{equation*}
where $D_k$ are projections and $N_k$ are nilpotent matrices, satisfying $D_kD_{k'}=\delta_{kk'}D_k$ and $D_kN_{k'}=N_{k'}D_k=\delta_{kk'}N_k$. Moreover, since $\mathcal{T}$ is a quantum channel, then $|\nu_k|\leq 1$ for $k=1,\dots, d^2$ and $N_k=0$ when $|\nu_k|=1$.\cite{fcs} Then we may separate the part of $T$, denoted by $\tilde{T}$, that corresponds to the peripheral spectrum of $T$ by defining 
\begin{equation*}
    \tilde{T}:=\sum_{k:\,|\nu_k|=1} (\nu_k D_k + N_k).
\end{equation*}
The matrix $\tilde{T}$ corresponds to some map $\tilde{\mathcal{T}}$ in the space $\mathcal{B}(\mathcal{K})$, which can be shown to be a quantum channel (see, for example, Lemma III.2 in Ref. \onlinecite{qcmi-decay}).

\begin{lemma}\label{lemma:supp_lemma}
    Consider a quantum channel $\mathcal{T}:\mathcal{B}(\mathcal{K})\rightarrow\mathcal{B}(\mathcal{K})$ and a quantum channel $\tilde{\mathcal{T}}:\mathcal{B}(\mathcal{K})\rightarrow\mathcal{B}(\mathcal{K})$, which is a peripheral part of $\mathcal{T}$ (as defined above). Let $C$ and $\tilde{C}$ be Choi matrices for $\mathcal{T}$ and $\tilde{\mathcal{T}}$, respectively. Then,
    \begin{equation*}
        \mathrm{ker}(\tilde{C})\subseteq\mathrm{ker}(C),
    \end{equation*}
    or, equivalently,
    \begin{equation}\label{eqn:supp_c_in_tc}
        \mathrm{supp}(C)\subseteq\mathrm{supp}(\tilde{C}).
    \end{equation}    
\end{lemma}
\begin{proof}

To start the proof we notice that, since $C=\tilde{C}+\Delta C$, it suffices to show that $\mathrm{supp}(\Delta C)\subseteq \mathrm{supp}(\tilde{C})$, or $\mathrm{ker}(\tilde{C})\subseteq\mathrm{ker}(\Delta C)$. We will prove the latter statement by obtaining a matrix inequality of the form
\begin{equation*}
    a_1\tilde{C} \leq \Delta C \leq a_2\tilde{C},
\end{equation*}
where $a_1,a_2\in\mathbb{R}$ are some constants. 

Since $C$ is a Choi matrix, it is positive, and we have immediately
\begin{equation}\label{eqn:delta_c_ineq1}
    \Delta C \geq -\tilde{C}.    
\end{equation} 

Now recall that
\begin{equation}
    C=\left(\mathrm{id}\otimes\mathcal{T}\right)\left(|+\rangle\langle +|\right),
\end{equation}
where $|+\rangle=\sum_{n=1}^d |n\rangle\otimes|n\rangle$ is a maximally entangled vector on $\mathcal{K}$, with $d:=\dim\mathcal{K}$ and $\{|n\rangle\}_{n=1}^d$ being some orthonormal basis on $\mathcal{K}$.
The state $|+\rangle\langle +|$ is defined on the space $\mathcal{B}(\mathcal{K})\otimes\mathcal{B}(\mathcal{K})$; each factor $\mathcal{B}(\mathcal{K})$ may be considered a Hilbert space with respect to Hilbert-Schmidt inner product. On each factor $\mathcal{B}(\mathcal{K})$ we introduce a so-called Schwinger basis of unitary matrices $Y_k$, with $k=1,2,\dots, d^2$, which satisfy \cite{schwinger}
\begin{align}
    \mathrm{tr}\left(Y^\dagger_{k'} Y_k\right)&=d \delta_{kk'}, \\ \nonumber
    Y_1&=\mathbb{1}_{d}.
\end{align}
One may think of the Schwinger basis as a generalization of the Pauli matrix basis for $\mathbb{C}^{2\times 2}$. 
It immediately follows that
\begin{equation}
    \mathrm{tr}\left( Y_k\right)=0,\text{ for } k=2,3,\dots,d^2.
\end{equation}
The state $|+\rangle\langle+|$ may then be expressed as \cite{Siewert_2022}
\begin{equation}
    |+\rangle\langle+|=\frac{1
}{d}\sum_{k=1}^{d^2} Y_k\otimes \bar{Y}_k.
\end{equation}
Then,
\begin{equation}
    C=\frac{1
}{d}\mathbb{1}_{d}\otimes\mathcal{T}(\mathbb{1}_{d})+\frac{1
}{d}\sum_{k=2}^{d}  Y_k\otimes\mathcal{T}(\bar{Y}_k),
\end{equation}
where we have explicitly separated the term corresponding to $k=1$. 

From Propositions 3.2 and 3.3 of Ref. \onlinecite{fcs} and Appendix I of Ref. \onlinecite{qcmi-decay} it follows that $Y_1=\mathbb{1}_{d}$ may be decomposed as $\mathbb{1}_{d}=\sum_{\alpha=1}^N\sum_{k_\alpha=1}^{p_\alpha}\Pi^\alpha_{k_\alpha}$ (in the notation introduced in Section \ref{subsec:generalization}), so this family of projectors $\Pi^\alpha_{k_\alpha}$ spans the subspace of $\mathcal{B}(\mathcal{K})$ corresponding to the peripheral spectrum of $\mathcal{T}$. This implies that $\mathcal{T}(\Pi^\alpha_{k_\alpha})=\tilde{\mathcal{T}}(\Pi^\alpha_{k_\alpha})$, hence $\mathcal{T}(\mathbb{1}_{d})=\tilde{\mathcal{T}}(\mathbb{1}_{d})$, allowing us to express $\tilde{C}$ and $\Delta C$ in terms of the Schwinger basis,
\begin{align}
    \tilde{C}&=\frac{1
}{d}\mathbb{1}_{d}\otimes\tilde{\mathcal{T}}(\mathbb{1}_{d})=\frac{1
}{d}\mathbb{1}_{d}\otimes\mathcal{T}(\mathbb{1}_{d}), \\ \nonumber
\Delta C &= \frac{1}{d}\sum_{k=2}^{d_M^2}Y_k\otimes\mathcal{T}(\bar{Y}_k).
\end{align}

We define a completely positive map $\mathcal{N}:\mathcal{B}(\mathcal{K})\rightarrow\mathcal{B}(\mathcal{K}
)$ by
\begin{equation}
    \mathcal{N}(X):=\sum_{k=2}^{d^2} Y_k X Y_k^\dagger.
\end{equation}
Using the relation \cite{schwinger} 
\begin{equation}\label{eqn:tr_ch_decomp}
    \mathrm{tr}(X)\mathbb{1}_{d}=\frac{1}{d}\sum_{k=1}^{d^2}Y_k X Y^\dagger_k,
\end{equation}
which is (up to normalization) a Kraus representation of the completely mixing quantum channel, we infer that
\begin{equation}
    \mathcal{N}(X)=d\mathrm{tr}(X)\mathbb{1}_{d}-X.
\end{equation}
Now consider action of $\mathcal{N}$ on each of the Schwinger matrices $Y_k$, recalling that $Y_1=\mathbb{1}_{d}$ and $\mathrm{tr}(Y_n)=0$ for $n=2,\dots,d^2$,
\begin{align}
    \mathcal{N}(Y_1)&=\sum_{k=2}^{d^2}Y_k Y_1 Y^\dagger_k=(d^2-1)Y_1, \\ \nonumber
    \mathcal{N}(Y_n)&=\sum_{k=2}^{d^2}Y_k Y_n Y^\dagger_k=-Y_n,\text{ for }n=2,\dots,d^2.
\end{align}
Since the map $\mathcal{N}$ is completely positive, the map $\mathcal{N}\otimes\mathrm{id}:\mathcal{B}(\mathcal{K})\otimes\mathcal{B}(\mathcal{K})\rightarrow\mathcal{B}(\mathcal{K}
)\otimes\mathcal{B}(\mathcal{K})$ is positive, and its action on the Choi matrix $C$ preserves its positivity, leading to the inequality
\begin{align}\label{eqn:delta_c_ineq2}
    0 & \leq (\mathcal{N}\otimes\mathrm{id})(C) \\ \nonumber
    &=\frac{1}{d}\sum_{k=1}^{d_M^2}\mathcal{N}(Y_k)\otimes\mathcal{T}(\bar{Y}_k)\\ \nonumber
    &=(d^2-1)\frac{1}{d} \mathbb{1}_{d}\otimes\mathcal{T}(\mathbb{1}_{d})-\frac{1}{d}\sum_{k=2}^{d^2}Y_k\otimes\mathcal{T}(\bar{Y}_k)\\ \nonumber
    &=(d^2-1)\tilde{C}-\Delta C.
\end{align}
Combining (\ref{eqn:delta_c_ineq1}) and (\ref{eqn:delta_c_ineq2}), we obtain
\begin{equation}
    -\tilde{C} \leq \Delta C \leq (d^2-1)\tilde{C}.
\end{equation}
For any $|\phi\rangle\in\mathrm{ker}(\tilde{C})$ we observe
\begin{equation}
    0 = -\langle\phi|\tilde{C}|\phi\rangle \leq \langle\phi|\Delta{C}|\phi\rangle \leq (d^2-1)\langle\phi|\tilde{C}|\phi\rangle = 0.
\end{equation}
The result $\langle \phi| \Delta C| \phi \rangle = 0,$ and the polarization identity applied to the complex subspace $\mathrm{ker}(\tilde{C}),$ implies that $\Delta C$ is the zero operator on $\mathrm{ker}(\Delta{C})$. Hence,
\begin{equation}
    \mathrm{ker}(\tilde{C}) \subseteq \mathrm{ker}(\Delta{C}),
\end{equation}
completing the proof.
\end{proof}

\noindent\textbf{Remark:}
For the quantum channel $\mathcal{T}=\mathcal{E}^{|B|}$ a stronger result holds,
\begin{equation}\label{eqn:supp_equal}
    \mathrm{supp}(C)=\mathrm{supp}(\tilde{C}).
\end{equation}
Indeed, Lemma \ref{lemma:supp_lemma} validates the decomposition (\ref{eqn:psi_decomp}), 
\begin{align*}
    |\psi_{ij}\rangle&=\left(\mathbb{1}_{M\otimes M}+\Gamma\right)|\tilde{\psi}_{ij}\rangle, 
\end{align*}
for any $(i,j)\in\mathcal{O}$, where $\|\Gamma\| \rightarrow 0$ as $|B| \rightarrow 0$, and vectors $|\tilde{\psi}_{ij}\rangle$ are mutually orthogonal. The equations (\ref{eqn:c_tilde_decomp_copy}), (\ref{eqn:c_decomp_copy}), and (\ref{eqn:psi_decomp}) together imply that for large enough $|B|$ the matrix $\mathbb{1}_{M\otimes M}+\Gamma$ is full-rank and invertible, hence for the quantum channel $\mathcal{E}^{|B|}$,  
\begin{equation}
    \mathrm{supp}(\tilde{C})=\mathrm{span}\{|\tilde{\psi}_{ij}\rangle\}=\mathrm{span}\{|\psi_{ij}\rangle\}\subseteq\mathrm{supp}(C).
\end{equation}
Combined with (\ref{eqn:supp_c_in_tc}), this implies (\ref{eqn:supp_equal}).

\section{Optimality of the decay rate}\label{appendix:tight-bound}
In this appendix we show that the decay rate $q=-2\ln\nu_{\mathrm{gap}}$ cannot be improved (at least without additional assumptions imposed on iuMPS). We prove this statement by providing an example of an injective iuMPS for which QCMI may be explicitly evaluated to be $I(A:C|B)=O\left(e^{-q|B|}\right)$. (Recall that for an injective iuMPS $K=0$ in Theorem \ref{thrm:theorem_1} and the polynomial factor in the bound for QCMI is absent.) 

For our example we set $\mathcal{H}_s=\mathbb{C}^8$ and $\mathcal{H}_M=\mathbb{C}^2$. We will treat $\mathcal{H}_s=\mathbb{C}^8$ as $\mathbb{C}^4\oplus\mathbb{C}^4$, and set orthonormal bases $\{|\xi_{ij}\rangle\}_{i,j=1}^2$ and $\{|\zeta_{ij}\rangle\}_{i,j=1}^2$ in the first and the second copies of $\mathbb{C}^4$, respectively. 
We denote Pauli matrices as $\sigma_x$, $\sigma_y$, and $\sigma_z$, and the eigenvalues of $\sigma_z$ as $\sigma_{z,1}:=1$ and $\sigma_{z,2}:=-1$. We set $\sigma=\frac{\mathbb{1}}{2}$. We also set some constant $0\leq \kappa < 1$. 

We construct the operator  
\begin{equation}\label{eqn:isometry-example}
    V=\sum_{i,j=1}^{2}\left(\sqrt{\sigma_i}|\xi_{ij}\rangle+\sqrt{\frac{\kappa}{2}\sigma_{z,i}\sigma_{z,j}}|\zeta_{ij}\rangle\right)\otimes|i\rangle\langle j|,
\end{equation}
which is an isometry,
\begin{align*}
    V^\dagger V &= \sum_{i,j,j'=1}^{2}\left(\sqrt{\sigma_i}\langle\xi_{ij'}|+\sqrt{\frac{\kappa}{2}\sigma_{z,i}\sigma_{z,j'}}\langle\zeta_{ij'}|\right)\\ \nonumber &\qquad\times\left(\sqrt{\sigma_i}|\xi_{ij}\rangle+\sqrt{\frac{\kappa}{2}\sigma_{z,i}\sigma_{z,j}}|\zeta_{ij}\rangle\right)\otimes|j'\rangle\langle j|\\ \nonumber
    &= \sum_{i,j=1}^{2}\left(\sigma_i+\frac{\kappa}{2}\sigma_{z,i}\sigma_{z,j}\right)|j\rangle\langle j| \\ \nonumber
    &= \sum_{j=1}^{2} |j\rangle\langle j| \\ \nonumber
    &= \mathbb{1}.
\end{align*}
We choose the constructed isometry (\ref{eqn:isometry-example}) to define the iuMPS in our example.

The isometry (\ref{eqn:isometry-example}) induces the quantum channel $\mathcal{E}:\mathcal{B}(\mathbb{C}^2)\rightarrow\mathcal{B}(\mathbb{C}^2)$, defined by
\begin{align*}
    \mathcal{E}(X) &= \mathrm{tr}_s(VXV^\dagger)\\ \nonumber
    &= \sum_{i,j=1}^{2} \sigma_i \langle j|X|j\rangle |i\rangle\langle i| + \sum_{i,j=1}^{2} \frac{\kappa}{2}\sigma_{z,i}\langle j|X|j\rangle\sigma_{z,j}|i\rangle\langle i|\\ \nonumber
    &=\mathrm{tr}(X)\sigma + \frac{\kappa}{2}\mathrm{tr}(\sigma_z X)\sigma_z,
\end{align*}
with eigenvectors and eigenvalues 
\begin{align*}
\mathcal{E}(\sigma) &=\sigma, \\ \nonumber
\mathcal{E}(\sigma_x) &=0, \\ \nonumber
\mathcal{E}(\sigma_y) &=0, \\ \nonumber
\mathcal{E}(\sigma_z) &=\kappa\sigma_z.
\end{align*}
It follows that for $\mathcal{E}$, the density operator $\sigma=\frac{\mathbb{1}}{2}$ is the fixed point and $\nu_{\mathrm{gap}}=\kappa$. 

Further we will need the expression for the $n$-fold composition of $\mathcal{E}$,
\begin{equation}\label{eqn:exmp_n_fold}
    \mathcal{E}^{n}(X)=\mathrm{tr}(X)\sigma + \frac{\kappa^n}{2}\mathrm{tr}(\sigma_z X)\sigma_z.
\end{equation}

To calculate the entropies in the expression for $I(A:C|B)$, we will use the relation (\ref{eqn:rho_r_isom}), from which it follows that
\begin{equation*}
    S(\rho_R)=S\left((\mathcal{E}^{|R|}\otimes \mathrm{id}_{M'})(|\sqrt{\sigma}\rangle\langle\sqrt{\sigma}|))\right),
\end{equation*}
where $|\sqrt{\sigma}\rangle=\sum_{i=1}^{2}\sqrt{\sigma_i}|i\rangle\otimes|i\rangle\in\mathcal{H}_M\otimes\mathcal{H}_{M'}$. For the sake of convenience, we denote
\begin{equation*}
    \rho'_R:=(\mathcal{E}^{|R|}\otimes \mathrm{id}_{M'})(|\sqrt{\sigma}\rangle\langle\sqrt{\sigma}|).
\end{equation*}
Using (\ref{eqn:exmp_n_fold}), we may calculate
\begin{widetext}
\begin{align*}
    \rho'_R &=(\mathcal{E}^{|R|}\otimes \mathrm{id}_{M'})(|\sqrt{\sigma}\rangle\langle\sqrt{\sigma}|) \\ \nonumber &= \left(\mathcal{E}^{|R|}\otimes \mathrm{id}_{M'}\right)\left(\sum_{i,j=1}^{2}\sqrt{\sigma_i\sigma_j}|i\rangle\langle j| \otimes |i\rangle\langle j|\right) \\ \nonumber
    &= \sigma\otimes\sum_{i,j=1}^{2}\sqrt{\sigma_i\sigma_j}\delta_{ij}|i\rangle\langle j|+\frac{\kappa^{|R|}}{2}\sigma_z\otimes\sum_{i,j=1}^{2}\sqrt{\sigma_i\sigma_j}\langle i|\sigma_z|j\rangle |i\rangle\langle j| \\ \nonumber
    &= \frac{1}{4}\left(\mathbb{1}\otimes\mathbb{1}+\kappa^{|R|}\sigma_z \otimes \sigma_z\right),
\end{align*}
\end{widetext}
where we have used $\sigma_1=\sigma_2=1/2$ to get to the last line. Since both matrices $\mathbb{1}\otimes\mathbb{1}$ and $\sigma_z \otimes \sigma_z$ are diagonal and ${(\sigma_z \otimes \sigma_z)^{2n}=\mathbb{1}\otimes\mathbb{1}}$, ${(\sigma_z \otimes \sigma_z)^{2n+1}=\sigma_z \otimes \sigma_z}$ for $n\in\mathbb{N}$, then, using Taylor expansion, we may explicitly calculate
\begin{widetext}
\begin{align*}
    \ln \rho'_R &=\ln\left(\frac{1}{4}\left(\mathbb{1}\otimes\mathbb{1}+\kappa^{|R|}\sigma_z \otimes \sigma_z\right)\right)\\ \nonumber
    &=\ln\left(\frac{1}{4}\mathbb{1}\otimes\mathbb{1}\right)+\ln\left(\mathbb{1}\otimes\mathbb{1}+\kappa^{|R|}\sigma_z\otimes\sigma_z\right) \\ \nonumber
    &= -2\ln2 \mathbb{1}\otimes\mathbb{1} + \frac{1}{2}\left(\ln(1+\kappa^{|R|})+\ln(1-\kappa^{|R|})\right) \mathbb{1}\otimes\mathbb{1} +\frac{1}{2}\left(\ln(1+\kappa^{|R|})-\ln(1-\kappa^{|R|})\right)\sigma_z\otimes\sigma_z.
\end{align*}
\end{widetext}
This allows us to obtain an explicit expression for $S(\rho_R)$,
\begin{widetext}
\begin{align*}
    S(\rho_R) &= S(\rho'_R) \\ \nonumber
    &=-\mathrm{tr}(\rho'_R\ln\rho'_R) \\ \nonumber
    &= \mathrm{tr}\left(\frac{1}{4}\left(\mathbb{1}\otimes\mathbb{1}+\kappa^{|R|}\sigma_z \otimes \sigma_z\right)\right. \\ \nonumber
    &\qquad\times\left.\left(2\ln2 \mathbb{1}\otimes\mathbb{1} - \frac{1}{2}\left(\ln(1+\kappa^{|R|})+\ln(1-\kappa^{|R|})\right) \mathbb{1}\otimes\mathbb{1}- \frac{1}{2}\left(\ln(1+\kappa^{|R|})-\ln(1-\kappa^{|R|})\right)\sigma_z\otimes\sigma_z\right)\right)\\ \nonumber
    &= 2\ln 2 - \frac{1}{2}\left(\ln(1+\kappa^{|R|})+\ln(1-\kappa^{|R|})\right) - \frac{\kappa^{|R|}}{2}\left(\ln(1+\kappa^{|R|})-\ln(1-\kappa^{|R|})\right). 
\end{align*}
\end{widetext}
Technically, we could use the above expression to calculate QCMI, but it is more convenient to make the Taylor expansion of $S(\rho_R)$ in powers of $\kappa$,
\begin{equation*}
    S(\rho_R)=2\ln 2 - \frac{\kappa^{2|R|}}{2} + O(\kappa^{4|R|}).
\end{equation*}
We may then calculate QCMI,
\begin{widetext}
\begin{align*}
    I(A:C|B) &= -\frac{\kappa^{2(|A|+|B|)}}{2}+O(\kappa^{4(|A|+|B|)})-\frac{\kappa^{2(|B|+|C|)}}{2}+O(\kappa^{4(|B|+|C|)})+\frac{\kappa^{2(|A|+|B |+|C|)}}{2}+O(\kappa^{4(|A|+|B |+|C|)})+\frac{\kappa^{2|B|}}{2} + O(\kappa^{4|B|}) \\ \nonumber
     &=  \frac{(1-\kappa^{2|A|})(1-\kappa^{2|B|})}{2}e^{2\ln\kappa|R|}+ O(\kappa^{4|B|}).
\end{align*}
\end{widetext}
Thus, for the chosen example of iuMPS, QCMI cannot be bounded by any function $Qe^{-q|B|}$ from the statement of Theorem \ref{thrm:theorem_1} (recall that $K=0$ for the considered iuMPS) with $q>-2\ln\nu_{\mathrm{gap}}$.

\bibliography{bibliography}

\end{document}